\documentclass[aps, prd, superscriptaddress, nofootinbib, amsmath, floatfix, notitlepage, longbibliography]{revtex4-2}
\usepackage[colorlinks=true, linkcolor=blue, citecolor=blue, urlcolor=blue]{hyperref}
\usepackage{xcolor}
\colorlet{ORANGE}{orange}
\usepackage{graphicx}
\usepackage{subfigure}
\usepackage{multirow}
\usepackage{slashed}
\usepackage{physics}
\usepackage[normalem]{ulem}
\usepackage{lineno}
\usepackage{subcaption}
\usepackage{verbatim}
\usepackage{todonotes}

\newcommand{\dhd}{{\textstyle d} \lower.03ex\hbox{\kern-0.38em$^{\scriptstyle-}$}\kern-0.05em{}}

\begin{document}
\title{Quark TMDPDFs from Large to Small $x$: A Unified Description of CSS, DGLAP, and Small-$x$ Double Logarithms}

\author{Swagato Mukherjee}
\affiliation{Physics Division, Brookhaven National Laboratory, Upton, New York 11973, USA}
\author{Vladimir V. Skokov}
\affiliation{Department of Physics and Astronomy, North Carolina State University, Raleigh, NC 27695, USA}
\author{Andrey Tarasov}
\email{ataraso@ncsu.edu}
\affiliation{Department of Physics and Astronomy, North Carolina State University, Raleigh, NC 27695, USA}
\affiliation{Center for Frontiers in Nuclear Science (CFNS) at Stony Brook University, Stony Brook, NY 11794, USA}
\author{Shaswat Tiwari}
\email{stiwari1@bnl.gov}
\affiliation{Physics Division, Brookhaven National Laboratory, Upton, New York 11973, USA}

\begin{abstract}
  We extend the full-kinematic background-field construction developed for gluon transverse-momentum dependent parton distribution functions (TMDPDFs) in Ref.~\cite{Mukherjee:2023snp} to quark TMDPDFs, providing a common framework for their large- and small-$x$ dynamics. We compute the leading-power quark TMD beam function at next-to-leading order using the background-field method. The background fields retain both longitudinal and transverse momenta, with neither collinear nor eikonal ordering imposed. We distinguish the renormalization scales that define the TMD operator from the factorization scales that separate its perturbative and non-perturbative modes. With standard TMD renormalization, the quark TMDPDF obeys Collins–Soper–Sterman (CSS) evolution, while the lower-scale separation is organized directly in terms of background-field TMD operators. We refer to this full-kinematic construction as the MSTT scheme.
  In the MSTT scheme, the leading term of the resulting operator expansion reproduces both the Dokshitzer-Gribov-Lipatov-Altarelli-Parisi (DGLAP) logarithm and the small-$x$ double logarithm in the appropriate limits, while retaining parts of the higher-twist and sub-eikonal corrections. We also formulate a rapidity-only scheme for quark TMDPDFs with distinct renormalization and factorization cutoffs. We derive its evolution equation, which reduces to DGLAP evolution in the collinear limit and to double-logarithmic quark evolution at small $x$, and establish its matching to the MSTT and standard TMD schemes.

\end{abstract}
\maketitle
\tableofcontents
\section{Introduction}
Quantum chromodynamics (QCD) factorization organizes high-energy scattering by separating dynamical modes associated with different kinematic scales. The short-distance dynamics is encoded in perturbative coefficient functions, while long-distance dynamics is represented by hadronic matrix elements of QCD operators. Schematically,
\begin{eqnarray}
  \mathcal{X} \propto \sum_k H_k(\sigma) \otimes \langle P_f|\mathcal{O}_k(\sigma)|P_i\rangle\,.
  \label{eq:cs-init-fact}
\end{eqnarray}
Here $\sigma$ denotes the upper renormalization-scale set; in the mode decomposition it also marks the boundary between the modes assigned to the hard coefficient $H_k(\sigma)$ and the quantum modes entering the renormalized operator $\mathcal{O}_k(\sigma)$. A second, lower factorization-scale set $\sigma'$ separates the perturbative quantum modes from the non-perturbative background modes. This distinction between the two mode separations, and between their associated scales, will be central below~\cite{Collins:2011zzd}.

The need for such a two-stage description is especially clear when one seeks a common treatment of large- and small-$x$ QCD. These limits are controlled by different orderings of parton emissions. At large $x$, transverse-momentum ordering leads to the Dokshitzer-Gribov-Lipatov-Altarelli-Parisi (DGLAP) evolution~\cite{Dokshitzer:1977sg,Gribov:1972ri,Altarelli:1977zs}. At small $x$, longitudinal or rapidity ordering leads to Balitsky-Fadin-Kuraev-Lipatov (BFKL) evolution~\cite{Fadin:1975cb,Kuraev:1976ge,Kuraev:1977fs,Balitsky:1978ic}, its nonlinear Balitsky-Kovchegov (BK) extension~\cite{Balitsky:1995ub,Balitsky:1997mk,Kovchegov:1999yj}, and, more generally, the Jalilian-Marian-Iancu-McLerran-Weigert-Leonidov-Kovner (JIMWLK) functional evolution~\cite{Jalilian-Marian:1997qno,Jalilian-Marian:1997ubg,Kovner:2000pt,Iancu:2000hn,Ferreiro:2001qy,Kovner:2013ona,Kovner:2014lca,Lublinsky:2016meo}. A unified description must interpolate between these distinct expansions without assuming either ordering from the outset.

This problem is directly relevant to the kinematic range of the future Electron-Ion Collider (EIC). A substantial part of the phenomenologically relevant EIC phase space lies away from the asymptotic limits in which either a strict eikonal approximation or a leading-twist collinear expansion is parametrically guaranteed to dominate. Comparative studies of Color Glass Condensate and collinear descriptions in EIC kinematics find sensitivity to the longitudinal momentum prescription and to beyond-eikonal effects~\cite{Mukherjee:2026yqe}. Arbitrary-$x$ operator analyses of dijet production likewise show that higher-twist, sub-eikonal, and transverse-momentum effects are needed to connect the eikonal and TMD regimes~\cite{Mukherjee:2026cte,Mukherjee:2026six,Kar:2026cnx}. These observations motivate a leading-power TMD formulation in which the background fields retain both longitudinal and transverse momenta, without imposing collinear or eikonal ordering.
The interplay between large- and small-$x$ dynamics has been studied in back-to-back dijet production and Sudakov broadening~\cite{Caucal:2021ent,Caucal:2022ulg,Caucal:2023nci,Caucal:2023fsf,Taels:2022tza,delCastillo:2021znl}, searches for saturation at moderate $x$, sub-eikonal corrections~\cite{Balitsky:2015qba,Altinoluk:2015gia,Balitsky:2016dgz,Agostini:2019hkj,Kovchegov:2021iyc,Chirilli:2021lif,Altinoluk:2022jkk}, and next-to-leading order small-$x$ evolution~\cite{Balitsky:2007feb,Balitsky:2013fea,Kovner:2013ona,Lappi:2015fma,Beuf:2014uia,Iancu:2015vea,Ducloue:2019ezk}. It is also central to small-$x$ spin physics, where the Color Glass Condensate must be extended beyond leading eikonal accuracy to incorporate large-$x$ dynamics~\cite{Kovchegov:2015pbl,Kovchegov:2016zex,Kovchegov:2021lvz,Cougoulic:2022gbk,Borden:2024bxa,Adamiak:2023yhz,JAMCollaborationSmall-xAnalysisGroup:2025tfa,McLerran:1993ni,McLerran:1993ka,McLerran:1994vd}.
Transverse-momentum dependent parton distribution functions (TMDPDFs) provide a natural operator framework for this problem because they retain both the longitudinal momentum fraction $x$ and the transverse separation $b_\perp$, the variable conjugate to the measured transverse momentum $q_\perp$. Their relation to small-$x$ dynamics and to gluon distributions has been explored in many studies~\cite{Dominguez:2011wm,Dominguez:2011br,Mueller:2012uf,Mueller:2013wwa,Boussarie:2020fpb,Boussarie:2021wkn,Boussarie:2021ybe,Boussarie:2023xun,Altinoluk:2024zom,Caucal:2025xxh,Metz:2011wb,Balitsky:2015qba,Balitsky:2016dgz,Zhou:2016tfe,Xiao:2017yya,Hentschinski:2021lsh,Altinoluk:2025ewj,Caucal:2024bae}. In the small-$b_\perp$ region, TMDPDFs can be matched onto collinear operators. At leading power, this gives the standard collinear matching expansion
\begin{eqnarray}
  &&f(x, b_\perp; \mu, \zeta) = C_1(x, L^{\mu'}_b, \mu, \zeta)\otimes f_1(x, \mu') + b^2_\perp C_2(x, L^{\mu'}_b, \mu, \zeta)\otimes f_2(x, \mu') + \dots\,,
  \label{eq:coll-exp}
\end{eqnarray}
where the upper scale set $\sigma=(\mu^2,\zeta)$ governs standard Collins-Soper-Sterman (CSS) evolution, while $\mu'$ is the relevant lower collinear factorization scale. When $\mu=\mu'$, Eq.~(\ref{eq:coll-exp}) reduces to the usual collinear matching formula~\cite{Collins:2011zzd,Boussarie:2023izj,Kang:2012em,Sun:2013hua,Scimemi:2019gge}. In the small-$b_\perp$ region its leading term is often written as
\begin{eqnarray}
  &&f(x, b_\perp; \mu, \zeta) = C_1(x, L^{\mu}_b, \mu, \zeta)\otimes f_1(x, \mu)~f_{NP}(x,b_\perp)\,,
  \label{eq:coll-exp-NP}
\end{eqnarray}
with higher-twist terms absorbed into the non-perturbative function $f_{NP}$. This form underlies much of TMD phenomenology and has also been used to constrain collinear PDFs~\cite{Barry:2025glq,Portela:2026wwn}.

The separate collinear and eikonal expansions are not uniformly adequate over the full TMD region. TMD factorization applies at small measured transverse momentum~\cite{Balitsky:2017flc,Balitsky:2017gis,Vladimirov:2023aot,Piloneta:2025jjb}, corresponding to transverse separations approaching the non-perturbative scale, $b_\perp\lesssim\Lambda_{\mathrm{QCD}}^{-1}$. The contributions from higher collinear twists there can be comparable to the leading term, and the sensitivity of observables to the functional form of $f_{NP}$ indicates that higher-twist dynamics is already relevant~\cite{Bacchetta:2025ara,Moos:2023yfa,Yang:2024drd,Granger:2026prq}. Conversely, in the small-$x$ regime no transverse-momentum ordering is imposed, and the eikonal expansion contains contributions from all collinear twists. Reconstructing the full TMDPDF from either expansion would therefore require resumming an infinite set of terms.

The construction considered here remains leading power in TMD factorization. The leading-power TMD operator is not expanded in $b_\perp$ or in the small-$x$ eikonal limit. Its finite-$b_\perp$ and arbitrary-$x$ dependence therefore contains terms that, when re-expanded in the collinear limit, would be identified as higher-twist collinear contributions, and when re-expanded in the high-energy $x\to0$ limit, would be identified as sub-eikonal corrections. We retain the portions of these contributions generated by the leading-power TMD operator, rather than constructing the complete evolution of independent higher-twist or sub-eikonal operators.

The background-field method provides the appropriate construction for this purpose. It implements the separation of QCD modes directly at the level of the Lagrangian by integrating out perturbative modes in the presence of background fields~\cite{Abbott:1980hw,Abbott:1981ke,Balitsky:2025bup}. Because the background fields need not be assigned a collinear or eikonal ordering, this method can retain their longitudinal and transverse momenta simultaneously. It has been applied to collinear factorization in the Bjorken limit $Q^2\to\infty$~\cite{Balitsky:1987bk}, high-energy rapidity factorization at small $x$~\cite{Balitsky:1995ub}, and TMD factorization for processes with small measured transverse momentum~\cite{Balitsky:2017flc,Balitsky:2017gis,Collins:2011zzd,Boussarie:2023izj}.

In the present formulation, the upper scale set $\sigma$ defines the boundary between the modes assigned to the hard coefficient and the quantum modes entering the renormalized TMD operator. The lower scale set $\sigma'$ then separates the perturbative quantum modes from the non-perturbative background modes. Integrating over the perturbative modes between these two scale sets gives the operator-level relation
\begin{eqnarray}
  && \langle P_f| \mathcal{O}_k(\sigma) |P_i\rangle = \sum_m C_{km}(\sigma, \sigma') \otimes \langle P_f| \tilde{\mathcal{O}}_m(\sigma') |P_i\rangle \,.
  \label{eq:fact-op}
\end{eqnarray}
Here $\mathcal{O}_k(\sigma)$ is defined and renormalized with respect to the upper scale set $\sigma$, while the background-field operator $\widetilde{\mathcal{O}}_m(\sigma')$ is defined at the lower scale set $\sigma'$ and renormalized there when required. Thus, $\sigma$ specifies the hard-quantum matching boundary, whereas $\sigma'$ specifies both the quantum-background separation and the renormalization of the background-field operator.

Equation~(\ref{eq:fact-op}) contains information beyond the evolution equations associated with the upper scales. The latter follow by differentiating with respect to $\sigma$, whereas the coefficient functions also depend on how the perturbative and background modes are separated at $\sigma'$. Since the upper and lower schemes can be chosen independently, this freedom can be used to expose different perturbative limits of the same initial operator.

We apply this construction to the leading-power quark TMD beam function at next-to-leading order. For the upper scale $\sigma$, the standard TMD scheme uses $\sigma=(\mu^2,\zeta)$ and is governed by CSS evolution. For the lower scale $\sigma'$, we use the background-field operator construction introduced for gluon TMDPDFs in Ref.~\cite{Mukherjee:2023snp}, denoted the MSTT scheme. Its operators are built from background fields whose longitudinal and transverse momenta are retained, without collinear or eikonal ordering. The lower scales are $\sigma'=(\mu',\rho)$, and the background fields are expanded directly in leading-power TMD operators rather than in collinear operators or operators of fixed eikonality. After the upper-scale renormalization, the finite remainder of the beam function is assigned to coefficient functions, while divergences associated with the lower-scale separation are absorbed into the background-field operators. For quark TMD operators at NLO this latter renormalization is trivial because the unexpanded result has no genuine infrared divergence, unlike the gluon case with its infrared rapidity divergence and associated BFKL dynamics~\cite{Mukherjee:2023snp}.

The MSTT prescription is not the only possible choice for the quantum-background separation at the lower scales. In the two-scale descriptions of the collinear and high-energy limits, transverse-momentum and rapidity dependence are controlled by separate parameters, $\mu'$ and $\rho$, in analogy with the upper-scale parameters $\mu$ and $\zeta$ in the CSS scheme. This motivates asking whether the lower-scale separation can instead be defined with a single longitudinal cutoff while retaining the large-$x$ and small-$x$ evolution associated with the two limiting descriptions.
This question is motivated by the rapidity-only scheme developed for gluon TMDPDFs in Refs.~\cite{Balitsky:2015qba,Balitsky:2016dgz}. There, a single longitudinal cutoff replaces the separate transverse-momentum and rapidity regulators, and the resulting evolution exhibits the large-$x$ DGLAP-like and small-$x$ BFKL/BK-type dynamics of gluon TMDPDFs. In that construction, however, the same scale controls the complete evolution, so no independent separation is made between the upper renormalization of the TMD operator and the lower quantum-background factorization.

Here we implement the rapidity-only scheme within the two-scale framework. At the upper scales $\sigma$, a single longitudinal cutoff regulates the ultraviolet divergences associated with the transverse-momentum and rapidity dependence of the TMD operator, providing a one-scale counterpart of CSS evolution. We also compute the soft function in this upper-rapidity cutoff scheme and show that, at NLO, its contribution is already contained in the renormalized beam function, eliminating the need for an additional soft factor. At the lower scales $\sigma'$, the upper-scale renormalization has already been performed. The rapidity cutoff is then applied to the remaining finite perturbative terms to define the separation between the quantum and background fields. The resulting lower-scale evolution is therefore distinct from the upper rapidity-only evolution: the $\sigma$ dependence contains only the TMD-renormalization dynamics, while the large- and small-$x$ evolution generated by the lower scheme is assigned to $\sigma'$. For quark TMDPDFs, this lower-scale evolution reduces to DGLAP evolution in the collinear limit and to the characteristic double-logarithmic quark evolution at small $x$.

Further, we examine the collinear and high-energy $x\to0$ limits at the $\sigma^\prime$ scale for the MSTT and rapidity-only schemes. In both cases, the leading operator reproduces the DGLAP logarithm of collinear matching and the double logarithm governing the small-$x$ quark TMDPDF in their respective limits. At finite $b_\perp$ and arbitrary $x$, it also retains portions of the higher-twist collinear and sub-eikonal structure generated by the leading-power TMD operator. The complete evolution of independent higher-twist or sub-eikonal operators is beyond the scope of this work. Matching the MSTT and rapidity-only schemes makes explicit how the same perturbative information is distributed between coefficient functions and background-field operators.

The remainder of the paper is organized as follows. Section~\ref{sec:definitions} introduces the background-field approach to QCD factorization. Section~\ref{sec:NLOpcalc} outlines the NLO calculation of Eq.~(\ref{eq:fact-op}), and Sec.~\ref{sec:quarkTMDPDFsNLO} computes the bare quark TMD beam function. Section~\ref{sec:renorm} specifies the upper-scale renormalization, including the standard TMD scheme in Sec.~\ref{sec:renorm} and the rapidity-only scheme in Sec.~\ref{sec:rap-onlyUV}; the corresponding soft function is computed in Sec.~\ref{sec:soft-rap-only}. Section~\ref{sec:low-fact} introduces the lower-scale schemes. We present the MSTT scheme in Sec.~\ref{sec:modified-TMD} and compare it with collinear, high-energy, and rapidity-only schemes in Secs.~\ref{sec:matching_collinear}, \ref{sec:matching_eikonal}, and \ref{sec:rapidity-only}, respectively.

\section{Background-field setup and NLO matching\label{sec:definitions}}
We define the quark TMD operator and the two-stage background-field matching used to compute its NLO coefficient.
Our central object is the leading-power TMDPDFs, defined through the matrix element
\begin{eqnarray}
  &&\Phi^{[\Gamma]}_f(x, b_\perp) = \int \frac{dz^-}{2\pi} e^{-2ixp^+z^-} \langle P, S|\mathcal{U}^{[\Gamma]}_f(z^- , -z^-, \frac{1}{2}b_\perp)|P, S\rangle\,,
  \label{eq:matrix-element-quark}
\end{eqnarray}
where the TMD operator, or beam function, for quarks of flavor $f$ is
\begin{eqnarray}
  &&\mathcal{U}^{[\Gamma]}_f(z^-_1, z^-_2, b_\perp) = \bar{q}_f(z^-_1, b_\perp)[z^-_1, \pm\infty]_{b}\mathcal{L}_{b}\Gamma\mathcal{L}^\dag_{-b}[\pm\infty, z^-_2]_{- b}q_f(z^-_2, - b_\perp)\,.
  \label{eq:def-operator}
\end{eqnarray}

Here $\Gamma$ is a $\gamma$-matrix projector that selects TMDPDFs of a given polarization type.

The light-cone Wilson lines in Eq.~(\ref{eq:def-operator}) are defined as
\begin{eqnarray}
  &&[x^-, y^-]_{z}\equiv \mathcal{P}\exp\Big(ig\int^{x^-}_{y^-} dz^- A_-(z^-, z_\perp)\Big)\,.
\end{eqnarray}

The direction of the Wilson lines is determined by the physical process: future-pointing ($+\infty$) for semi-inclusive deeply inelastic scattering (SIDIS) and past-pointing ($-\infty$) for Drell-Yan production.

Throughout this paper we focus on unpolarized quark TMDPDFs relevant for SIDIS, corresponding to future-pointing Wilson lines with projector $\Gamma = \gamma^+$. In this case the light-cone Wilson lines are joined at spatial infinity by semi-infinite transverse links originating at $b_\perp$ and $-b_\perp$, denoted $\mathcal{L}_{b}$ and $\mathcal{L}^\dag_{-b}$ in Eq.~(\ref{eq:def-operator}).

The matrix element (\ref{eq:matrix-element-quark}) is taken between target states of momentum $P$ and spin $S$. We use the light-cone momentum vector
\begin{eqnarray}
  &&p^\mu = P^\mu - \frac{n^\mu}{2}\frac{M^2}{n\cdot P}\,,
\end{eqnarray}
where $M$ is the target mass and $n_\mu$ is a light-like unit vector with $n=(n^+, n^-, n_\perp)=(0,1,0)$; a conjugate light-cone direction is $\bar{n}=(\bar{n}^+, \bar{n}^-, \bar{n}_\perp)=(1,0,0)$.
The TMDPDFs are obtained from the matrix element (\ref{eq:matrix-element-quark}) by projecting onto the relevant tensor structures,
\begin{eqnarray}
  \label{eq:Decomposition}
  &&\Phi^{[\gamma^+]}_f(x, b_\perp) = f_{1;f}(x, b_\perp) + i \epsilon^{\mu\nu}_\perp b_\mu s_{T\nu} M f^\perp_{1T;f}(x, b_\perp)\,,
  \label{eq:TMDgplus}
\end{eqnarray}
where $s^\mu_{T} = g^{\mu\nu}_T S_\nu$ is the transverse component of the target spin vector $S$. The function $f_1$ is the unpolarized TMDPDF describing the distribution of unpolarized quarks in an unpolarized hadron, while the Sivers function $f^\perp_{1T}$ describes the unpolarized quark distribution in a polarized hadron.

The bare TMD operator (\ref{eq:def-operator}) contains two UV-type divergences: the standard UV divergence, regulated by dimensional regularization, and the rapidity UV divergence. Both can be renormalized at the operator level, independently of the hadron states - a distinctive feature of TMD factorization arising from the separation between the hard process and vacuum soft-gluon exchanges. This yields a universal two-scale evolution for TMDPDFs governed by the pair of equations:
\begin{eqnarray}
  &&\frac{d}{d\ln\mu^2}\Phi^{[\Gamma]}_f(x, b_\perp; \mu, \zeta) = \frac{\gamma^f_\mu(\mu, \zeta)}{2}\Phi^{[\Gamma]}_f(x, b_\perp; \mu, \zeta)\,,
  \label{eq:CSS1}
\end{eqnarray}
\begin{eqnarray}
  &&\frac{d}{d\ln\zeta}\Phi^{[\Gamma]}_f(x, b_\perp; \mu, \zeta) = \frac{\gamma^f_\zeta(\mu, b_\perp)}{2}\Phi^{[\Gamma]}_f(x, b_\perp; \mu, \zeta)\,,
  \label{eq:CSS2}
\end{eqnarray}
where the second equation, known as the Collins-Soper equation~\cite{Collins:1981uk,Collins:1981va}, describes evolution with the rapidity scale $\zeta$. The anomalous dimensions $\gamma^f_\mu$ and $\gamma^f_\zeta$ take the form
\begin{eqnarray}
  &&\gamma^f_\mu(\mu, \zeta) = \Gamma^f_{\rm cusp}[\alpha_s(\mu)]\ln\frac{\mu^2}{\zeta} + \gamma^f_s[\alpha_s(\mu)]\,,
\end{eqnarray}
\begin{eqnarray}
  &&\gamma^f_\zeta(\mu, b_\perp) = -2\int^{\mu}_{1/b_\perp}\frac{d\mu'}{\mu'}\Gamma^f_{\rm cusp}[\alpha_s(\mu')]+ \gamma^f_r[\alpha_s(1/b_\perp)] \,,
\end{eqnarray}
where the cusp anomalous dimension $\Gamma^f_{\rm cusp}$~\cite{Henn:2019swt,vonManteuffel:2020vjv}, the soft anomalous dimension $\gamma^f_s$~\cite{Duhr:2022cob,Das:2019btv}, and the rapidity anomalous dimension $\gamma^f_r$~\cite{Duhr:2022yyp} are known to high orders.

\begin{figure}[htb]
  \begin{center}
    \includegraphics[width=170mm]{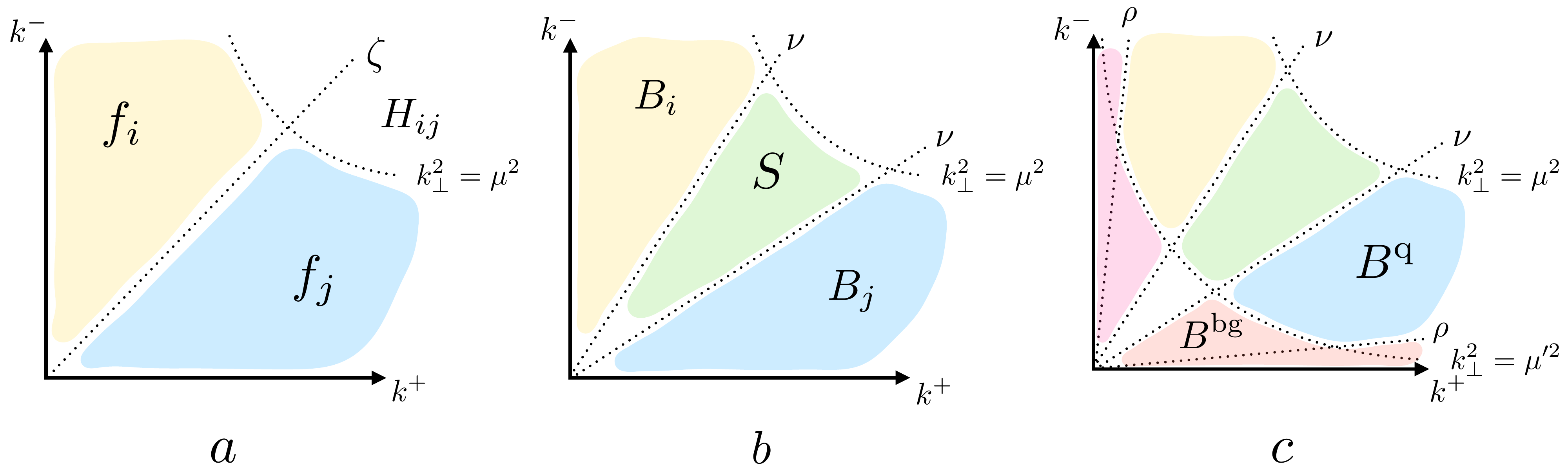}
  \end{center}
\caption{\label{fig:fig1}a) TMD factorization between the hard scattering $H$ and the hadron fields defined by the factorization scales $\mu^2$ and $\zeta$; b) The structure of the TMD factorization in the $\eta$ regulator scheme with scales $\nu$ and $\mu^2$; c) The structure of factorization in our approach. The factorization between ``quantum" $B^{\rm q}$ and ``background" $B^{\rm bg}$ fields is defined by scales $\rho$ and $\mu'^2$.}
\end{figure}

Our computation of TMDPDFs treats $\mu^2$ and $\zeta$ as the upper scales of standard TMD factorization: $\mu$ separates the hard scattering from the hadronic QCD fields, while $\zeta$ separates the hadron fields in rapidity. This separation is shown schematically in Fig.~\ref{fig:fig1}a. The UV divergences in the bare TMD operator (\ref{eq:def-operator}) are a direct consequence of this mode separation.

To compute TMDPDFs within this framework, we must formalize the mode separation defined by the TMD factorization scheme, which can be achieved within the background-field approach~\cite{Abbott:1980hw,Abbott:1981ke}, see also Ref.~\cite{Mukherjee:2023snp}.

To introduce the approach, we begin with the functional integral representation of a general physical observable $\mathcal{X}$:
\begin{eqnarray}
&&\mathcal{X} \propto \int \mathcal{D}C~ \Psi^\ast_{P_f}(\vec{C}(t_f)) \mathcal{H}(C) \Psi_{P_i}(\vec{C}(t_i)) e^{iS_{\rm QCD}(C)} \,,
\label{int:mel-form1}
\end{eqnarray}
where $\int \mathcal{D}C$ is a functional integral over QCD fields $C$.\footnote{For brevity, $C$ denotes both quark and gluon fields.} Here $\Psi_{P_i}$ and $\Psi^\ast_{P_f}$ are the wave functions of the initial and final hadron states, and $\mathcal{H}$ is an operator defining the hard process (e.g.\ virtual-photon production in Drell-Yan scattering).

Evaluating the functional integral (\ref{int:mel-form1}) in full generality is not possible due to the non-perturbative nature of QCD. For high-energy scattering observables, however, the presence of a hard scale (e.g.\ the photon virtuality $Q^2$ in DIS) renders part of the integral perturbatively calculable, giving rise to the factorization picture of scattering.

This factorization involves separating perturbative ($A$) and non-perturbative ($B$) field modes, which in the background-field approach is introduced by splitting the fields in Eq.~(\ref{int:mel-form1}) as
\begin{eqnarray}
&&C \to A + B\,,
\label{bfm-fields}
\end{eqnarray}
so that the functional integral in Eq.~(\ref{int:mel-form1}) factorizes as
\begin{eqnarray}
&&\int \mathcal{D}C e^{iS_{\rm QCD}(C)} \to \int \mathcal{D}B e^{iS_{\rm QCD}(B)} \int \mathcal{D}A e^{iS_{\rm bQCD}(A, B)}\,,
\label{eq:func-bfm}
\end{eqnarray}
where the background QCD action is
\begin{eqnarray}
&&S_{\rm bQCD}(A, B) = S_{\rm QCD}(A + B) - S_{\rm QCD}(B)\,.
\label{eq:bfactionA}
\end{eqnarray}

The fields $A$ in Eq.~(\ref{eq:func-bfm}) encode the perturbative part of the observable; we evaluate the integral $\int \mathcal{D}A$ perturbatively, which ultimately yields a hard coefficient $H$ in the factorization formula for the observable (\ref{eq:cs-init-fact}).

Crucially, the $A$-field integral is {\it not} evaluated in the vacuum: the $A$ fields interact with the hadronic background fields $B$ through the action (\ref{eq:bfactionA}), and the integral must be performed in a fixed $B$ configuration. The $B$ fields combine into QCD operators $\mathcal{O}_k(B)$, so the result of integrating out $A$ in Eq.~(\ref{int:mel-form1}) takes the form
\begin{eqnarray}
&&\int \mathcal{D}A e^{iS_{\rm bQCD}(A, B)}~ \mathcal{H}(A+B) = \sum_k H_k(\sigma) \otimes \mathcal{O}_k(B, \sigma)\,.
\label{eq:int-result}
\end{eqnarray}

The field splitting (\ref{bfm-fields}) is defined by the upper scale set $\sigma$, which delineates the different field modes. This set is specified by the chosen scheme; for standard TMD factorization it comprises $\mu^2$ and $\zeta$. Its dependence is explicit in Eq.~(\ref{eq:int-result}).

The remaining $B$-field integral in Eq.~(\ref{int:mel-form1}) produces matrix elements of the operators $\mathcal{O}_k(B)$ from Eq.~(\ref{eq:int-result}), giving the final factorized expression for the observable as
\begin{eqnarray}
&&\mathcal{X} \propto \sum_k H_k(\sigma) \otimes \langle P_f|\mathcal{O}_k(\hat{B})|P_i\rangle^{\sigma} \,,
\label{int:mel-afterA}
\end{eqnarray}
where $P_f$ and $P_i$ denote hadron momenta at $t_f\to \infty$ and $t_i\to-\infty$, respectively, and the matrix elements of the $B$ fields are
\begin{eqnarray}
&&\langle P_f | \mathcal{O}_k(\hat{B}) |P_i\rangle^{\sigma} = \int \mathcal{D}B~ \Psi^\ast_{P_f}(\vec{B}(t_f)) \mathcal{O}_k(B, \sigma) \Psi_{P_i}(\vec{B}(t_i)) e^{iS_{\rm QCD}(B)} \,.
\label{eq:mel-form1}
\end{eqnarray}
These matrix elements are then parametrized in terms of parton distribution functions appropriate to the chosen factorization scheme.

As noted above, the details of the mode separation (\ref{bfm-fields}) - and hence the structure of the factorization formula (\ref{int:mel-afterA}) - depend on the observable and the factorization scheme. The factorization scales $\sigma$ define kinematic boundaries between the modes: schematically, perturbative fields $A$ carry momenta $k > \sigma$, while the non-perturbative background fields $B$ carry $k < \sigma$.

For the standard TMD factorization, $\sigma = (\mu^2, \zeta)$: $\mu^2$ separates the hard-scattering mode $A$ from the non-perturbative fields $B$ in transverse momentum, while $\zeta$ separates the $B$-field modes of different hadrons in rapidity. This factorization structure is shown schematically in Fig.~\ref{fig:fig1}a.

In this case, the factorization formula (\ref{int:mel-afterA}) for production of a color-singlet state with invariant mass $Q$, rapidity $y$, and transverse momentum $q_\perp$ in unpolarized hadron-hadron scattering reads
\begin{eqnarray}
&&\frac{d\sigma}{dQdyd^2q_\perp} = \sum_{ij} H_{ij}(Q, \mu) \int d^2b_\perp e^{iq_\perp b_\perp} f_{1;i}(x_a, b_\perp; \mu, \zeta_a)f_{1;j}(x_b, b_\perp; \mu, \zeta_b) + O\left(\frac{q^2_\perp}{Q^2}\right)
\,,
\label{eq:intTMDff}
\end{eqnarray}

where $x_a = Qe^y/E_{\rm cm}$ and $x_b = Qe^{-y}/E_{\rm cm}$ are the parton momentum fractions, $E^2_{\rm cm} = 2 P^-_a P^+_b$ is the center-of-mass energy squared of the colliding hadrons, and the rapidity factorization scales
\begin{equation}
\zeta_a = 2(x_aP^-_a)^2, \ \ \ \ \ \zeta_b=2(x_bP^+_b)^2\,
\end{equation}
satisfy $\zeta_a\zeta_b=Q^4$.\footnote{Consequently, $\zeta_a$ and $\zeta_b$ reduce to a single factorization scale $\zeta$; its explicit form is not needed here.} In Eq.~(\ref{eq:intTMDff}), $f_{1;i(j)}$ are renormalized unpolarized TMDPDFs from the decomposition (\ref{eq:TMDgplus}) of the matrix element (\ref{eq:matrix-element-quark}).

An important technical aspect of the TMD formalism is that, while the dependence of TMDPDFs on $\mu^2$ and $\zeta$ is universal and governed by the evolution equations (\ref{eq:CSS1}) and (\ref{eq:CSS2}), the manner in which these scales are introduced - and hence the structure of the UV singularities in the bare operators - is scheme dependent.

In principle, one could introduce the factorization scales as explicit rigid cutoffs, appearing as bounds in the kinematic integrals, e.g.\ $\int^\infty_{\mu^2} dk^2_\perp$ for the $A$ fields and $\int^{\mu^2}_0 dk^2_\perp$ for the $B$ fields, with analogous cutoffs $\int^{\sqrt{\zeta}} dk^-$ separating the rapidity modes.
In practice, however, the rigid cutoff approach is cumbersome. To separate the fields, one can instead employ the $\eta$-regulator scheme of Refs.~\cite{Chiu:2012ir, Chiu:2011qc}, where $\mu^2$ is introduced through dimensional regularization with the standard UV regulator $\epsilon$.

The rapidity factorization between $B$ fields at scale $\zeta$ is then introduced via
\begin{eqnarray}
&&\int^\infty_0 \frac{d k^-}{k^-} \to 2^{\eta/2}  \Big(\frac{\nu}{\sqrt{\zeta}}\Big)^\eta (xP^+)^\eta \int^\infty_0 \frac{d k^-}{k^-}  |k^+ |^{-\eta} \,
\label{eq:eta-reg-new}
\end{eqnarray}
for the fields of the hadron of momenta $P^+$.\footnote{And similarly for the $\int dk^+$ integrals for the fields of the hadron of momenta $P^-$.}

An important feature of this approach (shared by most TMD factorization schemes)  is that the rapidity factorization between the fields of two hadrons introduces an overlapping region corresponding to the soft factor $S$, see Fig.~\ref{fig:fig1}b. Double counting of this overlap is avoided by first integrating over fields outside the overlapping region and then distributing the soft factor contribution equally between the TMDPDFs of both hadrons.

In the $\eta$-regulator scheme, this corresponds to first applying Eq.~(\ref{eq:eta-reg-new}) in the computation of the bare beam function, where $\nu$ is the rapidity factorization scale separating the $B$ fields outside the overlapping region from the soft factor, see Fig.~\ref{fig:fig1}b, and then adding the soft factor contribution computed from its definition as a vacuum matrix element
\begin{eqnarray}
&&\mathcal{S}(b_\perp)
= \frac{1}{N_c}\langle 0|{\rm tr}[S^\dag_{\bar{n}}(b_\perp)S_{n}(b_\perp) S^\dag_{n}(0_\perp) S_{\bar{n}}(0_\perp)]|0\rangle\,
\label{eq:soft-function}
\end{eqnarray}
constructed from the soft Wilson lines
\begin{eqnarray}
&&S_{n}(b_\perp) \equiv \mathcal{P}\exp\Big(ig\int^\infty_0 d(x\cdot \bar{n}) n\cdot A(x\cdot \bar{n}, b_\perp)\Big)\,,
\end{eqnarray}
and analogously for $S_{\bar{n}}(b_\perp)$.

The first step of this procedure yields the bare TMD operator (\ref{eq:def-operator}), also called the beam function, arising from integration over the $B$ fields in Fig.~\ref{fig:fig1}b, with the structure
\begin{eqnarray}
&&\mathcal{U}^{[\Gamma]}_f(z^-_1, z^-_2, b_\perp; \epsilon, \eta)\,,
\end{eqnarray}
which contains UV divergences renormalized by the universal factor
\begin{eqnarray}
&&Z_{UV} = Z^{-1}_2 Z^{TMD}_q\,,
\label{eq:qTMDz}
\end{eqnarray}
where $Z^{TMD}_q$ is a UV renormalization constant for the quark TMD operator.

Including the soft-factor contribution in the second step then yields the renormalized TMD operator
\begin{eqnarray}
&&\mathcal{U}^{[\Gamma]}_f(z^-_1, z^-_2, b_\perp; \mu, \zeta)
\label{eq:Urenorm}\\
&&= Z_{UV}(\mu,\epsilon,\mu/\sqrt{\zeta})\ \sqrt{\mathcal{S}(b_\perp; \epsilon, \eta)} \ \mathcal{U}^{[\Gamma]}_f(z^-_1, z^-_2, b_\perp; \epsilon, \eta)\,,
\nonumber
\end{eqnarray}

which enters the TMD factorization formula (\ref{eq:intTMDff}). Note that the $\eta$ dependence cancels in the product of the soft factor and the beam function, thus computing the renormalized TMDPDFs requires perturbative corrections to both.

While in the standard approach, rapidity divergences cancel in the product (\ref{eq:Urenorm}), in our analysis we will use a slightly different formulation in which the beam and soft functions are each separately rapidity-renormalized,
\begin{eqnarray}
&&\mathcal{U}^{[\Gamma]}_f(z^-_1, z^-_2, b_\perp; \mu, \nu/\sqrt{\zeta}) = Z_{B}(\mu, \epsilon, \nu/\sqrt{\zeta}, \eta)\ \mathcal{U}^{[\Gamma]}_f(z^-_1, z^-_2, b_\perp; \epsilon, \eta)\,
\label{eq:rrenbeam-def}
\end{eqnarray}
and
\begin{eqnarray}
&&\mathcal{S}(b_\perp; \mu, \nu) = Z_S(\mu, \epsilon, \mu/\nu, \eta)\ \mathcal{S}(b_\perp; \epsilon, \eta)\,,
\label{eq:rrensoft}
\end{eqnarray}
where $Z_{B}$ and $Z_S$ are corresponding renormalization constants.
With this formulation, both renormalized functions are finite, and the renormalized TMD operator (\ref{eq:Urenorm}) is given by
\begin{eqnarray}
&&\mathcal{U}^{[\Gamma]}_f(z^-_1, z^-_2, b_\perp; \mu, \zeta) = \sqrt{\mathcal{S}(b_\perp; \mu, \nu)} \ \mathcal{U}^{[\Gamma]}_f(z^-_1, z^-_2, b_\perp; \mu, \nu/\sqrt{\zeta})\,.
\label{eq:TMD-renorm}
\end{eqnarray}
This immediately implies the following relation between the renormalization constants:
\begin{eqnarray}
&&Z_{UV}(\mu,\epsilon,\mu/\sqrt{\zeta}) = \sqrt{Z_S(\mu, \epsilon, \mu/\nu, \eta)}\ Z_{B}(\mu, \epsilon, \nu/\sqrt{\zeta}, \eta)\,.
\label{eq:relationsZ}
\end{eqnarray}
The matching below is formulated for the beam function; the soft factor is restored multiplicatively through Eq.~(\ref{eq:TMD-renorm}).

\subsection{NLO matching to lower-scale background fields\label{sec:NLOpcalc}}
Starting from the $B$ fields defined at the upper scales $\sigma$, we integrate out the quantum modes between $\sigma$ and a lower set $\sigma'$ to obtain the NLO matching coefficient. We split the fields at the lower scales $\sigma'$ as
\begin{eqnarray}
B\to B^{\rm q} + B^{\rm bg}\,,
\label{eq:B-split}
\end{eqnarray}
where the ``quantum" fields $B^{\rm q}$ occupy the momentum range $\sigma' < k < \sigma$, and the ``background" fields $B^{\rm bg}$ have $k < \sigma'$.

Consequently, the functional integral over $B$ fields in the matrix element of Eq.~(\ref{int:mel-afterA}) factorizes analogously to Eq.~(\ref{eq:func-bfm}),
\begin{eqnarray}
&&\int \mathcal{D}B e^{iS_{\rm QCD}(B)} \to \int \mathcal{D}B^{\rm bg} e^{iS_{\rm QCD}(B^{\rm bg})} \int \mathcal{D}B^{\rm q}e^{iS_{\rm bQCD}(B^{\rm q}, B^{\rm bg})}\,,
\label{eq:func-bfm-forB}
\end{eqnarray}
where the background field action is
\begin{eqnarray}
&&S_{\rm bQCD}(B^{\rm q}, B^{\rm bg}) = S_{\rm QCD}(B^{\rm q} + B^{\rm bg}) - S_{\rm QCD}(B^{\rm bg})\,,
\label{eq:bfactionB}
\end{eqnarray}
c.f. Eq. (\ref{eq:bfactionA}).

The matrix element (\ref{eq:mel-form1}) in the factorization formula (\ref{int:mel-afterA}) then becomes
\begin{eqnarray}
&&\langle P_f | \mathcal{O}_k(\hat{B}) |P_i\rangle^{\sigma}
\nonumber\\
&&= \int \mathcal{D}B^{\rm bg} e^{iS_{\rm QCD}(B^{\rm bg})}~ \Psi^\ast_{P_f}(\vec{B}^{\rm bg}(t_f)) \Big( \int \mathcal{D}B^{\rm q}e^{iS_{\rm bQCD}(B^{\rm q}, B^{\rm bg})} \mathcal{O}_k(B^{\rm q} + B^{\rm bg}, \sigma)\Big) \Psi_{P_i}(\vec{B}^{\rm bg}(t_i)) \,.
\label{eq:mel-afterB}
\end{eqnarray}

\begin{figure}[htb]
\begin{center}
\includegraphics[width=90mm]{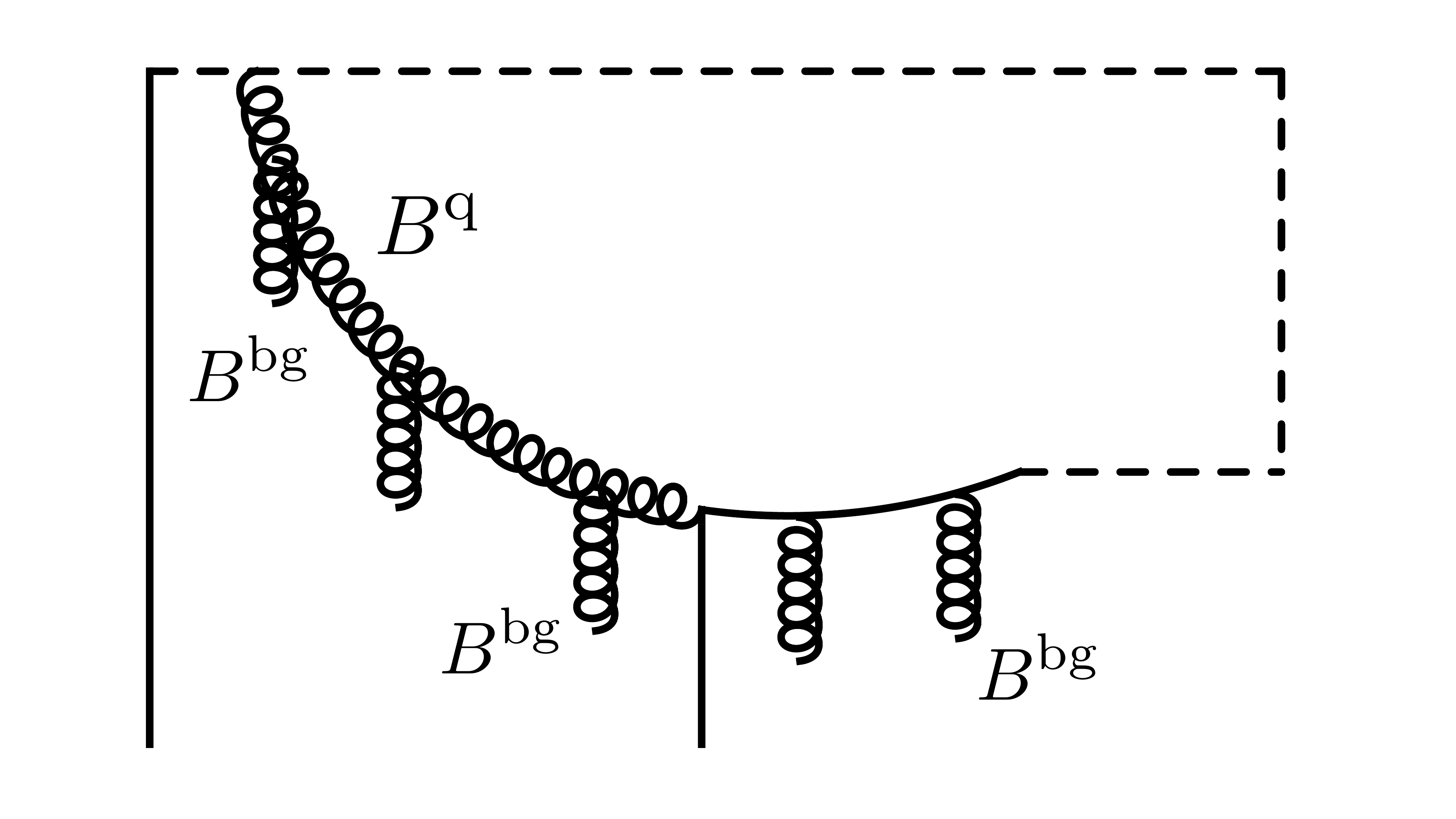}
\end{center}
\caption{\label{fig:fig2}A typical contribution to the functional integral over ``quantum" $B^{\rm q}$ fields at NLO in a fixed configuration of the background fields $B^{\rm bg}$.}
\end{figure}

The splitting (\ref{eq:B-split}) reduces the computation of the functional integral in Eq.~(\ref{eq:mel-form1}) to evaluating the $B^{\rm q}$ integral in a fixed $B^{\rm bg}$ background. Since $B^{\rm q}$ encodes quantum emission of the $B$ fields, this integral is evaluated perturbatively. A representative NLO diagram for the TMD operator (\ref{eq:def-operator}) is shown in Fig.~\ref{fig:fig2}.

In general, the result of this calculation takes the form
\begin{eqnarray}
&&\int \mathcal{D}B^{\rm q}e^{iS_{\rm bQCD}(B^{\rm q}, B^{\rm bg})} \mathcal{O}_k(B^{\rm q} + B^{\rm bg}, \sigma) = \sum_m C_{km}(\sigma, \sigma') \otimes \tilde{\mathcal{O}}_m(B^{\rm bg}, \sigma')\,,
\label{eq:Bq-fint}
\end{eqnarray}
where the perturbative coefficients $C_{km}(\sigma, \sigma')$ capture the scale dependence of the operators, and the operators $\tilde{\mathcal{O}}_m(B^{\rm bg}, \sigma')$ are built from the $B^{\rm bg}$ modes below the scales $\sigma'$, which are held fixed in the integral (\ref{eq:Bq-fint}).

Substituting (\ref{eq:Bq-fint}) into Eq.~(\ref{eq:mel-afterB}) and integrating over $B^{\rm bg}$ - a genuinely non-perturbative step - yields matrix elements of the operators $\tilde{\mathcal{O}}_m(B^{\rm bg}, \sigma')$:
\begin{eqnarray}
&& \int \mathcal{D}B^{\rm bg} e^{iS_{\rm QCD}(B^{\rm bg})}~ \Psi^\ast_{P_f}(\vec{B}^{\rm bg}(t_f)) \tilde{\mathcal{O}}_m(B^{\rm bg}, \sigma') \Psi_{P_i}(\vec{B}^{\rm bg}(t_i)) = \langle P_f | \tilde{\mathcal{O}}_m(\hat{B}^{\rm bg}) |P_i\rangle^{\sigma'}\,.
\label{eq:mel-forBbg}
\end{eqnarray}
These matrix elements are parametrized in terms of distribution functions defined at the scales $\sigma'$.

Combining Eqs.~(\ref{eq:Bq-fint}) and (\ref{eq:mel-forBbg}) gives the following result for the matrix element (\ref{eq:mel-form1}):
\begin{eqnarray}
&&\langle P_f | \mathcal{O}_k(\hat{B}) |P_i\rangle^{\sigma} = \sum_m C_{km}(\sigma, \sigma') \otimes \langle P_f | \tilde{\mathcal{O}}_m(\hat{B}^{\rm bg}) |P_i\rangle^{\sigma'} \,,
\label{eq:mel-afterB-result}
\end{eqnarray}
The upper and lower schemes may be chosen independently, determining how perturbative contributions are partitioned between $C_{km}$ and the lower-scale operators.
For collinear lower-scale operators, this relation yields the familiar DGLAP logarithms.
The arbitrary-$x$ construction extends the gluon analysis of Ref.~\cite{Mukherjee:2023snp} to quark TMDPDFs.

The factorization scales $\sigma$ and $\sigma'$ are kinematic cutoffs separating different field modes in our computation. In our approach, we often define these scales through the renormalization procedure. For instance, the transverse scales $\mu^2$ and $\mu'^2$ can be introduced via dimensional regularization, while the rapidity factorization scales $\eta$ and $\rho$, see Fig.~\ref{fig:fig1}c, can be introduced by regulating the divergent longitudinal integrals as
\begin{eqnarray}
&&\int^\infty_0 \frac{d k^-}{k^-} \to \rho^\eta \int^\infty_0 \frac{d k^-}{k^-}  |k^+ |^{-\eta} \,.
\label{eq:eta-reg-IR}
\end{eqnarray}
We will also consider factorization schemes with explicit kinematic cutoffs corresponding to $\sigma$ and $\sigma'$.

Within a given factorization scheme, we compute the perturbative coefficients $C_{km}(\sigma, \sigma')$ in Eq.~(\ref{eq:mel-afterB-result}) at NLO for the  leading quark TMDPDF operator in the background field expansion (\ref{eq:matrix-element-quark}). The calculation generates a hierarchy of operators at the $\sigma'$ scales, yielding an infinite series on the right-hand side of Eq.~(\ref{eq:mel-afterB-result}). We restrict the analysis to the leading-power contribution, leaving higher-order terms to future work.

\begin{figure}[htb]
\begin{center}
\includegraphics[width=140mm]{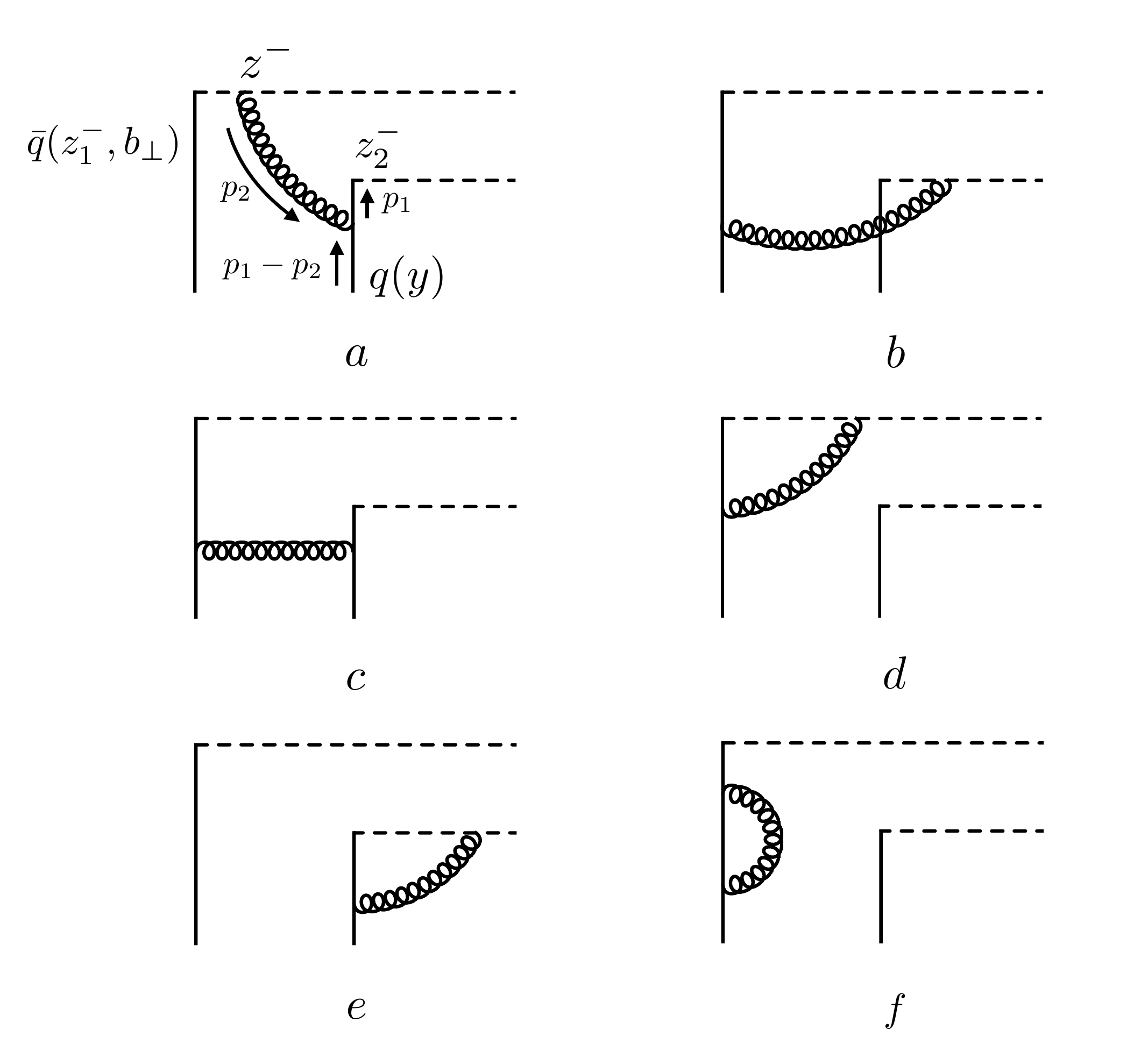}
\end{center}
\caption{\label{fig:fig3}NLO diagrams for computing the quark TMD operator (\ref{eq:def-operator}) in the background-Feynman gauge.}
\end{figure}

The corresponding leading coefficient function can be obtained from the diagrams in Fig.~\ref{fig:fig3}.
These diagrams describe quantum emission in a two-quark background. While reproducing gauge-invariant background-field operators in general requires an infinite set of interactions with background gluon fields, see e.g.\ Fig.~\ref{fig:fig2}, for the leading term of (\ref{eq:mel-afterB-result}), these interactions do not modify the perturbative coefficient and contribute only gauge factors to the background-field operator at the $\sigma'$ scales. Although the full set of interactions can be systematically included within the background-field approach, it suffices here to compute the leading diagrams in Fig.~\ref{fig:fig3} and insert the appropriate gauge factors into the final result. The computation of these diagrams is presented in the next section.

\section{Quark TMDPDFs at NLO \label{sec:quarkTMDPDFsNLO}}
In this section we compute the bare TMD operator (\ref{eq:def-operator}) at NLO within the background-field approach. The bare results for the diagrams in Fig.~\ref{fig:fig3} were obtained in Ref.~\cite{Mukherjee:2025aiw}; we refer the reader there for technical details. In Ref.~\cite{Mukherjee:2025aiw} the diagrams were computed in two gauges: the axial gauge and the background-Feynman gauge, the latter reducing to the standard Feynman gauge in the dilute limit. It was shown that in the background-Feynman gauge, under the condition $B^{\rm q}(+\infty n)=0$, the diagrams in Fig.~\ref{fig:fig3} fully determine the NLO correction to the quark TMD operator (\ref{eq:def-operator}).

While borrowing the results of Ref.~\cite{Mukherjee:2025aiw} would suffice, it is instructive to examine the diagram in Fig.~\ref{fig:fig3}e in some detail. In collinear matching, its contribution is naively trivial, since it is given by a scaleless transverse integral that vanishes in dimensional regularization. A more careful treatment, however, yields a cancellation $1/\epsilon_{UV} - 1/\epsilon_{IR} = 0$, signaling the importance of this diagram for the renormalization at both the upper $\sigma$ and lower scales $\sigma'$.
Specifically, the UV pole $1/\epsilon_{UV}$ of this diagram, along with the UV poles of the diagrams in Figs. \ref{fig:fig3}d and \ref{fig:fig3}f, determines the UV renormalization of TMDPDFs at the scales $\sigma = (\mu^2, \nu)$. Furthermore, in our factorization approach this diagram provides a non-trivial contribution relevant for our analysis of the interplay between large- and small-$x$ dynamics in TMDPDFs.

The initial expression for this diagram reads
\begin{eqnarray}
&&\mathcal{U}^{[\gamma^+]}_{f;{\rm  q(1) + bg}}(z^-_1, z^-_2, b_\perp)\Big|_{Fig.\ \ref{fig:fig3}e}
\label{eq:fig3e-init}\\
&&= -g^2 C_F \int_{z^-_2}^{\infty}dz^- \int d^4y \bar{q}_f(z^-_1, b_\perp) \gamma^+ (z^-_2, - b_\perp|\frac{i\slashed{p}}{p^2+i\epsilon}|y) \gamma^\mu q_f(y) n^\nu (z^-, - b_\perp|\frac{-ig_{\mu\nu}}{p^2+i\epsilon}|y)\,,
\nonumber
\end{eqnarray}
where $q_f$ ($\bar{q}_f$) is the background quark (anti-quark) field defined at the lower $\sigma'=(\mu'^2,\rho)$ scales. The label ``{\rm q(1) + bg}" indicates that the expression is built from fields below $\sigma$ and includes both the background fields at $\sigma'$ and the one-loop quantum emission.

In Eq.~(\ref{eq:fig3e-init}) we use Schwinger’s notation~\cite{Schwinger:1951nm}, where the coherent states $|x)$ and $|p)$ are eigenstates of the position and momentum operators,
\begin{eqnarray}
&&\hat{x}_\mu|x) = x_\mu|x);\ \ \ \hat{p}_\mu|p) = p_\mu|p)\,,
\end{eqnarray}
with normalization
\begin{eqnarray}
\int d^4x\, |x)( x| = 1\,;\ \ \ \int \dhd^4p \, |p) (p| = 1;\ \ \ ( x|p) = e^{ipx}\,,
\label{eq:bosonort}
\end{eqnarray}
where $\int \dhd^d p \equiv \int d^d p/(2\pi)^d$.

For an arbitrary function $f(\hat{p})$ of the momentum operator,
\begin{eqnarray}
&&(x|f(\hat{p})|y) = \int \dhd ^4p e^{-ip(x-y)}f(p)\,.
\label{eq:fopsw}
\end{eqnarray}
For brevity, the ``hat" notation for operators is suppressed throughout, as in Eq.~(\ref{eq:fig3e-init}).
Computation of Eq. (\ref{eq:fig3e-init}) is straightforward. After some trivial $\gamma$-matrix algebra, it becomes
\begin{eqnarray}
&&\mathcal{U}^{[\gamma^+]}_{f;{\rm  q(1) + bg}}(z^-_1, z^-_2, b_\perp)\Big|_{Fig.\ \ref{fig:fig3}e}
\label{eq:fig3e-short}\\
&&= - g^2 C_F \int_{z^-_2}^{\infty}dz^- \int d^4y \{\bar{q}_f(z^-_1,  b_\perp) \gamma^+ q_f(y) \} (z^-_2, - b_\perp|\frac{ 2p^+}{p^2+i\epsilon}|y) (y|\frac{1}{p^2+i\epsilon}|z^-, -b_\perp)\,.
\nonumber
\end{eqnarray}

Evaluating the quark and gluon propagators by taking $p^-$ poles gives
\begin{eqnarray}
&&(z^-_2, - b_\perp|\frac{ 2p^+}{p^2+i\epsilon}|y)
\label{eq:qprop1}\\
&&= \Big[ -i \theta( - y^+) \int^\infty_0 \frac{dp^+_1}{2\pi} + i \theta(y^+ ) \int^0_{-\infty} \frac{dp^+_1}{2\pi} \Big] e^{-ip^+_1 (z^-_2 - y^-)} \int \dhd^2 p_{1\perp} e^{-ip_{1\perp} (b_\perp + y_\perp) } e^{i\frac{p^2_{1\perp}}{2p^+_1} y^+}
\nonumber
\end{eqnarray}
and
\begin{eqnarray}
&&(y|\frac{1}{p^2+i\epsilon}|z^-, -b_\perp)
\label{eq:qprop2}\\
&&= \Big[ -i \theta(y^+ ) \int^\infty_0 \frac{dp^+_2}{2\pi} + i \theta( - y^+) \int^0_{-\infty} \frac{dp^+_2}{2\pi} \Big] \frac{1}{2p^+_2} e^{-ip^+_2 (y^- - z^-)} \int \dhd^2 p_{2\perp} e^{ip_{2\perp} (y_\perp + b_\perp)} e^{-i\frac{p^2_{2\perp}}{2p^+_2} y^+}\,.
\nonumber
\end{eqnarray}

Substituting Eqs.~(\ref{eq:qprop1}) and (\ref{eq:qprop2}) into Eq.~(\ref{eq:fig3e-short}) and integrating over $y^+$ (then reversing the sign of $p^+_2$) yields
\begin{eqnarray}
&&\mathcal{U}^{[\gamma^+]}_{f;{\rm  q(1) + bg}}(z^-_1, z^-_2, b_\perp)\Big|_{Fig.\ \ref{fig:fig3}e} = - ig^2 C_F \int_{z^-_2}^{\infty} dz^- \int dy^- \int d^2y_\perp \{\bar{q}_f(z^-_1,  b_\perp) \gamma^+ q_f(y^-, y_\perp) \}
\nonumber\\
&&\times \Big\{ \int^\infty_0 \frac{dp^+_1}{2\pi} e^{-ip^+_1 (z^-_2 - y^-)} \int_0^{\infty} \frac{dp^+_2}{2\pi} \frac{1}{2p^+_2} e^{ip^+_2 (y^- - z^-)} - \int^0_{-\infty} \frac{dp^+_1}{2\pi} e^{-ip^+_1 (z^-_2 - y^-)}  \int_{-\infty}^0 \frac{dp^+_2}{2\pi} \frac{1}{2p^+_2} e^{ip^+_2 (y^- - z^-)} \Big\}
\nonumber\\
&&\times \int \dhd^2 p_{1\perp} e^{-ip_{1\perp} (b_\perp + y_\perp) } \int \dhd^2 p_{2\perp} e^{ip_{2\perp} (y_\perp + b_\perp)} \frac{1}{ \frac{p^2_{1\perp}}{2p^+_1} + \frac{p^2_{2\perp}}{2p^+_2}}\,.
\end{eqnarray}

Introducing a new variable in place of $p_2^+$,
\begin{eqnarray}
&&u = \frac{p^+_1}{p^+_1 + p^+_2}\,,
\label{eq:u-first}
\end{eqnarray}
the expression becomes
\begin{eqnarray}
&&\mathcal{U}^{[\gamma^+]}_{f;{\rm  q(1) + bg}}(z^-_1, z^-_2, b_\perp)\Big|_{Fig.\ \ref{fig:fig3}e} = -\frac{g^2 C_F}{2\pi}  \int_{z^-_2}^{\infty}dz^- \int dy^- \int d^2y_\perp \{\bar{q}_f(z^-_1,  b_\perp) \gamma^+ q_f(y^-, y_\perp) \} \int_0^1 du
\\
&&\times \int^\infty_{-\infty} \frac{dp^+_1}{2\pi} \Big(\frac{\partial}{\partial y^-}e^{-i ( z^-_2 -  \frac{1}{u} y^- + \frac{1-u}{u} z^-) p^+_1 } \Big)
\int \dhd^2 p_{1\perp} e^{-ip_{1\perp} (b_\perp + y_\perp) } \int \dhd^2 p_{2\perp} e^{ip_{2\perp} (y_\perp + b_\perp)} \frac{1}{ (1-u) p^2_{1\perp} + u p^2_{2\perp} }\,.
\nonumber
\end{eqnarray}

Integrating by parts in $y^-$ and then integrating over $p^+_1$ and $y^-$ gives
\begin{eqnarray}
&&\mathcal{U}^{[\gamma^+]}_{f;{\rm  q(1) + bg}}(z^-_1, z^-_2, b_\perp)\Big|_{Fig.\ \ref{fig:fig3}e} = \frac{g^2 C_F}{2\pi} \int_0^1 du \frac{u}{1-u} \int_{z^-_2}^{\infty}dz^- \int d^2y_\perp \{\bar{q}_f(z^-_1,  b_\perp) \gamma^+ \frac{\partial}{\partial  z^-} q_f((1-u) z^- + uz^-_2 , y_\perp) \}
\nonumber\\
&&\times
\int \dhd^2 p_{1\perp} e^{-ip_{1\perp} (b_\perp + y_\perp) } \int \dhd^2 p_{2\perp} e^{ip_{2\perp} (y_\perp + b_\perp)} \frac{1}{ (1-u) p^2_{1\perp} + u p^2_{2\perp} }\,.
\end{eqnarray}

Integrating over $z^-$ and relabeling $p_{1\perp}\to -p_\perp$, $p_{2\perp}\to -k_\perp$, we arrive at
\begin{eqnarray}
&&\mathcal{U}^{[\gamma^+]}_{f;{\rm  q(1) + bg}}(z^-_1, z^-_2, b_\perp)\Big|_{Fig.\ \ref{fig:fig3}e} = -\frac{g^2 C_F}{2\pi} \int_0^1 du \frac{u}{1-u} \int d^2y_\perp \{\bar{q}_f(z^-_1,  b_\perp) \gamma^+ q_f( z^-_2 , y_\perp) \}
\label{eq:fig3etemp}\\
&&\times
\int \dhd^2 p_{\perp} e^{ip_{\perp} (b_\perp + y_\perp) } \int \dhd^2 k_{\perp} e^{-ik_{\perp} (y_\perp + b_\perp)} \frac{1}{ (1-u) p^2_{\perp} + u k^2_{\perp} }\,.
\nonumber
\end{eqnarray}

We restore gauge invariance of the right-hand-side quark operator by inserting the appropriate gauge factors as in Eq.~(\ref{eq:def-operator}). Shifting the quark operator of the r.h.s  in the transverse direction (valid for the forward matrix element (\ref{eq:matrix-element-quark})) and changing  $p_\perp \to p_\perp + k_\perp$ then give
\begin{eqnarray}
&&\mathcal{U}^{[\gamma^+]}_{f;{\rm  q(1) + bg}}(z^-, -z^-, \frac{1}{2}b_\perp;\epsilon)\Big|_{Fig.\ \ref{fig:fig3}e} = -\frac{g^2 C_F}{2\pi} \Big(\frac{\mu^2 e^{\gamma_E}}{4\pi}\Big)^\epsilon \int \dhd^{2-2\epsilon} p_{\perp} e^{i p_{\perp} b_\perp } \int_0^1 du \frac{u}{1-u}
\nonumber\\
&&\times  \int \dhd^{2-2\epsilon} k_{\perp} \frac{1}{ (1-u) (p+k)^2_{\perp} + u k^2_{\perp} } \int d^{2-2\epsilon}z_\perp e^{-i p_{\perp} z_\perp } \mathcal{U}^{[\gamma^+]}_{f;{\rm  bg}}(z^-, -z^-, \frac{1}{2}z_\perp)\,,
\label{eq:fig3e-af}
\end{eqnarray}
where $z_\perp = b_\perp - y_\perp$. The transverse integrals are regulated by dimensional regularization, with the bare coupling written as $g^2(\mu^2 e^{\gamma_E}/(4\pi))^\epsilon$.

From Eq.~(\ref{eq:fig3e-af}), the diagram in Fig.~\ref{fig:fig3}e contains a UV divergence at $k_\perp \to \infty$ and a UV rapidity divergence at $u\to1$, but no IR divergences. In particular, a potential IR divergence at $k_\perp\to0$ is regulated by the non-zero transverse momentum flow from the background field.

The UV divergent term in Eq.~(\ref{eq:fig3e-af}) reads
\begin{eqnarray}
&& -\frac{g^2 C_F}{8\pi^2} \frac{1}{\epsilon_{UV}}\int_0^1 du \frac{u}{1-u}  \mathcal{U}^{[\gamma^+]}_{f;{\rm  bg}}(z^-, -z^-, \frac{1}{2}b_\perp)\,.
\label{eq:fe-UV-diverge}
\end{eqnarray}
We isolate this UV divergence in Eq.~(\ref{eq:fig3e-af}) by writing
\begin{eqnarray}
&&\mathcal{U}^{[\gamma^+]}_{f;{\rm  q(1) + bg}}(z^-, -z^-, \frac{1}{2}b_\perp;\epsilon)\Big|_{Fig.\ \ref{fig:fig3}e} = -\frac{g^2 C_F}{2\pi} \Big(\frac{\mu^2 e^{\gamma_E}}{4\pi}\Big)^\epsilon\int_0^1 du \frac{u}{1-u} \int  \frac{\dhd^{2-2\epsilon} k_{\perp}}{k^2_\perp} \mathcal{U}^{[\gamma^+]}_{f;{\rm  bg}}(z^-, -z^-, \frac{1}{2}b_\perp)
\nonumber\\
&&-\frac{g^2 C_F}{2\pi} \Big(\frac{\mu^2 e^{\gamma_E}}{4\pi}\Big)^\epsilon \int \dhd^{2-2\epsilon} p_{\perp} e^{i p_{\perp} b_\perp }  \int_0^1 du \frac{u}{1-u} \int \dhd^{2-2\epsilon} k_{\perp} \Big( \frac{1}{(1-u) (p+k)^2_{\perp} + u k^2_{\perp} } - \frac{1}{ k^2_\perp }\Big)
\nonumber\\
&&\times \int d^{2-2\epsilon}z_\perp e^{-i p_{\perp}  z_\perp }\mathcal{U}^{[\gamma^+]}_{f;{\rm  bg}}(z^-, -z^-, \frac{1}{2}z_\perp)\,.
\label{eq:fig3e-tterms-UV}
\end{eqnarray}
The first term contains the UV divergence (\ref{eq:fe-UV-diverge}) and an IR divergence at $k_\perp\to 0$; they cancel as $1/\epsilon_{UV} - 1/\epsilon_{IR} = 0$, corresponding to the scaleless integral $\int d^2k_\perp / k^2_\perp$.

However, we retain the first term rather than dropping it, as it will be combined with a similar contribution from Fig.~\ref{fig:fig3}a. In that sum, only the UV divergence (\ref{eq:fe-UV-diverge}) survives, yielding
\begin{eqnarray}
&&\frac{g^2 C_F}{2\pi} \Big(\frac{\mu^2 e^{\gamma_E}}{4\pi}\Big)^\epsilon \int_0^1 du \frac{u}{1-u} \int  \frac{\dhd^2 k_{\perp}}{k^2_\perp} \Big(e^{ik_\perp b_\perp} - 1 \Big)\,.
\label{eq:typ-int-div}
\end{eqnarray}
The UV divergence at $k_\perp\to\infty$ is renormalized in the standard way. The term also contains a UV rapidity divergence at $u\to1$, which we regulate within the $\eta$-regulator scheme, Eq.~(\ref{eq:eta-reg-new}), via
\begin{eqnarray}
&&\int^1_0 \frac{du }{1-u} \to 2^{\eta/2}\Big(\frac{\nu}{\sqrt{\zeta}}\Big)^\eta \int^1_0 \frac{du}{1 - u}  \Big( \frac{1 - u}{u} \Big)^{-\eta} \,,
\label{eq:eta-sec}
\end{eqnarray}
where $\zeta = 2(xp^+)^2$ for a hadron moving in the $P^+$ direction.

Consider the second term of Eq.~(\ref{eq:fig3e-tterms-UV}). In the collinear matching procedure it is trivial, since the transverse momenta of the background fields are set to zero ($p_\perp = 0$).\footnote{Note the variable shift $p_\perp \to p_\perp + k_\perp$ after Eq.~(\ref{eq:fig3etemp}).} In our approach, however, this term yields a non-trivial contribution relevant to the large- and small-$x$ interplay.

The second term contains an IR divergence at $k_\perp\to0$, which cancels against the first term so that Eq.~(\ref{eq:fig3e-tterms-UV}) is IR finite. Integrating over $u$ in the second term, we find
\begin{eqnarray}
&&\mathcal{U}^{[\gamma^+]}_{f;{\rm  q(1) + bg}}(z^-, -z^-, \frac{1}{2}b_\perp;\epsilon)\Big|_{Fig.\ \ref{fig:fig3}e} = -\frac{g^2 C_F}{2\pi} \Big(\frac{\mu^2 e^{\gamma_E}}{4\pi}\Big)^\epsilon \int_0^1 du \frac{u}{1-u} \int  \frac{\dhd^{2-2\epsilon} k_{\perp}}{k^2_\perp} \mathcal{U}^{[\gamma^+]}_{f;{\rm  bg}}(z^-, -z^-, \frac{1}{2}b_\perp)
\nonumber\\
&&-\frac{g^2 C_F}{2\pi} \Big(\frac{\mu^2 e^{\gamma_E}}{4\pi}\Big)^\epsilon \int \dhd^{2-2\epsilon} p_{\perp} e^{i p_{\perp} b_\perp }  \int \frac{\dhd^{2-2\epsilon} k_{\perp}}{k^2_\perp} e^{-i k_{\perp} b_\perp } \Big( 1 + \frac{1 }{1 - k^2_\perp/p^2_\perp}\ln \frac{k^2_\perp}{p^2_\perp}  \Big)
\nonumber\\
&&\times\int d^{2-2\epsilon}z_\perp e^{-i (p_{\perp} - k_\perp) z_\perp }  \mathcal{U}^{[\gamma^+]}_{f;{\rm  bg}}(z^-, -z^-, \frac{1}{2}z_\perp)\,.
\label{eq:fig3e-result}
\end{eqnarray}
This is the final result for the diagram in Fig.~\ref{fig:fig3}e.

A similar structure was found in Ref.~\cite{Mukherjee:2023snp} for a gluon diagram of the same topology, except that the gluon diagram contained an IR rapidity divergence at $u \to 0$ which is absent in the present quark diagram.

The remaining diagrams in Fig.~\ref{fig:fig3} are computed analogously. In particular, the result for Fig.~\ref{fig:fig3}d coincides with that of Fig.~\ref{fig:fig3}e, Eq.~(\ref{eq:fig3e-result}).

For the diagrams in Figs.~\ref{fig:fig3}a and \ref{fig:fig3}b, Ref.~\cite{Mukherjee:2025aiw} gives
\begin{eqnarray}
&&\mathcal{U}^{[\gamma^+]}_{f;{\rm  q(1) + bg}}(z^-, -z^-, \frac{1}{2}b_\perp;\epsilon)\Big|_{Figs.\ \ref{fig:fig3}a,\ref{fig:fig3}b} = \frac{g^2 C_F}{2\pi} \Big(\frac{\mu^2 e^{\gamma_E}}{4\pi}\Big)^\epsilon \int \dhd^{2-2\epsilon} p_{\perp} e^{ip_{\perp} b_\perp }
\label{eq:fig3ab-init}\\
&&\times  \int_0^1 du \frac{u}{1-u} \int \dhd^{2-2\epsilon} k_{\perp} \frac{1}{ (1-u) p^2_{\perp} + u k^2_{\perp} } \int d^{2-2\epsilon}z_\perp e^{-i ( p_\perp - k_{\perp} ) z_\perp } \mathcal{U}^{[\gamma^+]}_{f;{\rm  bg}}( u z^- , -u z^- , \frac{1}{2}z_\perp)\,.
\nonumber
\end{eqnarray}
These diagrams contain only a UV rapidity divergence at $u\to 1$; all transverse integrals in Eq.~(\ref{eq:fig3ab-init}) are finite. The potential IR divergence at $k_\perp\to 0$ is regulated by the non-zero transverse momentum flow from the background fields, and the UV divergence at $k_\perp\to \infty$ is regulated by $b_\perp \neq 0$. There is no IR rapidity divergence at $u\to0$.

We isolate the UV rapidity divergence as
\begin{eqnarray}
&&\mathcal{U}^{[\gamma^+]}_{f;{\rm  q(1) + bg}}(z^-, -z^-, \frac{1}{2}b_\perp;\epsilon)\Big|_{Figs.\ \ref{fig:fig3}a,\ref{fig:fig3}b} = \frac{g^2 C_F}{2\pi} \Big(\frac{\mu^2 e^{\gamma_E}}{4\pi}\Big)^\epsilon \int_0^1 du \frac{u}{1-u} \int  \frac{\dhd^{2-2\epsilon} k_{\perp}}{ k^2_{\perp} } e^{i k_{\perp} b_\perp }  \mathcal{U}^{[\gamma^+]}_{f;{\rm  bg}}(  z^- , -z^- , \frac{1}{2}b_\perp)
\nonumber\\
&&+ \frac{g^2 C_F}{2\pi} \Big(\frac{\mu^2 e^{\gamma_E}}{4\pi}\Big)^\epsilon \int \dhd^{2-2\epsilon} p_{\perp} e^{ip_{\perp} b_\perp } \int_0^1 du \frac{u}{1-u} \int \dhd^{2-2\epsilon} k_{\perp} \frac{1}{ (1-u) p^2_{\perp} + u k^2_{\perp} } \int d^{2-2\epsilon}z_\perp e^{-i ( p_\perp - k_{\perp}) z_\perp }
\nonumber\\
&&\times\Big(\mathcal{U}^{[\gamma^+]}_{f;{\rm  bg}}( u z^- , -u z^- , \frac{1}{2}z_\perp) - \mathcal{U}^{[\gamma^+]}_{f;{\rm  bg}}( z^- , - z^- , \frac{1}{2}z_\perp)\Big) + \frac{g^2 C_F}{2\pi} \Big(\frac{\mu^2 e^{\gamma_E}}{4\pi}\Big)^\epsilon \int \dhd^{2-2\epsilon} p_{\perp} e^{ip_{\perp} b_\perp } \int_0^1 du \frac{u}{1-u} \int \dhd^{2-2\epsilon} k_{\perp}
\nonumber\\
&&\times \Big( \frac{ 1 }{ (1-u) p^2_{\perp} + u k^2_{\perp} } - \frac{1}{ k^2_{\perp} } \Big) \int d^{2-2\epsilon}z_\perp e^{-i (p_\perp - k_{\perp}) z_\perp } \mathcal{U}^{[\gamma^+]}_{f;{\rm  bg}}(  z^- , -z^- , \frac{1}{2}z_\perp)\,.
\label{eq:fig3ab-init-sep}
\end{eqnarray}
The first term has a UV divergence at $u \to 1$; its IR divergence at $k_\perp \to 0$ cancels against the last term, leaving Eq.~(\ref{eq:fig3ab-init-sep}) IR finite overall. The first term, combined with the analogous contributions from Figs.~\ref{fig:fig3}e and \ref{fig:fig3}d, reproduces the UV rapidity divergence in Eq.~(\ref{eq:typ-int-div}).

Integrating over $u$ in the last two terms of Eq.~(\ref{eq:fig3ab-init-sep}) gives
\begin{eqnarray}
&&\mathcal{U}^{[\gamma^+]}_{f;{\rm  q(1) + bg}}(z^-, -z^-, \frac{1}{2}b_\perp;\epsilon)\Big|_{Figs.\ \ref{fig:fig3}a,\ref{fig:fig3}b} = \frac{g^2 C_F}{2\pi} \Big(\frac{\mu^2 e^{\gamma_E}}{4\pi}\Big)^\epsilon \int_0^1 du \frac{u}{1-u} \int  \frac{\dhd^{2-2\epsilon} k_{\perp}}{ k^2_{\perp} } e^{i k_{\perp} b_\perp }  \mathcal{U}^{[\gamma^+]}_{f;{\rm  bg}}(  z^- , -z^- , \frac{1}{2}b_\perp)
\nonumber\\
&&+ \frac{g^2 C_F}{2\pi} \Big(\frac{\mu^2 e^{\gamma_E}}{4\pi}\Big)^\epsilon \int \dhd^{2-2\epsilon} p_{\perp} e^{ip_{\perp} b_\perp } \int_0^1 du \frac{u}{1-u} \int \dhd^{2-2\epsilon} k_{\perp} \frac{1}{ (1-u) p^2_{\perp} + u k^2_{\perp} } \int d^{2-2\epsilon}z_\perp e^{-i ( p_\perp - k_{\perp}) z_\perp }
\nonumber\\
&&\times\Big(\mathcal{U}^{[\gamma^+]}_{f;{\rm  bg}}( u z^- , -u z^- , \frac{1}{2}z_\perp) - \mathcal{U}^{[\gamma^+]}_{f;{\rm  bg}}( z^- , - z^- , \frac{1}{2}z_\perp)\Big) + \frac{g^2 C_F}{2\pi} \Big(\frac{\mu^2 e^{\gamma_E}}{4\pi}\Big)^\epsilon \int \dhd^{2-2\epsilon} p_{\perp} e^{ip_{\perp} b_\perp } \int  \frac{\dhd^{2-2\epsilon} k_{\perp}}{ k^2_{\perp} }
\nonumber\\
&&\times \Big(1 + \frac{1 }{1 - k^2_\perp/p^2_\perp}\ln \frac{k^2_\perp}{p^2_\perp} \Big) \int d^{2-2\epsilon}z_\perp e^{-i ( p_\perp - k_\perp) z_\perp } \mathcal{U}^{[\gamma^+]}_{f;{\rm  bg}}(  z^- , -z^- , \frac{1}{2}z_\perp)\,,
\label{eq:fig3ab-init-sep-finIR-result}
\end{eqnarray}
cf.\ Eq.~(\ref{eq:fig3e-result}). This is the final result for the diagrams in Figs.~\ref{fig:fig3}a and \ref{fig:fig3}b.

For the diagram in Fig.~\ref{fig:fig3}c we find
\begin{eqnarray}
&& \mathcal{U}^{[\gamma^+]}_{f;{\rm  q(1) + bg}}(z^-, -z^-, \frac{1}{2}b_\perp;\epsilon)\Big|_{Fig.\ \ref{fig:fig3}c} = \frac{ig^2 C_F (  1 - \epsilon )}{2\pi } \Big(\frac{\mu^2 e^{\gamma_E}}{4\pi}\Big)^\epsilon \int \dhd^{2 - 2\epsilon} p_{\perp} e^{ip_{\perp} b_\perp } \int^1_0 du
\nonumber\\
&&\times \int \dhd^{2 - 2\epsilon} k_{\perp} \frac{1}{( p^2_{\perp} (1-u)  + u k^2_{\perp} )^2 }\Big[ \frac{2ig^{\mu-}}{(2z^-)^2} \Big( \delta(1-u) -   \frac{ 4 p^2_{\perp}}{ p^2_{\perp} (1-u)  + u k^2_{\perp} } +  \frac{6 (1-u) p^4_{\perp} }{( p^2_{\perp} (1-u)  + u k^2_{\perp} )^2 } \Big)
\nonumber\\
&&+ \frac{g^{\mu k} p_k }{2z^-} \Big( 2  -  \frac{4 (1-u) p^2_{\perp}}{ p^2_{\perp} (1-u)  + u k^2_{\perp} } \Big) - i g^{\mu+} (1-u) p^2_\perp  \Big] \int d^{2 - 2\epsilon}z_\perp e^{-i ( p_\perp - k_{\perp} )z_\perp } \mathcal{U}^{[\gamma_\mu]}_{f;{\rm bg}}(u z^-, -u z^-, \frac{1}{2}z_\perp)\,,
\label{eq:fig3c-init}
\end{eqnarray}
where the transverse integrals are regulated by dimensional regularization.

Equation~(\ref{eq:fig3c-init}) contains the Fierz decomposition of the background quark bilinear and therefore mixes the initial $\gamma^+$ channel with the other Dirac projections. This work restricts the operator basis to the leading-power $\gamma^+$ background operator; we consequently discard the other Dirac-projection operators, rather than modifying the hard factor, and retain only
\begin{eqnarray}
&& \mathcal{U}^{[\gamma^+]}_{f;{\rm  q(1) + bg}}(z^-, -z^-, \frac{1}{2}b_\perp;\epsilon)\Big|_{Fig.\ \ref{fig:fig3}c} = \frac{g^2 C_F (  1 - \epsilon )}{2\pi } \Big(\frac{\mu^2 e^{\gamma_E}}{4\pi}\Big)^\epsilon \int \dhd^{2 - 2\epsilon} p_{\perp} e^{ip_{\perp} b_\perp } \int^1_0du (1-u)
\nonumber\\
&&\times \int \dhd^{2 - 2\epsilon} k_{\perp} \frac{ p^2_\perp}{( p^2_{\perp} (1-u)  + u k^2_{\perp} )^2 } \int d^{2 - 2\epsilon}z_\perp e^{-i ( p_\perp - k_{\perp} )z_\perp } \mathcal{U}^{[\gamma^+]}_{f;{\rm bg}}(u z^-, -u z^-, \frac{1}{2}z_\perp)\,.
\label{eq:fig3c-init-nmix}
\end{eqnarray}

For the diagram in Fig.~\ref{fig:fig3}f,
\begin{eqnarray}
&&\mathcal{U}^{[\gamma^+]}_{f;{\rm  q(1) + bg}} (z^-, -z^-, \frac{1}{2}b_\perp;\epsilon)\Big|_{Fig.\ \ref{fig:fig3}f} = -\frac{g^2 C_F(1-\epsilon)}{4\pi} \Big(\frac{\mu^2 e^{\gamma_E}}{4\pi}\Big)^\epsilon \int \dhd^{2-2\epsilon}p_{\perp} e^{i p_{\perp} b_\perp } \int^1_0 du
\nonumber\\
&&\times \int \dhd^{2-2\epsilon}k_{\perp} \frac{ 1 }{ (1-u) (p+k)^2_{\perp} + uk^2_{\perp}} \int d^{2-2\epsilon}z_\perp e^{-i p_{\perp} z_\perp} \mathcal{U}^{[\gamma^+]}_{f;{\rm bg}} (z^-, -z^-, \frac{1}{2}z_\perp )\,.
\label{eq:self-en-init}
\end{eqnarray}
The conjugate of Fig.~\ref{fig:fig3}f is not included; analogously to the LSZ method, it contributes to renormalization of the background-field operators at the $\sigma'$ scales.

From Eq.~(\ref{eq:self-en-init}), Fig.~\ref{fig:fig3}f is IR finite but contains a UV divergence, which we isolate as
\begin{eqnarray}
&&\mathcal{U}^{[\gamma^+]}_{f;{\rm  q(1) + bg}} (z^-, -z^-, \frac{1}{2}b_\perp;\epsilon)\Big|_{Figs.\ \ref{fig:fig3}f}
= -\frac{g^2 C_F}{16\pi^2}  \Big( \frac{1}{\epsilon} + L^{\mu}_b \Big) \mathcal{U}^{[\gamma^+]}_{f;{\rm  bg}}(z^-, -z^-, \frac{1}{2}b_\perp ) -\frac{g^2 C_F(1-\epsilon)}{4\pi} \Big(\frac{\mu^2 e^{\gamma_E}}{4\pi}\Big)^\epsilon
\nonumber\\
&&\times \int \dhd^{2-2\epsilon}p_{\perp} e^{i p_{\perp} b_\perp } \int \dhd^{2-2\epsilon}k_{\perp} \int^1_0 du \frac{ 1 }{ (1-u) (p+k)^2_{\perp} + uk^2_{\perp}} \int d^{2-2\epsilon}z_\perp e^{-i p_{\perp} z_\perp} \mathcal{U}^{[\gamma^+]}_{f;{\rm bg}} (z^-, -z^-, \frac{1}{2}z_\perp )
\nonumber\\
&&+\frac{g^2C_F}{4\pi} \Big(\frac{\mu^2 e^{\gamma_E}}{4\pi}\Big)^\epsilon \int \dhd^{2-2\epsilon}p_\perp e^{ip_\perp b_\perp } \int \frac{\dhd^{2-2\epsilon} k_\perp}{k^2_\perp} \Big(1 - e^{i k_\perp b_\perp } \Big) \int d^{2-2\epsilon}z_\perp e^{- ip_\perp z_\perp } \mathcal{U}^{[\gamma^+]}_{f;{\rm  bg}}(z^-, -z^-, \frac{1}{2}z_\perp )\,,
\label{eq:self-en-result}
\end{eqnarray}
where
\begin{eqnarray}
&&L^\mu_b \equiv \ln\Big(\frac{b^2_\perp \mu^2}{4e^{-2\gamma_E}}\Big)\,.
\label{eq:def-log}
\end{eqnarray}
The sum of the last two terms in Eq.~(\ref{eq:self-en-result}) is finite. Upon renormalization, the scale $\mu$ is identified with the upper cutoff of the divergent transverse integral  $\sigma$ in accordance with the background-field factorization of Sec.~\ref{sec:NLOpcalc}.

Summing Eqs.~(\ref{eq:fig3e-result}), (\ref{eq:fig3ab-init-sep-finIR-result}), (\ref{eq:fig3c-init-nmix}), and (\ref{eq:self-en-result}) gives the combined result for Fig.~\ref{fig:fig3}.
\begin{eqnarray}
&&\mathcal{U}^{[\gamma^+]}_{f;{\rm  q(1) + bg}}(z^-, -z^-, \frac{1}{2}b_\perp;\epsilon)\Big|_{Fig.\ \ref{fig:fig3}}
= \frac{g^2 C_F}{\pi} \Big(\frac{\mu^2 e^{\gamma_E}}{4\pi}\Big)^\epsilon \int_0^1 du \frac{u}{1-u} \int  \frac{\dhd^{2-2\epsilon} k_{\perp}}{ k^2_{\perp} } \Big(e^{i k_{\perp} b_\perp }  - 1 \Big)\mathcal{U}^{[\gamma^+]}_{f;{\rm  bg}}(  z^- , -z^- , \frac{1}{2}b_\perp)
\nonumber\\
&&-\frac{g^2C_F}{16\pi^2}  \Big( \frac{1}{\epsilon} + L^{\mu}_b - 1\Big)
\mathcal{U}^{[\gamma^+]}_{f;{\rm  bg}}(z^-, -z^-, \frac{1}{2}b_\perp )
+ \frac{g^2 C_F}{\pi} \Big(\frac{\mu^2 e^{\gamma_E}}{4\pi}\Big)^\epsilon \int^1_0 du \int d^{2 - 2\epsilon}z_\perp \mathcal{K}(u, z_\perp, b_\perp;\epsilon)
\nonumber\\
&&\times \mathcal{U}^{[\gamma^+]}_{f;{\rm bg}}(u z^-, -u z^-, \frac{1}{2}z_\perp)\,,
\label{eq:fig3bare-init}
\end{eqnarray}
where
\begin{eqnarray}
&&\mathcal{K}(u, z_\perp, b_\perp;\epsilon) \equiv \int \dhd^{2-2\epsilon} p_{\perp} e^{ip_{\perp} b_\perp } \Big( \int \dhd^{2-2\epsilon} k_{\perp} e^{-i ( p_\perp - k_{\perp}) z_\perp } \Big\{  \Big[ \frac{u}{1-u} \frac{1}{ (1-u) p^2_{\perp} + u k^2_{\perp} } \Big]_+
\nonumber\\
&&+ \delta(1-u) \frac{1}{k^2_\perp} \Big( 1 - e^{-i k_{\perp} b_\perp } \Big)\Big( 1 + \frac{1 }{1 - k^2_\perp/p^2_\perp}\ln \frac{k^2_\perp}{p^2_\perp} \Big) + \frac{  1 - \epsilon }{2 } \frac{(1-u) p^2_\perp}{( p^2_{\perp} (1-u)  + u k^2_{\perp} )^2 }
\nonumber\\
&& - \delta(1-u) \frac{1}{4 }\Big( -\frac{1-\epsilon}{p^2_\perp} e^{-i k_{\perp} b_\perp } \frac{1}{1 - k^2_{\perp}/p^2_\perp } \ln\frac{k^2_{\perp}}{p^2_{\perp}}
- \frac{1}{k^2_\perp} \Big(1 - e^{i k_\perp b_\perp } \Big) e^{-i k_{\perp} z_\perp } \Big)\Big\} - \frac{e^{-ip_\perp z_\perp}}{16\pi}\delta(1-u)\Big)\,.
\label{eq:kterms-ep}
\end{eqnarray}

The first two terms in Eq.~(\ref{eq:fig3bare-init}) contain UV divergences at $k_\perp \to \infty$, corresponding to the UV poles of Figs.~\ref{fig:fig3}d, \ref{fig:fig3}e, and \ref{fig:fig3}f discussed above. These are regulated by dimensional regularization, with the scale identified as the upper scale $\sigma$, i.e. the upper cutoff of the UV-divergent transverse integrals after renormalization.

The first term also contains a UV rapidity divergence at $u\to1$, regulated via the $\eta$-regulator scheme, Eq.~(\ref{eq:eta-sec}). Regularizing the UV-divergent integrals and expanding first in $\eta$ and then in $\epsilon$ gives
\begin{eqnarray}
&&\mathcal{U}^{[\gamma^+]}_{f;{\rm  q(1) + bg}}(z^-, -z^-, \frac{1}{2}b_\perp;\epsilon,\eta)\Big|_{Fig.\ \ref{fig:fig3}} = \frac{g^2 C_F}{8\pi^2} \Big\{ \Big(\frac{2}{\eta} + \ln\frac{2\nu^2}{\zeta} + \frac{3}{2}\Big)  \Big(\frac{1}{\epsilon_{UV}} + L^{\mu}_b\Big) + \frac{1}{2}\Big\}\mathcal{U}^{[\gamma^+]}_{f;{\rm  bg}}(z^-, -z^-, \frac{1}{2}b_\perp)
\nonumber\\
&&+ \frac{g^2 C_F}{\pi} \Big(\frac{\mu^2 e^{\gamma_E}}{4\pi}\Big)^\epsilon \int^1_0 du \int d^{2 - 2\epsilon}z_\perp \mathcal{K}(u, z_\perp, b_\perp;\epsilon) \mathcal{U}^{[\gamma^+]}_{f;{\rm bg}}(u z^-, -u z^-, \frac{1}{2}z_\perp)\,.
\label{eq:fig3bare-reg}
\end{eqnarray}
In the first term of Eq.~(\ref{eq:fig3bare-reg}), we explicitly label the pole as $1/\epsilon_{UV}$ to identify it as a UV divergence.

The diagrams in Fig.~\ref{fig:fig3} are IR finite, so Eq.~(\ref{eq:fig3bare-reg}) contains no IR divergences; the nonzero transverse-momentum flow from the background fields regulates any potential IR divergence in the last term. We can therefore take $\epsilon\to 0$ there and write
\begin{eqnarray}
&&\mathcal{U}^{[\gamma^+]}_{f;{\rm  q(1) + bg}}(z^-, -z^-, \frac{1}{2}b_\perp;\epsilon,\eta)\Big|_{Fig.\ \ref{fig:fig3}} = \frac{g^2 C_F}{8\pi^2} \Big\{ \Big(\frac{2}{\eta} + \ln\frac{2\nu^2}{\zeta} + \frac{3}{2}\Big)  \Big(\frac{1}{\epsilon_{UV}} + L^{\mu}_b\Big) +\frac{1}{2}\Big\}\mathcal{U}^{[\gamma^+]}_{f;{\rm  bg}}(z^-, -z^-, \frac{1}{2}b_\perp)
\nonumber\\
&&+ \frac{g^2 C_F}{\pi} \int^1_0 du \int d^{2}z_\perp \mathcal{K}(u, z_\perp, b_\perp) \mathcal{U}^{[\gamma^+]}_{f;{\rm bg}}(u z^-, -u z^-, \frac{1}{2}z_\perp)\,.
\label{eq:fig3bare-regf}
\end{eqnarray}
where
\begin{eqnarray}
&&\mathcal{K}(u, z_\perp, b_\perp) = \int \dhd^{2} p_{\perp} e^{ip_{\perp} b_\perp } \Big\{ \int \dhd^{2} k_{\perp} e^{-i ( p_\perp - k_{\perp} )z_\perp } \Big( \Big[ \frac{u}{1-u} \frac{1}{ (1-u) p^2_{\perp} + u k^2_{\perp} } \Big]_+ + \delta(1-u)  \frac{1}{ k^2_{\perp} }
\\
&&\times \Big(1 - e^{-i k_{\perp} b_\perp } \Big)\Big(1 + \frac{1 }{1 - k^2_\perp/p^2_\perp}\ln \frac{k^2_\perp}{p^2_\perp} \Big) + \frac{1}{2}\frac{ (1-u) p^2_\perp}{( p^2_{\perp} (1-u)  + u k^2_{\perp} )^2 } \Big)
- \frac{e^{-i p_{\perp} z_\perp }}{16\pi} \delta(1-u)\Big( \ln \frac{ 4e^{-2\gamma_E} }{p^2_\perp b^2_\perp } + 2\Big) \Big\} \,,
\nonumber
\end{eqnarray}
which is obtained by taking $\epsilon\to 0$ in Eq.~(\ref{eq:fig3bare-init}).

Equation~(\ref{eq:fig3bare-regf}) is the bare regularized result for the diagrams in Fig.~\ref{fig:fig3}. The separation between field modes is not yet fixed; within the background-field approach this is accomplished through the renormalization procedure introduced in the next section.

\section{Regularization of ultraviolet divergences at the upper scales $\sigma$\label{sec:renorm}}
The beam function in Eq.~(\ref{eq:fig3bare-regf}) is bare operator, and must be renormalized by regulating the UV divergences through the upper scale $\sigma$. In this section, we define the renormalized beam functions in both the standard CSS and rapidity-only schemes.

\subsection{CSS scheme}
Let us define renormalization at the $\sigma=(\mu^2,\nu)$ scales according to the standard TMD factorization scheme by applying the standard TMD renormalization constant for the beam function~(\ref{eq:rrenbeam-def}),
\begin{eqnarray}
&&\mathcal{B}^{[\gamma^+]}_{f}(z^-, -z^-, \frac{1}{2}b_\perp; \mu, \nu/\sqrt{\zeta})
= Z_B(\mu, \epsilon, \nu/\sqrt{\zeta}, \eta) \mathcal{U}^{[\gamma^+]}_{f;{\rm  q + bg}}(z^-, -z^-, \frac{1}{2}b_\perp;\epsilon,\eta)\,,
\label{eq:renorm-stepTMDstandard}
\end{eqnarray}
or, more generally,
\begin{eqnarray}
&&\mathcal{B}^{[\gamma^+]}_{f}(z^-, -z^-, \frac{1}{2}b_\perp; \mu, \nu/\sqrt{\zeta}) = \int^1_0 du \int d^2 z_\perp Z_B(u, z_\perp, b_\perp;\mu, \epsilon, \nu/\sqrt{\zeta}, \eta) \mathcal{U}^{[\gamma^+]}_{f;{\rm  q + bg}}(uz^-, -uz^-, \frac{1}{2}z_\perp;\epsilon,\eta)\,,
\label{eq:tmd-renorm}
\end{eqnarray}
where\footnote{This corresponds to the standard UV renormalization constant $$Z_{UV} = 1 - \frac{g^2 C_F}{8\pi^2} \Big( \frac{1}{\epsilon^2} + \frac{1}{\epsilon} \Big( \frac{3}{2} + \ln \frac{2 \mu^2}{\zeta} \Big)\Big) + \mathcal{O}(\alpha^2_s)$$ for TMDPDFs, and $$Z_S = 1 - \frac{g^2 C_F}{4\pi^2}\Big[ \frac{1}{\epsilon^2} + \frac{1}{\epsilon}  \ln\frac{\mu^2}{\nu^2} -\frac{2}{\eta}\Big(\frac{1}{\epsilon} + L^{\mu}_b\Big) \Big] + \mathcal{O}(\alpha^2_s)$$ for the soft function, see Eq. (\ref{eq:relationsZ}).}
\begin{eqnarray}
&&Z_B(u, z_\perp, b_\perp;\mu, \epsilon, \nu/\sqrt{\zeta}, \eta)
\nonumber\\
&&= \Big(1 - \frac{g^2 C_F}{8\pi^2} \Big( \frac{1}{\epsilon} \Big( \frac{3}{2} + \ln \frac{2\nu^2 }{\zeta} \Big) +  \frac{2}{\eta}\Big(\frac{1}{\epsilon} + L^{\mu}_b\Big) \Big)\Big)\delta(1-u)\delta^2(z_\perp-b_\perp) + \mathcal{O}(\alpha^2_s)\,.
\label{eq:renorm-factTMD}
\end{eqnarray}
This renormalization encompasses both the intrinsic renormalization of the theory and the scheme-imposed renormalization arising from the TMD factorization between the beam function, soft function, and hard factor, see Eq.~(\ref{eq:Urenorm}).

Applying Eq. (\ref{eq:renorm-factTMD}) to the $\sigma$-scale renormalization of Eq. (\ref{eq:fig3bare-regf}), we obtain\footnote{In computing Eq. (\ref{eq:fig3bare-renorm}), we take into account contribution of the leading term in ``quantum" fields $\mathcal{U}^{[\gamma^+]}_{f;{\rm q(0) + bg}} = \mathcal{U}^{[\gamma^+]}_{f; {\rm bg}}$, i.e. $$\mathcal{U}^{[\gamma^+]}_{f;{\rm q + bg}} =\mathcal{U}^{[\gamma^+]}_{f;{\rm q(0) + bg}} + \mathcal{U}^{[\gamma^+]}_{f;{\rm q(1) + bg}} + \dots \,.$$}
\begin{eqnarray}
&&\mathcal{B}^{[\gamma^+]}_{f}(z^-, -z^-, \frac{1}{2}b_\perp; \mu, \nu/\sqrt{\zeta}) = \mathcal{U}^{[\gamma^+]}_{f;{\rm bg}}(z^-, -z^-, \frac{1}{2}b_\perp) + \frac{g^2 C_F}{8\pi^2} \Big\{ \Big( \frac{3}{2} + \ln\frac{2\nu^2}{\zeta} \Big) L^{\mu}_b + \frac{1}{2}\Big\}
\label{eq:fig3bare-renorm}\\
&&\times \mathcal{U}^{[\gamma^+]}_{f;{\rm  bg}}(z^-, -z^-, \frac{1}{2}b_\perp) + \frac{g^2 C_F}{\pi}  \int^1_0 du \int d^{2}z_\perp \mathcal{K}(u, z_\perp, b_\perp) \mathcal{U}^{[\gamma^+]}_{f;{\rm bg}}(u z^-, -u z^-, \frac{1}{2}z_\perp) + \mathcal{O}(\alpha^2_s)\,.
\nonumber
\end{eqnarray}
The matrix element of this operator, combined with the soft-function contribution according to Eq.~(\ref{eq:TMD-renorm}), yields the TMDPDFs in Eq.~(\ref{eq:intTMDff}), and the logarithms are at UV scales $\sigma=(\mu^2,\nu)$.

\subsubsection{Soft function in the CSS scheme}
\begin{figure}[htb]
\begin{center}
\includegraphics[width=160mm]{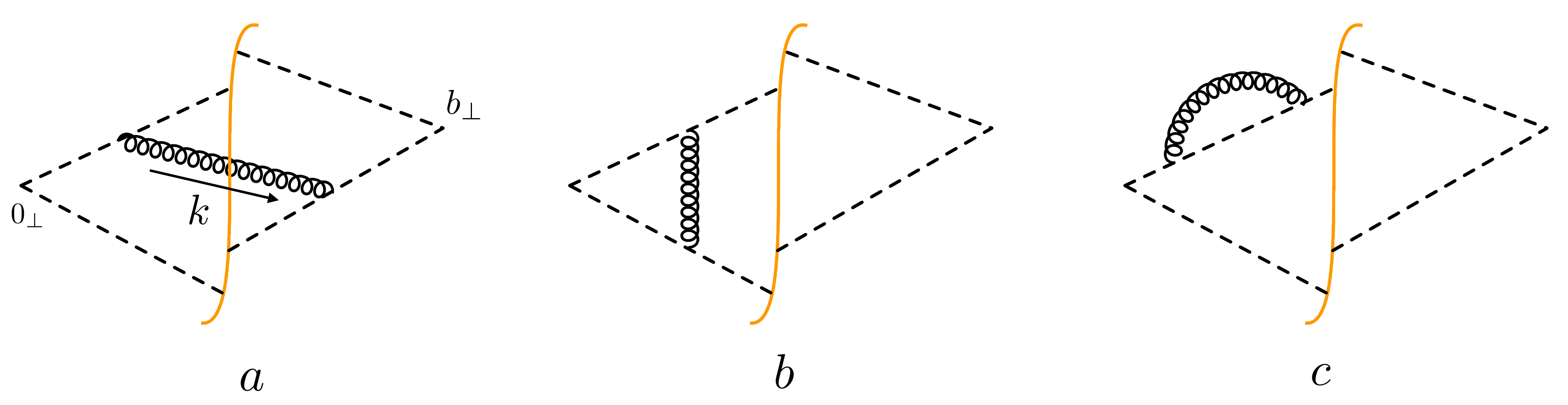}
\end{center}
\caption{\label{fig:soft}The soft function at the next-to-leading order. The conjugated diagrams are not explicitly shown.}
\end{figure}

The soft function (\ref{eq:soft-function}) can be computed straightforwardly at next-to-leading order in the $\eta$-regulator scheme. The next-to-leading contribution to the soft function is given by the diagrams in Fig.~\ref{fig:soft}.
Computing the diagrams in Fig.~\ref{fig:soft}, together with their complex conjugates, in Feynman gauge, we obtain the following expression for the bare soft function at the next-to-leading order:
\begin{eqnarray}
&&\mathcal{S}(b_\perp) = 1 + g^2 C_F \int \dhd^4k 2\pi\delta(k^2) \Big[ \frac{1}{k^+-i\epsilon} \frac{1}{k^-+i\epsilon} + \frac{1}{k^--i\epsilon} \frac{1}{k^+ + i\epsilon} \Big] e^{ik_\perp b_\perp}
\nonumber\\
&&- 2i g^2 C_F \int^\infty_{-\infty} \frac{dk^+}{2\pi} \int^\infty_{-\infty} \frac{dk^-}{2\pi} \int \dhd^2k_\perp \frac{1}{k^- + i\epsilon} \frac{1}{k^+ - i\epsilon}\frac{1}{2k^+ k^- - k^2_\perp + i\epsilon} + \mathcal{O}(\alpha^2_s)\,,
\end{eqnarray}
where the second term corresponds to the diagram in Fig.~\ref{fig:soft}a, while the last term represents the diagram in Fig.~\ref{fig:soft}b.
Using the $\delta$-function in the second term and taking the $k^+$ poles in the last term, we can carry out the integration over $k^+$ variable, which sets $k^+ = k^2_\perp/(2k^-)$. As a result, we obtain
\begin{eqnarray}
&&\mathcal{S}(b_\perp) = 1 + \frac{g^2 C_F}{\pi} \int^\infty_0 \frac{dk^-}{k^-} \int \frac{\dhd^2k_\perp}{k^2_\perp} \Big(  e^{ik_\perp b_\perp} -  1 \Big) + \mathcal{O}(\alpha^2_s) \,.
\label{eq:soft-kmin}
\end{eqnarray}
In terms of the variable
\begin{eqnarray}
&&u \equiv \frac{xp^+}{xp^+ + k^+} = \frac{xp^+}{xp^+ + \frac{k^2_\perp}{2k^- }}\,,
\label{var:u}
\end{eqnarray}
c.f. Eq. (\ref{eq:u-first}), we can rewrite the expression for the soft function as
\begin{eqnarray}
&&\mathcal{S}(b_\perp) = 1 + \frac{g^2 C_F}{\pi} \int^1_0 \frac{du}{u(1-u)} \int \frac{\dhd^2k_\perp}{k^2_\perp} \Big(  e^{ik_\perp b_\perp} -  1 \Big) + \mathcal{O}(\alpha^2_s) \,.
\label{eq:soft-u}
\end{eqnarray}
This result contains two types of divergences: the UV divergence of the transverse integral, which can be regulated by dimensional regularization, and the rapidity divergence of the integral over the variable $u$. In the $\eta$-regulator scheme, the rapidity divergence of the soft function can be regulated by inserting the factor
\begin{eqnarray}
&&\nu^\eta 2^{-\eta/2} |k^+ - k^-|^{-\eta} = \nu^\eta 2^{-\eta/2} \Big|\frac{k^2_\perp}{2k^-} - k^-\Big|^{-\eta}
\label{eq:reg-fact-eta}
\end{eqnarray}
in Eq. (\ref{eq:soft-kmin}), or equivalently
\begin{eqnarray}
&&\nu^\eta \Big| \sqrt{\zeta} \frac{1-u}{u} -  \frac{ k^2_\perp }{\sqrt{\zeta} } \frac{u}{1-u}\Big|^{-\eta}
\end{eqnarray}
in Eq. (\ref{eq:soft-u}), where $\zeta = 2(xp^+)^2$. Note that, after inserting this factor, both the rapidity UV divergence at $k^- \to \infty$, equivalently $u\to 1$, and the rapidity IR divergence at $k^- \to 0$, equivalently $u\to 0$ and $k^+\to \infty$, are regulated by the same regulator $\eta$ and the corresponding scale $\nu$. In the language of kinematic cutoffs, this means that the rapidity divergences appearing in the $k^-\to\infty$ and $k^+\to\infty$ limits are regulated by the same factorization scale $\nu$.
Inserting this factor in Eq. (\ref{eq:soft-kmin}):
\begin{eqnarray}
&&\mathcal{S}(b_\perp; \epsilon, \eta) = 1 + \frac{g^2 C_F}{\pi} \nu^\eta 2^{-\eta/2} \Big(\frac{\mu^2 e^{\gamma_E}}{4\pi}\Big)^\epsilon \int^\infty_0 \frac{dk^-}{k^-}  \Big|\frac{k^2_\perp}{2k^-} - k^-\Big|^{-\eta} \int \frac{\dhd^{2-2\epsilon}k_\perp}{k^2_\perp} \Big(  e^{ik_\perp b_\perp} -  1 \Big) + \mathcal{O}(\alpha^2_s) \,,
\end{eqnarray}
we can perform integration over $k^-$ and $k_\perp$ variables. Expanding first in $\eta$ and then in $\epsilon$, we find
\begin{eqnarray}
&&\mathcal{S}(b_\perp;\epsilon,\eta)
= 1 + \frac{ g^2 C_F }{4\pi^2} \Big[ \frac{1}{\epsilon^2} + (\frac{1}{\epsilon} + L^\mu_b)\Big( -\frac{2}{\eta}  +  \ln \frac{\mu^2}{\nu^2}  \Big) -  \frac{1}{2} (L^\mu_b)^2 - \frac{\pi ^2 }{12}\Big] + \mathcal{O}(\alpha^2_s) \,.
\label{ap:sf-fin}
\end{eqnarray}
Applying the renormalization factor for the soft function (\ref{eq:rrensoft}), we obtain
\begin{eqnarray}
&&\mathcal{S}(b_\perp; \mu, \nu) = Z_S(\mu, \epsilon, \mu/\nu, \eta) \mathcal{S}(b_\perp; \epsilon, \eta)
\nonumber\\
&&= 1 + \frac{ g^2 C_F }{4\pi^2} \Big(-  \frac{1}{2} (L^\mu_b)^2 + L^\mu_b  \ln \frac{\mu^2}{\nu^2}  - \frac{\pi ^2 }{12}\Big) + \mathcal{O}(\alpha^2_s)\,,
\label{eq:soft-ren}
\end{eqnarray}
where we used
\begin{eqnarray}
&&Z_S(\mu, \epsilon, \mu/\nu, \eta) = 1 - \frac{g^2 C_F}{4\pi^2}\Big[ \frac{1}{\epsilon^2} + \frac{1}{\epsilon}  \ln\frac{\mu^2}{\nu^2} -\frac{2}{\eta}\Big(\frac{1}{\epsilon} + L^{\mu}_b\Big) \Big] + \mathcal{O}(\alpha^2_s)\,.
\end{eqnarray}
Finally, combining the renormalized soft function in Eq.~(\ref{eq:soft-ren}) with the renormalized beam function in Eq.~(\ref{eq:fig3bare-renorm}), one straightforwardly obtains the expression for the renormalized TMD operator, see Eq.~(\ref{eq:TMD-renorm}):
\begin{eqnarray}
&&\mathcal{T}^{[\gamma^+]}_{f}(z^-, -z^-, \frac{1}{2}b_\perp; \mu, \zeta) = \mathcal{U}^{[\gamma^+]}_{f;{\rm bg}}(z^-, -z^-, \frac{1}{2}b_\perp)
\label{eq:fig3bare-renorm-result}\\
&&+ \frac{g^2 C_F}{8\pi^2} \Big\{ -  \frac{1}{2} (L^\mu_b)^2 + \Big( \frac{3}{2}  +  \ln\frac{2\mu^2}{\zeta} \Big) L^{\mu}_b + \frac{1}{2}  - \frac{\pi ^2 }{12} \Big\} \mathcal{U}^{[\gamma^+]}_{f;{\rm bg}}(z^-, -z^-, \frac{1}{2}b_\perp)
\nonumber\\
&&+ \frac{g^2 C_F}{\pi}  \int^1_0 du \int d^{2}z_\perp \mathcal{K}(u, z_\perp, b_\perp) \mathcal{U}^{[\gamma^+]}_{f;{\rm bg}}(u z^-, -u z^-, \frac{1}{2}z_\perp) + \mathcal{O}(\alpha^2_s)\,,
\nonumber
\end{eqnarray}
where the second term, which originates from the UV renormalization of the TMDPDFs, agrees with the known results, see e.g. Ref. \cite{Ebert:2019okf}. This leads to the standard two-scale evolution for TMDPDFs (\ref{eq:CSS1}) and (\ref{eq:CSS2}).

\subsection{Rapidity-only scheme\label{sec:rap-onlyUV}}
We now consider an alternative UV renormalization of the bare beam function in the rapidity-only scheme by imposing explicit cutoffs on the kinematic variables at the upper scale $\sigma$. The procedure differs from the CSS scheme multiplicative renormalization and defines a rapidity-only scheme, which is distinct from standard TMD factorization but related through the matching discussed below. To introduce the cutoff renormalization procedure, let us rewrite the bare result of computing the beam function at the next-to-leading order (\ref{eq:fig3bare-regf}), see also Eq. (\ref{eq:fig3bare-init}) and its derivation, as
\begin{eqnarray}
&&\mathcal{U}^{[\gamma^+]}_{f;{\rm  q(1) + bg}}(z^-, -z^-, \frac{1}{2}b_\perp;\epsilon)
= \mathcal{U}^{[\gamma^+]}_{f;{\rm bg}}(z^-, -z^-, \frac{1}{2}b_\perp) + \frac{g^2 C_F}{\pi} \Big(\frac{\mu^2 e^{\gamma_E}}{4\pi}\Big)^\epsilon \int_0^1 du \frac{u}{1-u} \int  \frac{\dhd^{2-2\epsilon} k_{\perp}}{ k^2_{\perp} } \Big(e^{i k_{\perp} b_\perp }  - 1 \Big)
\nonumber\\
&&\times \mathcal{U}^{[\gamma^+]}_{f;{\rm  bg}}(  z^- , -z^- , \frac{1}{2}b_\perp) -\frac{g^2C_F}{4\pi} \Big( \frac{\mu^{2} e^{\gamma_E}}{ 4\pi} \Big)^{\epsilon}\int \frac{\dhd^{2-2\epsilon} k_\perp}{k^2_\perp} \Big(1 - e^{i k_\perp b_\perp } \Big)
\mathcal{U}^{[\gamma^+]}_{f;{\rm  bg}}(z^-, -z^-, \frac{1}{2}b_\perp ) + \frac{g^2C_F}{16\pi^2}
\mathcal{U}^{[\gamma^+]}_{f;{\rm  bg}}(z^-, -z^-, \frac{1}{2}b_\perp )
\nonumber\\
&&+ \frac{g^2 C_F}{\pi} \int^1_0 du \int d^{2 }z_\perp \mathcal{K}(u, z_\perp, b_\perp) \mathcal{U}^{[\gamma^+]}_{f;{\rm bg}}(u z^-, -u z^-, \frac{1}{2}z_\perp) + \mathcal{O}(\alpha^2_s) \,,
\label{eq:fig3-uv-new}
\end{eqnarray}
Here the second and third terms are UV-divergent integrals. Under standard TMD renormalization they generate CSS evolution. This section instead regulates them in the rapidity-only scheme. In contrast to the lower rapidity-only scheme discussed in Sec.~\ref{sec:rapidity-only}, whose cutoff partitions finite quantum and background contributions at $\sigma'$, the present cutoff defines ultraviolet TMD-operator dynamics at the upper scale $\sigma$.
Before introducing the rapidity-only scheme, we
combine the second and the third term in Eq.~(\ref{eq:fig3-uv-new})

to get
\begin{eqnarray}
&&\mathcal{U}^{[\gamma^+]}_{f;{\rm  q(1) + bg}}(z^-, -z^-, \frac{1}{2}b_\perp;\epsilon)
= \mathcal{U}^{[\gamma^+]}_{f;{\rm bg}}(z^-, -z^-, \frac{1}{2}b_\perp)
\nonumber\\
&&-\frac{g^2C_F}{2\pi}  \Big( \frac{\mu^{2} e^{\gamma_E}}{ 4\pi} \Big)^{\epsilon} \int^1_0 du \frac{1+u^2}{1-u} \int \frac{\dhd^{2-2\epsilon} k_\perp}{k^2_\perp} \Big(1 - e^{i k_\perp b_\perp } \Big) \mathcal{U}^{[\gamma^+]}_{f;{\rm  bg}}(z^-, -z^-, \frac{1}{2}b_\perp ) + \frac{g^2C_F}{16\pi^2}
\mathcal{U}^{[\gamma^+]}_{f;{\rm  bg}}(z^-, -z^-, \frac{1}{2}b_\perp )
\nonumber\\
&&+ \frac{g^2 C_F}{\pi} \int^1_0 du \int d^{2}z_\perp \mathcal{K}(u, z_\perp, b_\perp) \mathcal{U}^{[\gamma^+]}_{f;{\rm bg}}(u z^-, -u z^-, \frac{1}{2}z_\perp) + \mathcal{O}(\alpha^2_s) \,.
\label{eq:fig3-uv-new-form2}
\end{eqnarray}
Here the  second term represents the UV-divergent part of the expression, the third term is a finite contribution originating from the UV-divergent diagram in Fig.~\ref{fig:fig3}f, and the last term collects the remaining finite contributions arising in the computation of the beam function.
As we will see below, the cutoff regulator in the rapidity-only renormalization scheme regulates both the rapidity divergence of the integral over the variable $u$ and the divergence of the transverse-momentum integrals. For this reason, we can remove dimensional regularization from the UV-divergent integral. Replacing the variable $u$ with the longitudinal variable
\begin{eqnarray}
&&k^- \equiv \frac{u}{1 - u} \frac{k^2_\perp}{2x p^+}\,,
\label{eq:long-var}
\end{eqnarray}
we can rewrite the Fourier transform of Eq.~(\ref{eq:fig3-uv-new-form2}) as
\begin{eqnarray}
&&\int \frac{dz^-}{2\pi} e^{-2ix p^+z^-} \mathcal{U}^{[\gamma^+]}_{f;{\rm  q(1) + bg}}(z^-, -z^-, \frac{1}{2}b_\perp)
= \int \frac{dz^-}{2\pi} e^{-2ix p^+z^-} \mathcal{U}^{[\gamma^+]}_{f;{\rm bg}}(z^-, -z^-, \frac{1}{2}b_\perp) -\frac{g^2C_F}{2\pi} \int^\infty_0 \frac{dk^-}{k^-} \int \frac{\dhd^{2} k_\perp}{k^2_\perp}
\nonumber\\
&&\times \Big(1 - e^{i k_\perp b_\perp } \Big) \frac{2xp^+ k^-}{(2xp^+ k^- + k^2_\perp)^2} \frac{(2xp^+ k^- + k^2_\perp)^2 + (2xp^+ k^-)^2}{2xp^+ k^- + k^2_\perp}
\int \frac{dz^-}{2\pi} e^{-2ix p^+z^-} \mathcal{U}^{[\gamma^+]}_{f;{\rm  bg}}(z^-, -z^-, \frac{1}{2}b_\perp )
\nonumber\\
&&+ \frac{g^2C_F}{16\pi^2}
\int \frac{dz^-}{2\pi} e^{-2ix p^+z^-} \mathcal{U}^{[\gamma^+]}_{f;{\rm  bg}}(z^-, -z^-, \frac{1}{2}b_\perp ) + \frac{g^2 C_F}{\pi} \int^1_0 \frac{du}{u} \int d^{2}z_\perp \mathcal{K}(u, z_\perp, b_\perp)
\nonumber\\
&&\times \int \frac{dz^-}{2\pi} e^{-2i\frac{x}{u} p^+ z^-} \mathcal{U}^{[\gamma^+]}_{f;{\rm bg}}( z^-, - z^-, \frac{1}{2}z_\perp) + \mathcal{O}(\alpha^2_s) \,,
\label{eq:fig3-uv-new-km}
\end{eqnarray}
where the second term contains the UV-divergent contribution of the original expression, while the remaining terms are finite. Note that the only divergence of the UV-divergent term appears in the limit of $k^-\to\infty$, while the integral over the transverse-momentum variable remains finite. From this, we conclude that the rapidity UV divergence in Eq.~(\ref{eq:fig3-uv-new-km}), which arises in the limit $k^- \to \infty$, accounts for both UV divergences in Eq.~(\ref{eq:fig3-uv-new-form2}), associated with the limits $u \to 1$ and $k_\perp \to \infty$.
The rapidity-only scheme at the upper scale $\sigma$ can now be introduced through the replacement
\begin{eqnarray}
&&\int_0^\infty \frac{dk^-}{k^-}\to \int_0^\sigma \frac{dk^-}{k^-}\,,
\label{eq:UV-rep}
\end{eqnarray}
with the remaining part of the integral absorbed into the renormalization of the field operator (\ref{eq:fig3-uv-new-km}), giving rise to the renormalized rapidity-only TMDPDFs:
\begin{eqnarray}
&&\mathcal{R}^{[\gamma^+]}_{f}(z^-, -z^-, \frac{1}{2}b_\perp;\sigma)\,.
\end{eqnarray}
Note that, in this formulation, the cutoff parameter $\sigma$ plays the role of the factorization scale in the UV region for the quantum field modes and regulates all UV divergences that arise in the computation of the beam function.
Introducing the rapidity cutoff in the UV-divergent terms as
\begin{eqnarray}
&& -\frac{g^2C_F}{2\pi} \int^\sigma_0 \frac{dk^-}{k^-} \int \frac{\dhd^{2} k_\perp}{k^2_\perp} \Big(1 - e^{i k_\perp b_\perp } \Big) \frac{2xp^+ k^-}{(2xp^+ k^- + k^2_\perp)^2} \frac{(2xp^+ k^- + k^2_\perp)^2 + (2xp^+ k^-)^2}{2xp^+ k^- + k^2_\perp}
\nonumber\\
&&\times \int \frac{dz^-}{2\pi} e^{-2ix p^+z^-} \mathcal{U}^{[\gamma^+]}_{f;{\rm  bg}}(z^-, -z^-, \frac{1}{2}b_\perp )
\end{eqnarray}
and integrating over $k^-$ in these terms, we obtain
\begin{eqnarray}
&&\int \frac{dz^-}{2\pi} e^{-2ix p^+z^-} \mathcal{R}^{[\gamma^+]}_{f}(z^-, -z^-, \frac{1}{2}b_\perp;\sigma)
= \int \frac{dz^-}{2\pi} e^{-2ix p^+z^-} \mathcal{U}^{[\gamma^+]}_{f;{\rm bg}}(z^-, -z^-, \frac{1}{2}b_\perp) -\frac{g^2C_F}{2\pi} \int \frac{\dhd^{2} k_\perp}{k^2_\perp} \Big(1 - e^{i k_\perp b_\perp } \Big)
\nonumber\\
&&\times \Big(2 \ln \frac{2xp^+\sigma + k^2_\perp}{k^2_\perp} + \frac{ 2 k^2_\perp}{ 2xp^+\sigma+k^2_\perp } - \frac{ (k^2_\perp)^2 }{2 (2xp^+\sigma+k^2_\perp)^2}-\frac{3}{2} \Big)
\int \frac{dz^-}{2\pi} e^{-2ix p^+z^-} \mathcal{U}^{[\gamma^+]}_{f;{\rm  bg}}(z^-, -z^-, \frac{1}{2}b_\perp )
\nonumber\\
&&+ \frac{g^2C_F}{16\pi^2}
\int \frac{dz^-}{2\pi} e^{-2ix p^+z^-} \mathcal{U}^{[\gamma^+]}_{f;{\rm  bg}}(z^-, -z^-, \frac{1}{2}b_\perp ) + \frac{g^2 C_F}{\pi} \int^1_0 \frac{du}{u} \int d^{2}z_\perp \mathcal{K}(u, z_\perp, b_\perp)
\nonumber\\
&&\times \int \frac{dz^-}{2\pi} e^{-2i\frac{x}{u} p^+ z^-} \mathcal{U}^{[\gamma^+]}_{f;{\rm bg}}( z^-, - z^-, \frac{1}{2}z_\perp) + \mathcal{O}(\alpha^2_s) \,.
\label{eq:fig3-uv-new-regUV}
\end{eqnarray}
From this result, one can directly derive an evolution equation for the rapidity-only TMDPDFs:
\begin{eqnarray}
&&\sigma\frac{d}{d\sigma}\int \frac{dz^-}{2\pi} e^{-2ix p^+z^-} \mathcal{R}^{[\gamma^+]}_{f}(z^-, -z^-, \frac{1}{2}b_\perp;\sigma) = -\frac{g^2C_F}{2\pi} \int \frac{\dhd^{2} k_\perp}{k^2_\perp} \Big(1 - e^{i k_\perp b_\perp } \Big)
\label{eq:RFTMDevol}\\
&&\times \frac{2xp^+\sigma}{(2xp^+\sigma + k^2_\perp)^2} \frac{(2xp^+\sigma+k^2_\perp)^2 + (2xp^+\sigma)^2 }{ 2xp^+\sigma+k^2_\perp}
\int \frac{dz^-}{2\pi} e^{-2ix p^+z^-} \mathcal{R}^{[\gamma^+]}_{f}(z^-, -z^-, \frac{1}{2}b_\perp; \sigma) +\mathcal{O}(\alpha^2_s)\,.
\nonumber
\end{eqnarray}
Because the rapidity-only regulator controls both the rapidity divergence and the divergence of the transverse-momentum integrals, the rapidity-only scheme involves a single evolution equation, rather than the two equations of standard TMD factorization.

\subsubsection{Soft function in the rapidity-only scheme\label{sec:soft-rap-only}}
In the previous section, we derived an evolution equation for the rapidity-only TMDPDFs (\ref{eq:RFTMDevol}), which is analogous to the two-scale evolution, see Eqs. (\ref{eq:CSS1}) and (\ref{eq:CSS2}), within the standard TMD factorization approach.
As discussed in Sec.~\ref{sec:definitions}, standard two-scale TMD evolution involves the soft factor, Eq.~(\ref{eq:Urenorm}). The rapidity-only evolution equation (\ref{eq:RFTMDevol}) does not include it as a separate contribution, which is unnecessary in the rapidity-only scheme~\cite{Balitsky:2015qba}. We verify this by computing the soft factor in that scheme and comparing it with the beam function (\ref{eq:fig3-uv-new-regUV}).
Let us start with the expression for the soft factor (\ref{eq:soft-kmin}), which can be rewritten in the following form:
\begin{eqnarray}
&&\mathcal{S}(b_\perp) = 1 + \frac{g^2 C_F}{\pi} \int^\infty_0 \frac{dk^-}{k^-} \frac{2xp^+ k^-}{2xp^+ k^- + k^2_\perp} \int \frac{\dhd^2k_\perp}{k^2_\perp} \Big(  e^{ik_\perp b_\perp} -  1 \Big)
\nonumber\\
&&+ \frac{g^2 C_F}{\pi} \int_0^\infty \frac{dk^+}{k^+}\frac{k^+}{xp^+ + k^+} \int \frac{\dhd^2k_\perp}{k^2_\perp} \Big(  e^{ik_\perp b_\perp} -  1 \Big) + \mathcal{O}(\alpha^2_s) \,,
\label{eq:soft-fact-last}
\end{eqnarray}
where $k^+ = k^2_\perp/(2k^-)$. It is easy to see that the second term in this expression contains a UV rapidity divergence in the limit $k^-\to\infty$. At the same time, the last term contains a UV rapidity divergence in the limit $k^+ \to \infty$, which corresponds to $k^-\to0$. Expressed in terms of the $u$ variable (\ref{var:u}), these two limits correspond, respectively, to the UV divergence as $u\to1$ and the IR divergence as $u\to0$, see Eq. (\ref{eq:soft-u}).
The expression for the soft factor in Eq. (\ref{eq:soft-fact-last}) also contains a UV divergence in the limit $k_\perp\to\infty$. However, for the discussion below, we do not need to consider this divergence explicitly.
The rapidity divergencies in the soft factor need to be regulated. In the $\eta$-regulator scheme, this is achieved by including both $k^+$ and $k^-$ variables in the regulator factor (\ref{eq:reg-fact-eta}). Now, within the rapidity-only regularization, let us impose cutoff regulators $\sigma^-$ and $\sigma^+$ for the momentum variables $k^-$ and $k^+$, respectively. Note that the upper cutoff of the $k^+$ variable imposes a lower cutoff for the $k^-$, and vice versa, since the two variables are related by $k^- = k^2_\perp/(2k^+)$. This leads to the following regulated expression for the soft factor
\begin{eqnarray}
&&\mathcal{S}(b_\perp) = 1 + \frac{g^2 C_F}{\pi}  \int \frac{\dhd^2k_\perp}{k^2_\perp} \Big(  e^{ik_\perp b_\perp} -  1 \Big) \int^{\sigma^-}_{\frac{k^2_\perp}{2\sigma^+}} dk^- \frac{1 }{k^- + \frac{k^2_\perp}{2xp^+}}
\nonumber\\
&&+ \frac{g^2 C_F}{\pi}  \int \frac{\dhd^2k_\perp}{k^2_\perp} \Big(  e^{ik_\perp b_\perp} -  1 \Big) \int_{\frac{k^2_\perp}{2\sigma^-}}^{\sigma^+} dk^+\frac{1}{k^+ + xp^+} + \mathcal{O}(\alpha^2_s) \,.
\end{eqnarray}
Integrating over the $k^-$ and $k^+$ variables, it is easy to obtain the following form of the soft factor
\begin{eqnarray}
&&\mathcal{S}(b_\perp) = 1 + \frac{g^2 C_F}{\pi} \int \frac{\dhd^2k_\perp}{k^2_\perp} \Big(  e^{ik_\perp b_\perp} -  1 \Big) \ln \frac{ 2 \sigma^+ \sigma^- }{k^2_\perp} + \mathcal{O}(\alpha^2_s) \,.
\end{eqnarray}
Note that the soft factor has to be distributed evenly between the two interacting hadrons moving along the $P^+$ and $P^-$ directions. Therefore, we can write
\begin{eqnarray}
&&\mathcal{S}(b_\perp) = 1 + \frac{g^2 C_F}{2\pi}  \int \frac{\dhd^2k_\perp}{k^2_\perp} \Big(  e^{ik_\perp b_\perp} -  1 \Big) \Big( \ln \frac{2 (\sigma^+)^2}{k^2_\perp} + \ln \frac{2 (\sigma^-)^2}{k^2_\perp} \Big) + \mathcal{O}(\alpha^2_s) \,,
\end{eqnarray}
and attribute
\begin{eqnarray}
&&-\frac{g^2 C_F}{2\pi} \int \frac{\dhd^2k_\perp}{k^2_\perp} \Big(1 - e^{ik_\perp b_\perp} \Big) \ln \frac{2 \sigma^2}{k^2_\perp}
\label{eq:softcont-rap}
\end{eqnarray}
to the soft-factor contribution associated with the hadron moving along the $P^+$ direction. In the last equation, we have chosen the cutoff parameters for the two hadrons to be equal, $\sigma^+ = \sigma^- = \sigma$. Similarly, in the $\eta$-regulator scheme, the same regulator $\eta$ regulates both rapidity divergences $u\to1$ and $u\to0$ in the soft factor.
Our goal is to show that the soft contribution in Eq.~(\ref{eq:softcont-rap}) is already included in the rapidity-only beam function, Eq.~(\ref{eq:fig3-uv-new-regUV}); adding it to Eq.~(\ref{eq:RFTMDevol}) would double count it.
To demonstrate this, let us consider the rapidity-only regularized expression for the beam function (\ref{eq:fig3-uv-new-regUV}) in the Sudakov regime, where $x$ is large and all transverse momenta are small, $p_\perp\sim k_\perp$, such that $2xp^+\sigma \gg k^2_\perp$. In this limit, the beam function takes the form
\begin{eqnarray}
&&\int \frac{dz^-}{2\pi} e^{-2ix p^+z^-} \mathcal{R}^{[\gamma^+]}_{f}(z^-, -z^-, \frac{1}{2}b_\perp;\sigma)\Big|_{\rm Sud.}
= \int \frac{dz^-}{2\pi} e^{-2ix p^+z^-} \mathcal{U}^{[\gamma^+]}_{f;{\rm bg}}(z^-, -z^-, \frac{1}{2}b_\perp) -\frac{g^2C_F}{2\pi} \int \frac{\dhd^{2} k_\perp}{k^2_\perp} \Big(1 - e^{i k_\perp b_\perp } \Big)
\nonumber\\
&&\times \Big( \ln \frac{2\sigma^2 }{k^2_\perp} + \ln \frac{\zeta}{k^2_\perp}  + \frac{ 2 k^2_\perp}{ 2xp^+\sigma } - \frac{ (k^2_\perp)^2 }{2 (2xp^+\sigma)^2}-\frac{3}{2} \Big)
\int \frac{dz^-}{2\pi} e^{-2ix p^+z^-} \mathcal{U}^{[\gamma^+]}_{f;{\rm  bg}}(z^-, -z^-, \frac{1}{2}b_\perp )
\nonumber\\
&&+ \frac{g^2C_F}{16\pi^2}
\int \frac{dz^-}{2\pi} e^{-2ix p^+z^-} \mathcal{U}^{[\gamma^+]}_{f;{\rm  bg}}(z^-, -z^-, \frac{1}{2}b_\perp ) + \frac{g^2 C_F}{\pi} \int^1_0 \frac{du}{u} \int d^{2}z_\perp \mathcal{K}(u, z_\perp, b_\perp)
\nonumber\\
&&\times \int \frac{dz^-}{2\pi} e^{-2i\frac{x}{u} p^+ z^-} \mathcal{U}^{[\gamma^+]}_{f;{\rm bg}}( z^-, - z^-, \frac{1}{2}z_\perp) + \mathcal{O}(\alpha^2_s) \,,
\label{eq:fig3-uv-new-regUV-sud}
\end{eqnarray}
where, as usual, we have introduced a variable $\zeta = 2(xp^+)^2$.
Comparing Eqs.~(\ref{eq:softcont-rap}) and (\ref{eq:fig3-uv-new-regUV-sud}), the rapidity-only scheme already includes the soft contribution in the regularized beam function.
The rapidity-only evolution equation, Eq.~(\ref{eq:RFTMDevol}), is therefore the one-scale counterpart of the standard CSS equations, Eqs.~(\ref{eq:CSS1}) and~(\ref{eq:CSS2}).
Note that, in the Sudakov regime, the evolution equation for the rapidity-only TMDPDFs in Eq. (\ref{eq:RFTMDevol}) takes the form
\begin{eqnarray}
&&\sigma\frac{d}{d\sigma}\int \frac{dz^-}{2\pi} e^{-2ix p^+z^-} \mathcal{R}^{[\gamma^+]}_{f}(z^-, -z^-, \frac{1}{2}b_\perp;\sigma)\Big|_{\rm Sud.} = -\frac{g^2C_F}{\pi} \int^{2xp^+\sigma} \frac{\dhd^{2} k_\perp}{k^2_\perp} \Big(1 - e^{i k_\perp b_\perp } \Big)
\\
&&\times \int \frac{dz^-}{2\pi} e^{-2ix p^+z^-} \mathcal{R}^{[\gamma^+]}_{f}(z^-, -z^-, \frac{1}{2}b_\perp; \sigma) +\mathcal{O}(\alpha^2_s)\,,
\nonumber
\end{eqnarray}
where the upper limit of the integral is fixed by the condition $2xp^+\sigma \gg k_\perp^2$, which defines the Sudakov regime of the emission. This leads to the estimate
\begin{eqnarray}
&&\sigma\frac{d}{d\sigma}\int \frac{dz^-}{2\pi} e^{-2ix p^+z^-} \mathcal{R}^{[\gamma^+]}_{f}(z^-, -z^-, \frac{1}{2}b_\perp;\sigma)\Big|_{\rm Sud.}
\label{eq:sud-evol-ronly}\\
&&= -\frac{\alpha_s C_F}{\pi} \ln \Big(2xp^+\sigma b^2_\perp\Big) \int \frac{dz^-}{2\pi} e^{-2ix p^+z^-} \mathcal{R}^{[\gamma^+]}_{f}(z^-, -z^-, \frac{1}{2}b_\perp;\sigma) + \mathcal{O}(\alpha^2_s) \,,
\nonumber
\end{eqnarray}
which is consistent with double-logarithmic CSS evolution and is again obtained without a separate soft-factor contribution. The evolution equation contains one logarithm, $\ln(2xp^+\sigma b_\perp^2)$; integrating it with respect to $\ln\sigma$ generates the corresponding double logarithm, with the remaining $\sigma$-independent term fixed by matching.

\subsubsection{Finite terms in the rapidity-only scheme}
Like standard UV renormalization, the rapidity-only scheme leaves an ambiguity in the finite contribution included in the renormalization of the bare beam function in Eq.~(\ref{eq:fig3-uv-new-km}).
Indeed, in the derivation presented in the previous two sections, we fixed the finite part of the renormalization in a particular scheme-dependent way: the replacement in Eq. (\ref{eq:UV-rep}) was introduced only in the UV-divergent terms of Eq. (\ref{eq:fig3-uv-new-km}), while the finite terms of the bare result were kept unchanged.
Alternatively, one could apply the replacement (\ref{eq:UV-rep}) uniformly in all terms of the bare result. A different rapidity-only TMDPDF would then result, with Eq.~(\ref{eq:RFTMDevol}) receiving an additional contribution from the finite terms of Eq.~(\ref{eq:fig3-uv-new-km}). This prescription has been used for gluon TMDPDFs in the rapidity-only scheme~\cite{Balitsky:2015qba,Balitsky:2016dgz,Balitsky:2026lop}.
Therefore, from our analysis we conclude that the evolution properties of such TMDPDFs would go beyond the double-logarithmic evolution of TMDPDFs, see Eqs. (\ref{eq:fig3bare-renorm-result}) and (\ref{eq:sud-evol-ronly}), and include contribution of the finite terms into the evolution with the rapidity cutoff parameter $\sigma$.
The evolution properties generated by the finite terms of the bare result are, in fact, quite remarkable. As we will see in the next section, these terms lead to the evolution dynamics consistent with both the large-$x$ DGLAP evolution in the collinear limit of $b_\perp\to0$, and high-energy evolution in the limit of small-$x$. However, these terms have nothing to do with the UV renormalization of TMDPDFs per se. For this reason, we do not include their contribution in our rapidity cutoff procedure for the UV renormalization of TMDPDFs, similarly to the standard TMD approach.
The finite terms, however, are relevant to the IR structure of TMDPDFs, which can be accessed through the factorization at the $\sigma'$ scales corresponding to the lower bound of the quantum fields.

\section{Regularization of infrared divergences at the lower scales $\sigma^\prime$\label{sec:low-fact}}
In the previous section, NLO renormalization of the bare beam function defined the scheme at the upper scale $\sigma$. It can be performed in standard TMD factorization, Eq.~(\ref{eq:fig3bare-renorm-result}), with the two-scale evolution of Eqs.~(\ref{eq:CSS1}) and~(\ref{eq:CSS2}), or in the rapidity-only scheme, Eq.~(\ref{eq:fig3-uv-new-regUV}), with the single-scale evolution of Eq.~(\ref{eq:RFTMDevol}). Both describe the same double-logarithmic UV structure.
With the UV renormalization of the obtained results completed, we focus on the IR sector to define the quantum-background separation.

\subsection{MSTT scheme\label{sec:modified-TMD}}

Although UV-renormalized quark TMDPDFs in Eqs.~(\ref{eq:fig3bare-renorm-result}) and (\ref{eq:fig3-uv-new-regUV}) have no genuine IR divergences due to nonzero transverse momentum of the background field, let us adopt the simplest possible scheme, defining the renormalized background-field operators by
\begin{eqnarray}
&&\mathcal{M}^{[\gamma^+]}_{f}(z^-, -z^-, \frac{1}{2}b_\perp; \mu', \nu') = \mathcal{U}^{[\gamma^+]}_{f;{\rm  bg}}(z^-, -z^-, \frac{1}{2}b_\perp ) + \mathcal{O}(\alpha^2_s)\,,
\label{eq:MSTT-def-dist}
\end{eqnarray}
c.f. Eq. (\ref{eq:rrenbeam-def}), where the scales $\sigma' = (\mu', \nu')$ define the separation between the background fields and the quantum-field modes integrated out in our calculation.
As a result, the UV-renormalized TMDPDFs in Eqs.~(\ref{eq:fig3bare-renorm-result}) and (\ref{eq:fig3-uv-new-regUV}) take the form
\begin{eqnarray}
&&\mathcal{T}^{[\gamma^+]}_{f}(z^-, -z^-, \frac{1}{2}b_\perp; \mu, \zeta) = \mathcal{M}^{[\gamma^+]}_{f}(z^-, -z^-, \frac{1}{2}b_\perp; \mu', \nu' )
\label{eq:fig3bare-renorm-result-fin}\\
&&+ \frac{g^2 C_F}{8\pi^2} \Big\{ -  \frac{1}{2} (L^\mu_b)^2 + \Big( \frac{3}{2}  +  \ln\frac{2\mu^2}{\zeta} \Big) L^{\mu}_b + \frac{1}{2}  - \frac{\pi ^2 }{12} \Big\} \mathcal{M}^{[\gamma^+]}_{f}(z^-, -z^-, \frac{1}{2}b_\perp; \mu', \nu' )
\nonumber\\
&&+ \frac{g^2 C_F}{\pi}  \int^1_0 du \int d^{2}z_\perp \mathcal{K}(u, z_\perp, b_\perp) \mathcal{M}^{[\gamma^+]}_{f}(uz^-, -uz^-, \frac{1}{2}z_\perp; \mu', \nu' ) + \mathcal{O}(\alpha^2_s)\,,
\nonumber
\end{eqnarray}
and
\begin{eqnarray}
&&\int \frac{dz^-}{2\pi} e^{-2ix p^+z^-} \mathcal{R}^{[\gamma^+]}_{f}(z^-, -z^-, \frac{1}{2}b_\perp;\sigma)
= \int \frac{dz^-}{2\pi} e^{-2ix p^+z^-} \mathcal{M}^{[\gamma^+]}_{f}(z^-, -z^-, \frac{1}{2}b_\perp; \mu', \nu' )
\label{eq:fig3-uv-new-regUV-fin}\\
&&-\frac{g^2C_F}{2\pi} \int \frac{\dhd^{2} k_\perp}{k^2_\perp} \Big(1 - e^{i k_\perp b_\perp } \Big) \Big(2 \ln \frac{2xp^+\sigma + k^2_\perp}{k^2_\perp} + \frac{ 2 k^2_\perp}{ 2xp^+\sigma+k^2_\perp } - \frac{ (k^2_\perp)^2 }{2 (2xp^+\sigma+k^2_\perp)^2}-\frac{3}{2} \Big)
\nonumber\\
&&\times \int \frac{dz^-}{2\pi} e^{-2ix p^+z^-}\mathcal{M}^{[\gamma^+]}_{f}(z^-, -z^-, \frac{1}{2}b_\perp; \mu', \nu' ) + \frac{g^2C_F}{16\pi^2}
\int \frac{dz^-}{2\pi} e^{-2ix p^+z^-} \mathcal{M}^{[\gamma^+]}_{f}(z^-, -z^-, \frac{1}{2}b_\perp; \mu', \nu' )
\nonumber\\
&&+ \frac{g^2 C_F}{\pi} \int^1_0 \frac{du}{u} \int d^{2}z_\perp \mathcal{K}(u, z_\perp, b_\perp) \int \frac{dz^-}{2\pi} e^{-2i\frac{x}{u} p^+ z^-} \mathcal{M}^{[\gamma^+]}_{f}(z^-, -z^-, \frac{1}{2}z_\perp; \mu', \nu' ) + \mathcal{O}(\alpha^2_s) \,
\nonumber
\end{eqnarray}
correspondingly. At the order considered, the background-field operator in Eq.~(\ref{eq:MSTT-def-dist}) has trivial dependence on the lower scale $\sigma'$ and does not evolve. The MSTT scheme assigns the finite terms in Eqs.~(\ref{eq:fig3bare-renorm-result}) and (\ref{eq:fig3-uv-new-regUV}) to quantum-field coefficients. In the gluon case, by contrast, a genuine IR divergence produces nontrivial lower-scale evolution~\cite{Mukherjee:2023snp}.
One is free to choose a different scheme at the lower scale $\sigma'$, redistributing finite terms between quantum-field modes and background-field operators. Below we illustrate this freedom with collinear factorization, high-energy factorization, and the rapidity-only scheme.
Compared with those factorization schemes, the MSTT scheme assigns the perturbative information obtained in deriving the bare result to the quantum-field contribution only. Its coefficients therefore contain the most complete perturbative information available at the present order. The perturbative coefficients in the MSTT scheme incorporate the dynamics relevant at both large and small $x$, providing a smooth connection between the two limits.
To demonstrate this explicitly, we need to compare the MSTT expansion (\ref{eq:fig3bare-renorm-result-fin}) and (\ref{eq:fig3-uv-new-regUV-fin}) with other factorization schemes defined at the $\sigma'$ scales.

\subsubsection{Collinear limit\label{sec:matching_collinear}}
The collinear limit of the MSTT scheme can be examined by choosing a background field obeying collinear ordering at the $\sigma'$ scales, i.e. the background field transfers large longitudinal but vanishing transverse momenta.

Starting from the UV-renormalized Eq.~(\ref{eq:fig3bare-renorm-result}), it is useful to rewrite the finite $\mathcal{K}$ terms explicitly in the following form:
\begin{eqnarray}
&&\mathcal{T}^{[\gamma^+]}_{f}(z^-, -z^-, \frac{1}{2}b_\perp; \mu, \zeta) = \mathcal{U}^{[\gamma^+]}_{f;{\rm bg}}(z^-, -z^-, \frac{1}{2}b_\perp)
+ \frac{g^2 C_F}{8\pi^2} \Big\{ -  \frac{1}{2} (L^\mu_b)^2 + \Big( \frac{3}{2}  +  \ln\frac{2\mu^2}{\zeta} \Big) L^{\mu}_b + \frac{1}{2}  - \frac{\pi ^2 }{12} \Big\}
\nonumber\\
&&\times \mathcal{U}^{[\gamma^+]}_{f;{\rm bg}}(z^-, -z^-, \frac{1}{2}b_\perp) + \frac{g^2 C_F}{\pi} \int^1_0 du \int d^{2}z_\perp \int \dhd^{2} p_{\perp} e^{ip_{\perp} b_\perp } \int \dhd^{2} k_{\perp} e^{-i ( p_\perp - k_{\perp}) z_\perp } \Big\{ \Big[ \frac{u}{1-u} \frac{1}{ (1-u) p^2_{\perp} + u k^2_{\perp} } \Big]_+
\nonumber\\
&&\times \mathcal{U}^{[\gamma^+]}_{f;{\rm bg}}(u z^-, -u z^-, \frac{1}{2}z_\perp)
+ \frac{u}{1-u} \frac{1}{ k^2_{\perp} } \Big( 1 - e^{i k_{\perp} b_\perp }  \Big) e^{-i k_{\perp} z_\perp } \mathcal{U}^{[\gamma^+]}_{f;{\rm  bg}}(  z^- , -z^- , \frac{1}{2}z_\perp) + \frac{u}{1-u}  \Big(1 - e^{-i k_{\perp} b_\perp } \Big)
\nonumber\\
&&\times \frac{1}{ (1-u) p^2_{\perp} + u k^2_{\perp} }   \mathcal{U}^{[\gamma^+]}_{f;{\rm  bg}}( z^- , - z^- , \frac{1}{2}z_\perp)
+ \frac{  1 }{2 } \frac{(1-u) p^2_\perp}{( p^2_{\perp} (1-u)  + u k^2_{\perp} )^2 } \mathcal{U}^{[\gamma^+]}_{f;{\rm bg}}(u z^-, -u z^-, \frac{1}{2}z_\perp) - \frac{1}{4 }  e^{-i k_{\perp} b_\perp }
\nonumber\\
&&\times \frac{1}{(1-u)p^2_\perp + u k^2_\perp} \mathcal{U}^{[\gamma^+]}_{f;{\rm bg}}( z^-, - z^-, \frac{1}{2}z_\perp)
+ \frac{\delta(1-u)}{4 } \frac{1}{k^2_\perp} \Big(1 - e^{i k_\perp b_\perp } \Big) e^{-i k_{\perp} z_\perp } \mathcal{U}^{[\gamma^+]}_{f;{\rm bg}}( z^-, - z^-, \frac{1}{2}z_\perp)  \Big\}
+\mathcal{O}(\alpha^2_s)\,,
\end{eqnarray}
where the various terms follow directly from the derivation in Sec. \ref{sec:quarkTMDPDFsNLO}. This expression simplifies to
\begin{eqnarray}
&&\mathcal{T}^{[\gamma^+]}_{f}(z^-, -z^-, \frac{1}{2}b_\perp; \mu, \zeta) = \mathcal{U}^{[\gamma^+]}_{f;{\rm bg}}(z^-, -z^-, \frac{1}{2}b_\perp)
+ \frac{g^2 C_F}{8\pi^2} \Big\{ -  \frac{1}{2} (L^\mu_b)^2 + \Big( \frac{3}{2}  +  \ln\frac{2\mu^2}{\zeta} \Big) L^{\mu}_b + \frac{1}{2}  - \frac{\pi ^2 }{12} \Big\}
\label{eq:fig3bare-renormIRform2}\\
&&\times \mathcal{U}^{[\gamma^+]}_{f;{\rm bg}}(z^-, -z^-, \frac{1}{2}b_\perp) + \frac{g^2 C_F}{\pi} \int^1_0 du \int d^{2}z_\perp \int \dhd^{2} p_{\perp} e^{ip_{\perp} b_\perp } \int \dhd^{2} k_{\perp} e^{-i ( p_\perp - k_{\perp}) z_\perp } \Big\{ \frac{1}{2}\Big( \frac{  1 + u^2}{1-u} - \bar{u}\Big)
\nonumber\\
&&\times \frac{1}{ (1-u) p^2_{\perp} + u k^2_{\perp} } \mathcal{U}^{[\gamma^+]}_{f;{\rm bg}}(u z^-, -u z^-, \frac{1}{2}z_\perp)
- \frac{1}{2}\frac{ 1 + u^2}{1-u} \frac{1}{ (1-u) p^2_{\perp} + u k^2_{\perp} }  e^{-i k_{\perp} b_\perp }  \mathcal{U}^{[\gamma^+]}_{f;{\rm  bg}}( z^- , - z^- , \frac{1}{2}z_\perp)
\nonumber\\
&&+ \frac{1}{2} \frac{1+u^2}{1-u} \frac{1}{k^2_\perp} \Big(1 - e^{i k_\perp b_\perp } \Big) e^{-i k_{\perp} z_\perp } \mathcal{U}^{[\gamma^+]}_{f;{\rm bg}}( z^-, - z^-, \frac{1}{2}z_\perp) + \frac{ 1 }{2 } \frac{(1-u) p^2_\perp}{( p^2_{\perp} (1-u)  + u k^2_{\perp} )^2 } \mathcal{U}^{[\gamma^+]}_{f;{\rm bg}}(u z^-, -u z^-, \frac{1}{2}z_\perp) \Big\}
+\mathcal{O}(\alpha^2_s)\,.
\nonumber
\end{eqnarray}

To implement impose collinear background field  at the $\sigma'$ scales, we expand the bare TMD operators in terms of bare collinear operators:
\begin{eqnarray}
&&\mathcal{U}^{[\gamma^+]}_{f;{\rm bg}}\big( z^-, -z^-, b_\perp\big) = \mathcal{O}^{[\gamma^+]}_{f;{\rm bg}}( z^-, -z^-) + b_k \mathcal{O}^{k[\gamma^+]}_{f;{\rm bg}}( z^-, -z^-) + \dots\,,
\label{eq:collinear-exp}
\end{eqnarray}
where the first bare collinear operator
\begin{eqnarray}
&&\mathcal{O}^{[\gamma^+]}_{f;{\rm bg}}(z^-, -z^-) = \bar{q}_f(z^-, 0_\perp)[z^-, \infty]_{0}\gamma^+[\infty, -z^-]_{0}q_f(-z^-, 0_\perp)\,,
\label{eq:def-col-operator}
\end{eqnarray}
c.f. Eq. (\ref{eq:def-operator}), and the higher-twist collinear operators are defined by
\begin{eqnarray}
&&\mathcal{O}^{k[\gamma^+]}_{{f;{\rm bg}}}( z^-, -z^-) \equiv \frac{\partial}{\partial b_k}\mathcal{U}^{[\gamma^+]}_{f;{\rm bg}}\big( z^-, -z^-, b_\perp\big) \Big|_{b_\perp=0}\,,
\end{eqnarray}
etc. In momentum space the expansion reads
\begin{eqnarray}
&&\int d^2 z_\perp e^{-ip_{\perp} z_\perp } \mathcal{U}^{[\gamma^+]}_{f;{\rm bg}}(z^-, -z^-, \frac{1}{2}z_\perp)
\label{eq:collinear-exp-mom}\\
&&=  (2\pi)^2 \delta^2(p_\perp) \mathcal{O}^{[\gamma^+]}_{f;{\rm bg}}( z^-, -z^-) + \frac{i}{2}  \frac{\partial}{\partial p_k} (2\pi)^2 \delta^2(p_\perp) \mathcal{O}^{k[\gamma^+]}_{f;{\rm bg}}( z^-, -z^-) + \dots\,.
\nonumber
\end{eqnarray}
For each term in Eq.~(\ref{eq:collinear-exp-mom}) $p_\perp=0$ for the background-field momenta is enforced, which is precisely the collinear factorization condition.

Although Eq.~(\ref{eq:collinear-exp}) can formally be constructed at any $b_\perp$, the expansion converges only in the small-$b_\perp$ regime, $b_\perp \ll \Lambda^{-1}_{QCD}$, where background transverse momenta are small - the natural domain of collinear factorization.

Eq.~(\ref{eq:collinear-exp}) is an expansion of the TMD operator in transverse coordinate space, which translates into an expansion of Eq.~(\ref{eq:fig3bare-renormIRform2}) in $b_\perp$. The result is the collinear matching equation for TMDPDFs, Eq.~(\ref{eq:coll-exp}), with the leading term dominating in the region $b_\perp \ll \Lambda^{-1}_{QCD}$.

Substituting Eq.~(\ref{eq:collinear-exp}) into Eq.~(\ref{eq:fig3bare-renormIRform2}) and retaining the leading term of the expansion gives
\begin{eqnarray}
&&\mathcal{T}^{[\gamma^+]}_{f}(z^-, -z^-, \frac{1}{2}b_\perp; \mu, \zeta) = \mathcal{O}^{[\gamma^+]}_{f;{\rm bg}}(z^-, -z^-)
+ \frac{g^2 C_F}{8\pi^2} \Big\{ -  \frac{1}{2} (L^\mu_b)^2 + \Big( \frac{3}{2}  +  \ln\frac{2\mu^2}{\zeta} \Big) L^{\mu}_b + \frac{1}{2}  - \frac{\pi ^2 }{12} \Big\}
\label{eq:fig3col-ltw}\\
&&\times \mathcal{O}^{[\gamma^+]}_{f;{\rm bg}}(z^-, -z^-)
+ \frac{g^2(\mu') C_F}{2\pi}  \Big(\frac{\mu'^2 e^{\gamma_E}}{4\pi}\Big)^\epsilon \int^1_0 du  \int  \frac{\dhd^{2-2\epsilon} k_{\perp}}{ k^2_{\perp} } e^{ik_{\perp} b_\perp }  \Big[\frac{  1 + u^2}{1-u}\Big]_+ \mathcal{O}^{[\gamma^+]}_{f;{\rm bg}}(u z^-, -u z^-)
\nonumber\\
&&+ \mathcal{O}(b_\perp) +\mathcal{O}(\alpha^2_s)\,,
\nonumber
\end{eqnarray}
where dimensional regularization introduces $\mu'$, the renormalization scale of the collinear background-field operators; in the collinear scheme at the lower scale, $\sigma'\equiv\mu'$.

The integral over $k_\perp$ in the last term of Eq.~(\ref{eq:fig3col-ltw}) is IR divergent; regulating with dimensional regularization gives
\begin{eqnarray}
&&\mathcal{T}^{[\gamma^+]}_{f}(z^-, -z^-, \frac{1}{2}b_\perp; \mu, \zeta) = \mathcal{O}^{[\gamma^+]}_{f;{\rm bg}}(z^-, -z^-)
+ \frac{g^2 C_F}{8\pi^2} \Big\{ -  \frac{1}{2} (L^\mu_b)^2 + \Big( \frac{3}{2}  +  \ln\frac{2\mu^2}{\zeta} \Big) L^{\mu}_b + \frac{1}{2}  - \frac{\pi ^2 }{12} \Big\}
\label{eq:fig3col-ltw-expanded}\\
&&\times \mathcal{O}^{[\gamma^+]}_{f;{\rm bg}}(z^-, -z^-)
- \frac{g^2(\mu') C_F}{8\pi^2} \Big( \frac{1}{\epsilon} + L^{\mu'}_b \Big) \int^1_0 du \Big[\frac{  1 + u^2}{1-u}\Big]_+ \mathcal{O}^{[\gamma^+]}_{f;{\rm bg}}(u z^-, -u z^-) + \mathcal{O}(b_\perp) +\mathcal{O}(\alpha^2_s)\,.
\nonumber
\end{eqnarray}
Although the original expression~(\ref{eq:fig3bare-renormIRform2}) is finite, with all integrals regulated by the nonzero transverse-momentum flow of the background fields, the $1/\epsilon$ singularity in the above equation arise solely from the $p_\perp=0$ condition imposed by the collinear expansion~(\ref{eq:collinear-exp}).
This divergence must be absorbed into the background-field collinear operators, yielding renormalized operators $\mathcal{O}^{[\gamma^+]}_{f;{\rm bg}}(z^-, -z^-; \mu')$ at scale $\mu'$. This can be achieved by multiplicative renormalization of the collinear operators, cf.\ Eq.~(\ref{eq:MSTT-def-dist}):
\begin{eqnarray}
&&\mathcal{C}^{[\gamma^+]}_{f}(z^-, -z^-; \mu') = \int^1_0 du Z_{C}(u;\mu'^2) \mathcal{O}^{[\gamma^+]}_{f;{\rm  bg}}(uz^-, -uz^-)\,.
\label{eq:col-renorm}
\end{eqnarray}
As in the case of TMD operators, renormalization of the bare collinear operators introduces dependence on the factorization scale and specifies which modes are included in the renormalized operators, see Eq.~(\ref{eq:col-renorm}).
The renormalization constant for the leading-twist collinear operator
\begin{eqnarray}
&&Z_{C}(u;\mu')= \delta(1-u) - \frac{g^2(\mu') C_F}{8\pi^2} \frac{1}{\epsilon} \Big[\frac{  1 + u^2}{1-u}\Big]_+ +\mathcal{O}(\alpha^2_s)\,.
\label{eq:col-Z}
\end{eqnarray}
Substituting Eq. (\ref{eq:col-Z}) into Eq. (\ref{eq:col-renorm}) we get
\begin{eqnarray}
&&\mathcal{C}^{[\gamma^+]}_{f}(z^-, -z^-; \mu') = \mathcal{O}^{[\gamma^+]}_{f;{\rm  bg}}(uz^-, -uz^-) - \frac{g^2(\mu') C_F}{8\pi^2} \frac{1}{\epsilon} \int^1_0 du \Big[\frac{  1 + u^2}{1-u}\Big]_+ \mathcal{O}^{[\gamma^+]}_{f;{\rm  bg}}(uz^-, -uz^-) +\mathcal{O}(\alpha^2_s)\,.
\label{eq:col-ren}
\end{eqnarray}
Introducing renormalized collinear operators of the background fields (\ref{eq:col-ren}) in Eq. (\ref{eq:fig3col-ltw-expanded}), we obtain
\begin{eqnarray}
&&\mathcal{T}^{[\gamma^+]}_{f}(z^-, -z^-, \frac{1}{2}b_\perp; \mu, \zeta) = \mathcal{C}^{[\gamma^+]}_{f}(z^-, -z^-;\mu')
+ \frac{g^2 C_F}{8\pi^2} \Big\{ -  \frac{1}{2} (L^\mu_b)^2 + \Big( \frac{3}{2}  +  \ln\frac{2\mu^2}{\zeta} \Big) L^{\mu}_b + \frac{1}{2}  - \frac{\pi ^2 }{12} \Big\}
\nonumber\\
&&\times \mathcal{C}^{[\gamma^+]}_{f}(z^-, -z^-;\mu') - \frac{g^2 C_F}{8\pi^2} L^{\mu'}_b \int^1_0 du \Big[\frac{  1 + u^2}{1-u}\Big]_+ \mathcal{C}^{[\gamma^+]}_{f}(u z^-, -u z^-; \mu') + \mathcal{O}(b_\perp) + \mathcal{O}(\alpha^2_s)\,,
\label{eq:fig3col-ltw-expande-renorm}
\end{eqnarray}
where the collinear distributions are renormalized at the IR scale $\mu'$ of the background modes. The right-hand side of Eq.~(\ref{eq:fig3col-ltw-expande-renorm}) is an expansion in powers of $b_\perp$ with the general form of the collinear matching formula~(\ref{eq:coll-exp}).
Taking the matrix element~(\ref{eq:matrix-element-quark}) of Eq.~(\ref{eq:fig3col-ltw-expande-renorm}) and noting that the matrix elements on the right-hand side correspond to the unpolarized collinear PDF
\begin{eqnarray}
&&f_{1;f}(x, \mu') = \int \frac{dz^-}{2\pi} e^{-2ixp^+z^-} \langle P, S|\mathcal{C}^{[\gamma^+]}_f( z^- , - z^-;\mu')|P, S\rangle\,,
\end{eqnarray}
we obtain
\begin{eqnarray}
&&f_{1;f}(x, b_\perp; \mu, \zeta) = f_{1;f}(x,\mu') + \frac{g^2 C_F}{8\pi^2} \Big\{- \frac{1}{2}(L^{\mu}_b)^2 + \Big( \frac{3}{2} + \ln\frac{2\mu^2}{\zeta} \Big) L^{\mu}_b + \frac{1}{2} - \frac{\pi^2}{12} \Big\}f_{1;f}(x, \mu')
\nonumber\\
&& - \frac{g^2 C_F}{8\pi^2} L^{\mu'}_b \int^1_0 \frac{du}{u} \Big[\frac{  1 + u^2}{1-u}\Big]_+ f_{1;f}(\frac{x}{u}, \mu') + \mathcal{O}(\alpha^2_s)\,.
\label{eq:fig3col-ltw-expande-renorm-dist}
\end{eqnarray}
Setting $\mu' = \mu$, we reproduce the well-known NLO collinear matching result for the unpolarized TMDPDF at leading power, up to scheme-dependent finite terms.

Similarly, starting from the UV-renormalized result~(\ref{eq:fig3-uv-new-regUV}) in the using rapidity-only scheme and imposing collinear background field at the $\sigma'$ scales, we obtain
\begin{eqnarray}
&&\int \frac{dz^-}{2\pi} e^{-2ix p^+z^-} \mathcal{R}^{[\gamma^+]}_{f}(z^-, -z^-, \frac{1}{2}b_\perp;\sigma)
= \int \frac{dz^-}{2\pi} e^{-2ix p^+z^-} \mathcal{C}^{[\gamma^+]}_{f}(z^-, -z^-; \mu') -\frac{g^2C_F}{2\pi} \int \frac{\dhd^{2} k_\perp}{k^2_\perp} \Big(1 - e^{i k_\perp b_\perp } \Big)
\nonumber\\
&&\times \Big(2 \ln \frac{2xp^+\sigma + k^2_\perp}{k^2_\perp} + \frac{ 2 k^2_\perp}{ 2xp^+\sigma+k^2_\perp } - \frac{ (k^2_\perp)^2 }{2 (2xp^+\sigma+k^2_\perp)^2}-\frac{3}{2} \Big)
\int \frac{dz^-}{2\pi} e^{-2ix p^+z^-} \mathcal{C}^{[\gamma^+]}_{f}(z^-, -z^-; \mu' )
\nonumber\\
&&+ \frac{g^2C_F}{16\pi^2}
\int \frac{dz^-}{2\pi} e^{-2ix p^+z^-} \mathcal{C}^{[\gamma^+]}_{f}(z^-, -z^-; \mu' ) - \frac{g^2 C_F}{8\pi^2} L^{\mu'}_b \int^1_0 \frac{du}{u} \Big[\frac{  1 + u^2}{1-u}\Big]_+ \int \frac{dz^-}{2\pi} e^{-2i\frac{x}{u} p^+z^-} \mathcal{C}^{[\gamma^+]}_{f}( z^-, -z^-; \mu')
\nonumber\\
&&+ \mathcal{O}(b_\perp) + \mathcal{O}(\alpha^2_s) \,.
\label{eq:fig3-uv-new-regUV-col}
\end{eqnarray}

Comparing Eqs. (\ref{eq:fig3col-ltw-expande-renorm}) and (\ref{eq:fig3-uv-new-regUV-col}) with their MSTT counterparts Eqs. (\ref{eq:fig3bare-renorm-result-fin}) and (\ref{eq:fig3-uv-new-regUV-fin}), we establish the relation between the MSTT scheme and collinear factorization:
\begin{eqnarray}
&&\mathcal{M}^{[\gamma^+]}_{f}(z^-, -z^-, \frac{1}{2}b_\perp; \mu', \nu' ) + \frac{g^2 C_F}{\pi}  \int^1_0 du \int d^{2}z_\perp \mathcal{K}(u, z_\perp, b_\perp) \mathcal{M}^{[\gamma^+]}_{f}(uz^-, -uz^-, \frac{1}{2}z_\perp; \mu', \nu' )
\nonumber\\
&&= \mathcal{C}^{[\gamma^+]}_{f}(z^-, -z^-;\mu') - \frac{g^2 C_F}{8\pi^2} L^{\mu'}_b \int^1_0 du \Big[\frac{  1 + u^2}{1-u}\Big]_+ \mathcal{C}^{[\gamma^+]}_{f}(u z^-, -u z^-; \mu') + \mathcal{O}(b_\perp) + \mathcal{O}(\alpha^2_s)\,.
\label{eq:MSTT-col-match}
\end{eqnarray}
Thus, in the small-$b_\perp$ limit the convolution of the MSTT perturbative coefficient with the corresponding background-field operator reduces to the standard collinear matching structure of TMDPDF.
The matching between the MSTT and the collinear expansion extends beyond the leading contribution, and MSTT partially incorporates contributions from all collinear twists.

To see this we substituting the higher-order terms of expansion~(\ref{eq:collinear-exp}) into Eq.~(\ref{eq:fig3bare-renormIRform2}) systematically reconstructs the higher-twist terms in the collinear matching formula~(\ref{eq:coll-exp}), thereby providing the structure of the non-perturbative function $f_{NP}$ in Eq.~(\ref{eq:coll-exp-NP}).
\begin{eqnarray}
&&\mathcal{T}^{[\gamma^+]}_{f}(z^-, -z^-, \frac{1}{2}b_\perp; \mu, \zeta) = \mathcal{C}^{[\gamma^+]}_{f}(z^-, -z^-;\mu') + \frac{b_k}{2} \mathcal{C}^{k[\gamma^+]}_{f}(z^-, -z^-;\mu')
\label{eq:fig3col-ltwp-leading-dist-unpol-T}\\
&&+ \frac{g^2 C_F}{8\pi^2} \Big\{ -  \frac{1}{2} (L^\mu_b)^2 + \Big( \frac{3}{2}  +  \ln\frac{2\mu^2}{\zeta} \Big) L^{\mu}_b + \frac{1}{2}  - \frac{\pi ^2 }{12} \Big\} \Big(\mathcal{C}^{[\gamma^+]}_{f}(z^-, -z^-;\mu') + \frac{b_k}{2} \mathcal{C}^{k[\gamma^+]}_{f}(z^-, -z^-;\mu')\Big)
\nonumber\\
&&- \frac{g^2 C_F}{8\pi^2} L^{\mu'}_b \int_0^1 du \Big[\frac{ 1 + u^2}{1-u}\Big]_+ \Big( \mathcal{C}^{[\gamma^+]}_{f}(uz^-, -uz^-;\mu') + \frac{b_k}{2} \mathcal{C}^{k[\gamma^+]}_{f}(uz^-, -uz^-;\mu') \Big)
\nonumber\\
&& + \frac{g^2 C_F}{8\pi^2} L^{\mu'}_b \int_0^1 du  (1 + u^2 - \bar{u}u) \frac{ b_k }{ 2} \mathcal{C}^{k[\gamma^+]}_{f}(uz^-, -uz^-;\mu') + O(b^2_\perp) + \mathcal{O}(\alpha^2_s)\,.
\nonumber
\end{eqnarray}

Similarly, for the UV-renormalized result (\ref{eq:fig3-uv-new-regUV}) derived using the rapidity-only scheme at the $\sigma$ scales, we obtain
\begin{eqnarray}
&&\int \frac{dz^-}{2\pi} e^{-2ix p^+z^-} \mathcal{R}^{[\gamma^+]}_{f}(z^-, -z^-, \frac{1}{2}b_\perp;\sigma)
= \int \frac{dz^-}{2\pi} e^{-2ix p^+z^-} \Big(\mathcal{C}^{[\gamma^+]}_{f}(z^-, -z^-;\mu') + \frac{b_k}{2} \mathcal{C}^{k[\gamma^+]}_{f}(z^-, -z^-;\mu')\Big)
\label{eq:fig3col-ltwp-leading-dist-unpol-T-rap}\\
&&- \frac{g^2C_F}{2\pi} \int \frac{\dhd^{2} k_\perp}{k^2_\perp} \Big(1 - e^{i k_\perp b_\perp } \Big) \Big(2 \ln \frac{2xp^+\sigma + k^2_\perp}{k^2_\perp} + \frac{ 2 k^2_\perp}{ 2xp^+\sigma+k^2_\perp } - \frac{ (k^2_\perp)^2 }{2 (2xp^+\sigma+k^2_\perp)^2}-\frac{3}{2} \Big)
\nonumber\\
&&\times \int \frac{dz^-}{2\pi} e^{-2ix p^+z^-} \Big(\mathcal{C}^{[\gamma^+]}_{f}(z^-, -z^-;\mu') + \frac{b_k}{2} \mathcal{C}^{k[\gamma^+]}_{f}(z^-, -z^-;\mu')\Big) + \frac{g^2C_F}{16\pi^2}
\int \frac{dz^-}{2\pi} e^{-2ix p^+z^-} \Big(\mathcal{C}^{[\gamma^+]}_{f}(z^-, -z^-;\mu')
\nonumber\\
&&+ \frac{b_k}{2} \mathcal{C}^{k[\gamma^+]}_{f}(z^-, -z^-;\mu')\Big) - \frac{g^2 C_F}{8\pi^2} L^{\mu'}_b \int^1_0 \frac{du}{u} \Big[\frac{  1 + u^2}{1-u}\Big]_+ \int \frac{dz^-}{2\pi} e^{-2i\frac{x}{u} p^+z^-} \Big(\mathcal{C}^{[\gamma^+]}_{f}(z^-, -z^-;\mu') + \frac{b_k}{2}
\nonumber\\
&&\times\mathcal{C}^{k[\gamma^+]}_{f}(z^-, -z^-;\mu')\Big) + \frac{g^2 C_F}{8\pi^2} L^{\mu'}_b \int_0^1 \frac{du}{u}  (1 + u^2 - \bar{u}u) \int \frac{dz^-}{2\pi} e^{-2i\frac{x}{u} p^+z^-} \frac{ b_k }{ 2} \mathcal{C}^{k[\gamma^+]}_{f}(z^-, -z^-;\mu') + O(b^2_\perp) + \mathcal{O}(\alpha^2_s) \,.
\nonumber
\end{eqnarray}

Taking the matrix element~(\ref{eq:matrix-element-quark}) of Eq.~(\ref{eq:fig3col-ltwp-leading-dist-unpol-T}), the leading order terms in $b_\perp$ reproduce Eq.~(\ref{eq:fig3col-ltw-expande-renorm-dist}) . The next-to-leading terms in $b_\perp$ give the following matching relation between the Sivers function and the twist-3 collinear distribution:
\begin{eqnarray}
&& f^\perp_{1T;f}(x, b_\perp; \mu, \zeta) = \pi T_f(-x,0, x; \mu') + \frac{g^2 C_F}{8\pi^2} \Big\{ - \frac{1}{2}(L^{\mu}_b)^2
+ \Big( \frac{3}{2} + \ln\frac{2\mu^2}{\zeta} \Big) L^{\mu}_b +\frac{1}{2} - \frac{\pi^2}{12}\Big\} \pi T_f(-x,0, x; \mu')
\nonumber\\
&& - \frac{g^2 C_F}{8\pi^2} L^{\mu'}_b \int_0^1 \frac{du}{u} \Big[\frac{ 1 + u^2}{1-u}\Big]_+ \pi T_f(-\frac{x}{u},0, \frac{x}{u}; \mu') + \frac{g^2 C_F}{8\pi^2} L^{\mu'}_b \int_0^1 \frac{du}{u}  (1 + u^2 - u \bar{u}) \pi T_f(-\frac{x}{u},0, \frac{x}{u}; \mu')
\nonumber\\
&& + O(b^2_\perp) + \mathcal{O}(\alpha^2_s)\,,
\label{eq:Sivers-match}
\end{eqnarray}
where the twist-3 collinear PDF is
\begin{eqnarray}
&&2i\pi M \epsilon^{k\nu}_\perp s_{T\nu} T_f(-x,0, x) = \int \frac{dz^-}{2\pi} e^{-2ixp^+z^-} \langle P, S|\mathcal{O}^{k[\gamma^+]}_{f}( z^- , - z^-)|P, S\rangle\,.
\end{eqnarray}
Comparing Eq.~(\ref{eq:Sivers-match}) with the known matching result for the Sivers function, we find only partial agreement in the logarithmic terms.\footnote{As we mentioned above our construction neglects non-logarithmic finite terms originating from the $\sim\epsilon$ terms in the standard treatment within the dimensional regularization. Contribution of these terms can be understood as part of the factorization scheme definition, prescribing separation of the quantum and background field modes.} This discrepancy has a clear origin: Eq.~(\ref{eq:Sivers-match}) was derived in the dilute approximation, in which the contribution of the gluon background field is reduced to the Wilson-line gauge factors required to ensure the gauge invariance of the corresponding background-field operators. More involved structures of the background gluon field, in particular those involving explicit insertions of the gluon field-strength tensor, are neglected in this approximation. Such terms give rise to twist-3 collinear correlation functions and therefore contribute to the collinear matching expansion of the Sivers distribution. The MSTT result (\ref{eq:fig3bare-renormIRform2}) contains full dynamics defined by the leading-twist matching kernel in Eq.~(\ref{eq:coll-exp}), see Eq. (\ref{eq:fig3col-ltw-expande-renorm-dist}). Beyond leading twist it captures the corresponding dynamics only partially, even though contributions from all collinear twists are effectively taken into account. A complete description of the twist-3 collinear dynamics requires going beyond the dilute approximation and including higher-power TMD operators arising from diagrams such as that shown in Fig.~\ref{fig:fig2} - a calculation that is feasible within our background-field approach.
To make this explicit, let us compare Eqs.~(\ref{eq:fig3col-ltwp-leading-dist-unpol-T}) and (\ref{eq:fig3col-ltwp-leading-dist-unpol-T-rap}) with their counterparts obtained within the MSTT factorization scheme imposed at the $\sigma'$ scales, Eqs.~(\ref{eq:fig3bare-renorm-result-fin}) and (\ref{eq:fig3-uv-new-regUV-fin}). This yields
\begin{eqnarray}
&&\mathcal{M}^{[\gamma^+]}_{f}(z^-, -z^-, \frac{1}{2}b_\perp; \mu', \nu' ) + \frac{g^2 C_F}{\pi}  \int^1_0 du \int d^{2}z_\perp \mathcal{K}(u, z_\perp, b_\perp) \mathcal{M}^{[\gamma^+]}_{f}(uz^-, -uz^-, \frac{1}{2}z_\perp; \mu', \nu' )
\label{eq:MSTT-col-match-tw3}\\
&& = \mathcal{C}^{[\gamma^+]}_{f}(z^-, -z^-;\mu') + \frac{b_k}{2} \mathcal{C}^{k[\gamma^+]}_{f}(z^-, -z^-;\mu') - \frac{g^2 C_F}{8\pi^2} L^{\mu'}_b \int_0^1 du \Big[\frac{ 1 + u^2}{1-u}\Big]_+ \Big( \mathcal{C}^{[\gamma^+]}_{f}(uz^-, -uz^-;\mu')
\nonumber\\
&&+ \frac{b_k}{2} \mathcal{C}^{k[\gamma^+]}_{f}(uz^-, -uz^-;\mu') \Big)  + \frac{g^2 C_F}{8\pi^2} L^{\mu'}_b \int_0^1 du  (1 + u^2 - \bar{u}u) \frac{ b_k }{ 2} \mathcal{C}^{k[\gamma^+]}_{f}(uz^-, -uz^-;\mu') + O(b^2_\perp) + \mathcal{O}(\alpha^2_s)\,,
\nonumber
\end{eqnarray}
c.f. Eq. (\ref{eq:MSTT-col-match}).
We find that the MSTT description of leading-power quark TMDPDFs, represented by the left-hand side of Eq.~(\ref{eq:MSTT-col-match-tw3}), is consistent with the structure obtained from collinear factorization beyond leading twist. In particular, it incorporates effects associated with all collinear twists, as shown on the right-hand side of Eq.~(\ref{eq:MSTT-col-match-tw3}), although the corresponding higher-twist dynamics is reproduced only partially.
Thus, the lowest-order MSTT term obtained in our computation within the dilute approximation provides a broader description of the TMDPDF structure than the leading-twist term of the collinear expansion.

\subsubsection{High-energy limit\label{sec:matching_eikonal}}

In this section we will show the perturbative part of the MSTT-scheme TMDPDF includes the large rapidity logarithms at small $x$. For high-energy factorization in the small Bjorken-$x_B$ region the longitudinal momentum fraction $x$ carried by the background field is small. Consequently, the background field obeys ordering in longitudinal momenta instead of the transverse-momenta ordering of the collinear case. The high-energy limit of the MSTT scheme can be studied by choosing a background field at the scale $\sigma'$ that transfers large transverse but vanishing longitudinal momenta.

Consider the following formal expansion of the Fourier transform of the bare TMD background-field operators appearing in Eqs. (\ref{eq:fig3bare-renorm-result}) and (\ref{eq:fig3-uv-new-regUV}), where the renormalization scheme at the upper scales has already been fixed:
\begin{eqnarray}
&&\int \frac{dz^-}{2\pi} e^{-2i xp^+ z^-} \mathcal{U}^{[\gamma^+]}_{f;{\rm bg}}(z^-, -z^-, \frac{1}{2}b_\perp)
\label{eq:eikonal-expansion-Fourier}\\
&&= \int \frac{dy^-}{2\pi} \mathcal{U}^{[\gamma^+]}_{f;{\rm bg}}(y^-, -y^-, \frac{1}{2}b_\perp) - 2i xp^+  \int \frac{dy^-}{2\pi} y^- \mathcal{U}^{[\gamma^+]}_{f;{\rm bg}}(y^-, -y^-, \frac{1}{2}b_\perp) + \dots\,.
\nonumber
\end{eqnarray}
Expansion~(\ref{eq:eikonal-expansion-Fourier}) converges when the longitudinal momentum of the active quark $xp^+$ is much smaller than the target longitudinal momentum $p^+$. The leading term is the eikonal contribution and the rest are subeikonal corrections. The corresponding expansion for the background-field beam function in coordinate space reads
\begin{eqnarray}
&&\mathcal{U}^{[\gamma^+]}_{f;{\rm bg}}(z^-, -z^-, \frac{1}{2}b_\perp)
\label{eq:eikonal-expansion}\\
&&= 2\pi \delta(z^-) \int \frac{dy^-}{2\pi} \mathcal{U}^{[\gamma^+]}_{f;{\rm bg}}(y^-, -y^-, \frac{1}{2}b_\perp) - 2\pi \delta'(z^-) \int \frac{dy^-}{2\pi} y^- \mathcal{U}^{[\gamma^+]}_{f;{\rm bg}}(y^-, -y^-, \frac{1}{2}b_\perp) + \dots\,.
\nonumber
\end{eqnarray}
The expansion (\ref{eq:eikonal-expansion}), where the $\delta$-functions and their derivatives effectively localize the background fields near $z^-=0$, provides the corresponding shock-wave description of the eikonal expansion.

To implement high-energy factorization at the $\sigma'$ scales, we substitute the eikonal expansion~(\ref{eq:eikonal-expansion-Fourier}) into the $\sigma$-renormalized result~(\ref{eq:fig3bare-renorm-result}) in its form (\ref{eq:fig3bare-renormIRform2}), retaining only the leading term:
\begin{eqnarray}
&&\int \frac{dz^-}{2\pi} e^{-2ixp^+z^-} \mathcal{T}^{[\gamma^+]}_{f}(z^-, -z^-, \frac{1}{2}b_\perp; \mu, \nu) = \int \frac{dz^-}{2\pi} \mathcal{U}^{[\gamma^+]}_{f;{\rm bg}}(z^-, -z^-, \frac{1}{2}b_\perp)
\label{eq:fig3bare-renormIRform2-rap}\\
&&+ \frac{g^2 C_F}{8\pi^2} \Big\{ -  \frac{1}{2} (L^\mu_b)^2 + \Big( \frac{3}{2}  +  \ln\frac{2\mu^2}{\zeta} \Big) L^{\mu}_b + \frac{1}{2}  - \frac{\pi ^2 }{12} \Big\} \int \frac{dz^-}{2\pi} \mathcal{U}^{[\gamma^+]}_{f;{\rm bg}}(z^-, -z^-, \frac{1}{2}b_\perp)
\nonumber\\
&&+ \frac{g^2 C_F}{\pi} \Big(\frac{\mu^2 e^{\gamma_E}}{4\pi}\Big)^\epsilon \Big(\frac{\rho}{xp^+}\Big)^\xi \int^1_0 du \Big( \frac{1 - u}{u} \Big)^{-\xi} \int d^{2-2\epsilon}z_\perp \int \dhd^{2-2\epsilon} p_{\perp} e^{ip_{\perp} b_\perp } \int \dhd^{2-2\epsilon} k_{\perp} e^{-i ( p_\perp - k_{\perp}) z_\perp }
\nonumber\\
&&\times \Big\{ \frac{1}{2}\Big( \frac{  1 + u^2}{1-u} - \bar{u}\Big) \frac{1}{ (1-u) p^2_{\perp} + u k^2_{\perp} } \frac{1}{u}
- \frac{1}{2}\frac{ 1 + u^2}{1-u} \frac{1}{ (1-u) p^2_{\perp} + u k^2_{\perp} }  e^{-i k_{\perp} b_\perp }
\nonumber\\
&&+ \frac{1}{2} \frac{1+u^2}{1-u} \frac{1}{k^2_\perp} \Big(1 - e^{i k_\perp b_\perp } \Big) e^{-i k_{\perp} z_\perp } + \frac{ 1 }{2 } \frac{(1-u) p^2_\perp}{( p^2_{\perp} (1-u)  + u k^2_{\perp} )^2 } \frac{1}{u} \Big\} \int \frac{dz^-}{2\pi} \mathcal{U}^{[\gamma^+]}_{f;{\rm bg}}(z^-, -z^-, \frac{1}{2}z_\perp) + \mathcal{O}(x)
+\mathcal{O}(\alpha^2_s)\,.
\nonumber
\end{eqnarray}
Eq.~(\ref{eq:fig3bare-renormIRform2-rap}) contains infrared rapidity divergence ($\sim\int du/u$) in the otherwise finite terms of Eq.~(\ref{eq:fig3bare-renormIRform2}). It arises from the strict longitudinal-momentum ordering imposed by the expansion~(\ref{eq:eikonal-expansion}) at the scale $\sigma'$. We regulate this infrared rapidity divergence using the $\eta$-regulator scheme, introducing in Eq.~(\ref{eq:fig3bare-renormIRform2-rap}) the replacement
\begin{eqnarray}
&&\int^1_0du \to \Big(\frac{\rho}{xp^+}\Big)^\xi \int^1_0 du \Big( \frac{1 - u}{u} \Big)^{-\xi}\,.
\label{eq:rIR-reg}
\end{eqnarray}
Here, $\rho$ is interpreted as the lower rapidity factorization scale, while $\xi$ is the corresponding regulator of the rapidity divergence at $u\to 0$. The eikonal expansion (\ref{eq:eikonal-expansion-Fourier}) also generates infrared divergences in the transverse-momentum integrals of Eq.~(\ref{eq:fig3bare-renormIRform2}), which we regulate using dimensional regularization.

Isolating the infrared divergent terms, we rewrite Eq.~(\ref{eq:fig3bare-renormIRform2-rap}) as
\begin{eqnarray}
&&\int \frac{dz^-}{2\pi} e^{-2ixp^+z^-} \mathcal{T}^{[\gamma^+]}_{f}(z^-, -z^-, \frac{1}{2}b_\perp; \mu, \nu) = \int \frac{dz^-}{2\pi} \mathcal{U}^{[\gamma^+]}_{f;{\rm bg}}(z^-, -z^-, \frac{1}{2}b_\perp) + \frac{g^2 C_F}{8\pi^2} \Big\{ -  \frac{1}{2} (L^\mu_b)^2
\label{eq:fig3bare-renormIRform2-rap-rew}\\
&&+ \Big( \frac{3}{2}  +  \ln\frac{2\mu^2}{\zeta} \Big) L^{\mu}_b + \frac{1}{2}  - \frac{\pi ^2 }{12} \Big\} \int \frac{dz^-}{2\pi} \mathcal{U}^{[\gamma^+]}_{f;{\rm bg}}(z^-, -z^-, \frac{1}{2}b_\perp) + \frac{g^2 C_F}{2\pi}  \Big(\frac{\mu^2 e^{\gamma_E}}{4\pi}\Big)^\epsilon \Big(\frac{\rho}{xp^+}\Big)^\xi \int d^{2-2\epsilon}z_\perp
\nonumber\\
&&\times \int \dhd^{2-2\epsilon} p_{\perp} e^{ip_{\perp} b_\perp } \int \dhd^{2-2\epsilon} k_{\perp} e^{-i ( p_\perp - k_{\perp}) z_\perp } \int^1_0 \frac{du}{u} \Big( \frac{1 - u}{u} \Big)^{-\xi} \frac{(1-u) p^2_\perp}{( p^2_{\perp} (1-u)  + u k^2_{\perp} )^2 }  \int \frac{dz^-}{2\pi} \mathcal{U}^{[\gamma^+]}_{f;{\rm bg}}( z^-, - z^-, \frac{1}{2}z_\perp)
\nonumber\\
&&  + \frac{g^2 C_F}{\pi} \int \dhd^{2} p_{\perp} e^{ip_{\perp} b_\perp }
\int d^{2}z_\perp e^{-i ( p_\perp - k_{\perp}) z_\perp } \Big\{ \int^1_0 du   \int \dhd^{2} k_{\perp}  \frac{1}{ (1-u) p^2_{\perp} + u k^2_{\perp} }
+ \int  \frac{\dhd^2 k_{\perp}}{ k^2_{\perp} } \Big(1 - e^{-i k_{\perp} b_\perp } \Big)
\nonumber\\
&&\times \Big(1 + \frac{1 }{1 - k^2_\perp/p^2_\perp}\ln \frac{k^2_\perp}{p^2_\perp} \Big) - \frac{1}{8\pi} \Big(\frac{1}{2}\ln \frac{ 4e^{-2\gamma_E} }{p^2_\perp b^2_\perp } + 1 \Big) e^{-i k_\perp z_\perp }
\Big\} \int \frac{dz^-}{2\pi} \mathcal{U}^{[\gamma^+]}_{f;{\rm bg}}( z^-, - z^-, \frac{1}{2}z_\perp)  + \mathcal{O}(x) + \mathcal{O}(\alpha^2_s)\,.
\nonumber
\end{eqnarray}
where the divergent terms,
\begin{eqnarray}
&&\mathcal{I}_c \equiv \frac{g^2 C_F}{2\pi}  \Big(\frac{\mu^2 e^{\gamma_E}}{4\pi}\Big)^\epsilon \Big(\frac{\rho}{xp^+}\Big)^\xi \int d^{2-2\epsilon}z_\perp \int \dhd^{2-2\epsilon} p_{\perp} e^{ip_{\perp} b_\perp }
\nonumber\\
&& \times \int \dhd^{2-2\epsilon} k_{\perp} e^{-i ( p_\perp - k_{\perp}) z_\perp } \int^1_0 \frac{du}{u} \Big( \frac{1 - u}{u} \Big)^{-\xi} \frac{(1-u) p^2_\perp}{( p^2_{\perp} (1-u)  + u k^2_{\perp} )^2 }
\int \frac{dz^-}{2\pi} \mathcal{U}^{[\gamma^+]}_{f;{\rm bg}}( z^-, - z^-, \frac{1}{2}z_\perp)\,,
\end{eqnarray}
arise solely from the diagram in Fig. \ref{fig:fig3}c.
Integrating over $u$ gives
\begin{eqnarray}
&&\mathcal{I}_c = \frac{g^2 C_F }{2\pi } \Big(\frac{\mu^2 e^{\gamma_E}}{4\pi}\Big)^\epsilon \Big(\frac{\rho}{xp^+}\Big)^\xi \Gamma(2-\xi)\Gamma(\xi) \int d^{2 - 2\epsilon}z_\perp \int \dhd^{2 - 2\epsilon} p_{\perp} e^{ip_{\perp} b_\perp }
\\
&&\times \int \dhd^{2 - 2\epsilon} k_{\perp} \frac{1}{p^2_\perp} \Big(\frac{k^2_\perp}{p^2_\perp}\Big)^{-\xi} \Big[ e^{i k_{\perp} z_\perp } \Big(  e^{-i p_\perp z_\perp } -  1 \Big)
+  e^{i k_{\perp} z_\perp } \Big] \int \frac{dz^-}{2\pi} \mathcal{U}^{[\gamma^+]}_{f;{\rm bg}}(z^-, -z^-, \frac{1}{2}z_\perp)\,,
\nonumber
\end{eqnarray}
where the last term contains a divergent transverse-momentum integral.
Performing the $p_\perp$ integration in this term,
\begin{eqnarray}
&&\mathcal{I}_c = \frac{g^2 C_F }{2\pi } \Big(\frac{\mu^2 e^{\gamma_E}}{4\pi}\Big)^\epsilon \Big(\frac{\rho}{xp^+}\Big)^\xi   \Gamma(2-\xi)\Gamma(\xi) \Big[ \int \dhd^{2-2\epsilon} p_{\perp} e^{ip_{\perp} b_\perp } \int \dhd^{2-2\epsilon} k_{\perp} \frac{1}{p^2_\perp} \Big(\frac{k^2_\perp}{p^2_\perp}\Big)^{-\xi} \int d^{2}z_\perp e^{i k_{\perp} z_\perp } \Big(  e^{-i p_\perp z_\perp } -  1 \Big)
\nonumber\\
&&+ \frac{1}{(4\pi)^{1-\epsilon}} \frac{ \Gamma (\xi-\epsilon) }{\Gamma (1-\xi)} \Big(\frac{b^2_\perp}{4}\Big) ^{ \epsilon- \xi} \int \dhd^{2-2\epsilon } k_{\perp} \int d^{2-2\epsilon}z_\perp e^{i k_{\perp} z_\perp } (k^2_\perp)^{-\xi} \Big] \int \frac{dz^-}{2\pi} \mathcal{U}^{[\gamma^+]}_{f;{\rm bg}}(z^-, -z^-, \frac{1}{2}z_\perp)\,,
\label{eq:match-B-exp-int-reg}
\end{eqnarray}
and then expanding first in $\xi$ and then in $\epsilon$ gives
\begin{eqnarray}
&&\mathcal{I}_c
= - \frac{g^2 C_F }{8\pi^2}  \int \dhd^{2 } k_{\perp} \int d^{2}z_\perp e^{i k_{\perp} z_\perp }  \Big[ \frac{1}{\epsilon^2} + \frac{1}{\epsilon}   \Big( \ln\frac{\rho}{xp^+} - \ln\frac{ k^2_\perp}{\mu^2} - 1 \Big) + \frac{1}{\xi} \Big( \frac{1}{\epsilon} + L^\mu_b \Big) - \frac{1}{2}( L^\mu_b)^2
\label{eq:div-term-smallx}\\
&& + L^\mu_b \Big( \ln\frac{\rho}{xp^+} - \ln\frac{ k^2_\perp}{\mu^2}- 1\Big) - \frac{\pi^2}{12} + 4\pi\int \dhd^{2} p_{\perp} e^{ip_{\perp} b_\perp } \frac{1}{p^2_\perp} \ln \frac{k^2_\perp}{p^2_\perp} \Big(  e^{-i p_\perp z_\perp } -  1 \Big) \Big] \int \frac{dz^-}{2\pi} \mathcal{U}^{[\gamma^+]}_{f;{\rm bg}}(z^-, -z^-, \frac{1}{2}z_\perp)\,.
\nonumber
\end{eqnarray}
Equation~(\ref{eq:div-term-smallx}) contains a double-logarithmic dependence on $\sigma'$:
\begin{eqnarray}
&&\sim - \frac{g^2 C_F }{8\pi^2 } L^{\mu}_b \ln \frac{\rho}{xp^+} \int \dhd^{2 } k_{\perp} \int d^{2}z_\perp e^{i k_{\perp} z_\perp } \int \frac{dz^-}{2\pi} \mathcal{U}^{[\gamma^+]}_{f;{\rm bg}}(z^-, -z^-, \frac{1}{2}z_\perp)\,.
\label{eq:devolv}
\end{eqnarray}
Although this double-logarithmic dependence on $\sigma'$ resembles the double-logarithmic dependence on $\sigma$ of the CSS evolution, they have fundamentally different physical origins. While the CSS evolution arise from the UV vacuum physics, those in Eq.~(\ref{eq:div-term-smallx}) originate from the small-$x$ dynamics of the background target.

These divergences must be absorbed by renormalizing of the background-field operators in Eq. (\ref{eq:fig3bare-renormIRform2-rap-rew}) at the $\sigma'$ scales
\begin{eqnarray}
&&\mathcal{W}^{[\gamma^+]}_{f}(b_\perp; \mu', \rho)
= \int d^2 z_\perp Z_W(z_\perp, b_\perp; \mu', \rho) \int \frac{dz^-}{2\pi} \mathcal{U}^{[\gamma^+]}_{f;{\rm bg}}(z^-, -z^-, \frac{1}{2}z_\perp)\,,
\label{eq:tmd-renorm-small-x}
\end{eqnarray}
where the renormalization constant,
\begin{eqnarray}
&&Z_W(z_\perp, b_\perp; \mu', \rho) = \delta^2(z_\perp - b_\perp)
\nonumber\\
&&- \frac{g^2(\mu') C_F }{8\pi^2 } \int \dhd^{2 } k_{\perp} e^{i k_{\perp} z_\perp } \Big[ \frac{1}{\epsilon^2} + \frac{1}{\epsilon} \Big( \ln \frac{\rho}{xp^+} - \ln \frac{ k^2_\perp}{ \mu'^2 }  - 1 \Big) + \frac{1}{\xi} \Big( \frac{1}{\epsilon} + L^{\mu'}_b \Big) \Big] + \mathcal{O}(\alpha^2_s)\,.
\end{eqnarray}
Equation (\ref{eq:tmd-renorm-small-x}) is analogous to Eq. (\ref{eq:col-renorm}) introduced in the previous sections; it effectively defines the factorization between quantum and background field modes at the $\sigma'$ scales according to the high-energy factorization scheme.
At the accuracy considered here, the renormalization constant can be rewritten as
\begin{eqnarray}
&&Z_W(z_\perp, b_\perp; \mu', \rho) = \delta^2(z_\perp - b_\perp)
\nonumber\\
&&- \frac{g^2(\mu) C_F }{8\pi^2 } \Big(\frac{\mu^2}{\mu'^2}\Big)^\epsilon \int \dhd^{2 } k_{\perp} e^{i k_{\perp} z_\perp } \Big[ \frac{1}{\epsilon^2} + \frac{1}{\epsilon} \Big( \ln \frac{\rho}{xp^+} - \ln \frac{ k^2_\perp}{ \mu'^2 }  - 1 \Big) + \frac{1}{\xi} \Big( \frac{1}{\epsilon} + L^{\mu'}_b \Big) \Big] + \mathcal{O}(\alpha^2_s)\,.
\end{eqnarray}
Substituting renormalized background field operator (\ref{eq:tmd-renorm-small-x}) into Eq. (\ref{eq:fig3bare-renormIRform2-rap-rew}), we obtain
\begin{eqnarray}
&&\int \frac{dz^-}{2\pi} e^{-2ixp^+z^-} \mathcal{T}^{[\gamma^+]}_{f}(z^-, -z^-, \frac{1}{2}b_\perp; \mu, \nu) = \mathcal{W}^{[\gamma^+]}_{f}(b_\perp; \mu', \rho) + \frac{g^2 C_F}{8\pi^2} \Big\{ -  \frac{1}{2} (L^\mu_b)^2
\label{eq:fig3bare-renormIRform2-rap-rew-ren}\\
&&+ \Big( \frac{3}{2}  +  \ln\frac{2\mu^2}{\zeta} \Big) L^{\mu}_b + \frac{1}{2}  - \frac{\pi ^2 }{12} \Big\} \mathcal{W}^{[\gamma^+]}_{f}(b_\perp; \mu', \rho) - \frac{g^2 C_F }{8\pi^2}  \int \dhd^{2 } k_{\perp} \int d^{2}z_\perp e^{i k_{\perp} z_\perp }  \Big[  - \frac{1}{2} (L^{\mu'}_b)^2
\nonumber\\
&&  + L^{\mu'}_b \Big( \ln\frac{\rho}{xp^+} - \ln\frac{ k^2_\perp}{\mu'^2}- 1 \Big)
- \frac{\pi^2}{12} + 4\pi\int \dhd^{2} p_{\perp} e^{ip_{\perp} b_\perp } \frac{1}{p^2_\perp} \ln \frac{k^2_\perp}{p^2_\perp} \Big(  e^{-i p_\perp z_\perp } -  1 \Big) \Big] \mathcal{W}^{[\gamma^+]}_{f}(z_\perp; \mu', \rho)
\nonumber\\
&&  + \frac{g^2 C_F}{\pi} \int \dhd^{2} p_{\perp} e^{ip_{\perp} b_\perp }
\int d^{2}z_\perp e^{-i ( p_\perp - k_{\perp}) z_\perp } \Big\{ \int^1_0 du   \int \dhd^{2} k_{\perp}  \frac{1}{ (1-u) p^2_{\perp} + u k^2_{\perp} }
+ \int  \frac{\dhd^2 k_{\perp}}{ k^2_{\perp} } \Big(1 - e^{-i k_{\perp} b_\perp } \Big)
\nonumber\\
&&\times \Big(1 + \frac{1 }{1 - k^2_\perp/p^2_\perp}\ln \frac{k^2_\perp}{p^2_\perp} \Big) - \frac{1}{8\pi} \Big(\frac{1}{2}\ln \frac{ 4e^{-2\gamma_E} }{p^2_\perp b^2_\perp } + 1 \Big) e^{-i k_\perp z_\perp }
\Big\} \mathcal{W}^{[\gamma^+]}_{f}(z_\perp; \mu', \rho)  + \mathcal{O}(x) + \mathcal{O}(\alpha^2_s)\,.
\nonumber
\end{eqnarray}
An analogous result can be straightforwardly derived for rapidity-only UV renormalized Eq.~(\ref{eq:fig3-uv-new-regUV}).

Taking the matrix elements of the operators in Eq.~(\ref{eq:fig3bare-renormIRform2-rap-rew-ren}) and employing the parameterization~(\ref{eq:TMDgplus}) for the TMDPDFs, together with an analogous decomposition of the matrix element,
\begin{eqnarray}
&&\int \frac{dz^-}{2\pi} \langle P, S|   \mathcal{U}^{[\gamma^+]}_{f;{\rm bg}}(z^-, -z^-, \frac{1}{2}b_\perp; \mu', \rho)|P, S\rangle = \mathcal{D}_{f}(b_\perp; \mu', \rho) + i \epsilon^{\mu\nu}_\perp b_\mu s_{T\nu} M \mathcal{H}_{f}(b_\perp; \mu', \rho)\,,
\label{eq:WW-mel}
\end{eqnarray}
where the background-field operator on the left-hand side of the equation is defined according to the high-energy factorization scheme, one can establish a relation between the standard TMDPDFs and the leading-eikonal small-$x$ unintegrated quark distributions. For instance, in the unpolarized case, we readily obtain
\begin{eqnarray}
&&f_{1;f}(x, b_\perp; \mu, \zeta) = \mathcal{D}_{f}(b_\perp; \mu', \rho) + \frac{g^2 C_F}{8\pi^2} \Big\{ -  \frac{1}{2} (L^\mu_b)^2 + \Big( \frac{3}{2}  +  \ln\frac{2\mu^2}{\zeta} \Big) L^{\mu}_b + \frac{1}{2}  - \frac{\pi ^2 }{12} \Big\} \mathcal{D}_{f}(b_\perp; \mu', \rho)
\\
&& - \frac{g^2 C_F }{8\pi^2}  \int \dhd^{2 } k_{\perp} \int d^{2}z_\perp e^{i k_{\perp} z_\perp }  \Big[  - \frac{1}{2} (L^{\mu'}_b)^2  + L^{\mu'}_b \Big( \ln\frac{\rho}{xp^+} - \ln\frac{ k^2_\perp}{\mu'^2}- 1 \Big)
- \frac{\pi^2}{12} + 4\pi\int \dhd^{2} p_{\perp} e^{ip_{\perp} b_\perp } \frac{1}{p^2_\perp} \ln \frac{k^2_\perp}{p^2_\perp}
\nonumber\\
&&  \times \Big(  e^{-i p_\perp z_\perp } -  1 \Big) \Big] \mathcal{D}_{f}(z_\perp; \mu', \rho) + \frac{g^2 C_F}{\pi} \int \dhd^{2} p_{\perp} e^{ip_{\perp} b_\perp }
\int d^{2}z_\perp e^{-i ( p_\perp - k_{\perp}) z_\perp } \Big\{ \int^1_0 du   \int \dhd^{2} k_{\perp}  \frac{1}{ (1-u) p^2_{\perp} + u k^2_{\perp} }
\nonumber\\
&&  + \int  \frac{\dhd^2 k_{\perp}}{ k^2_{\perp} } \Big(1 - e^{-i k_{\perp} b_\perp } \Big) \Big(1 + \frac{1 }{1 - k^2_\perp/p^2_\perp}\ln \frac{k^2_\perp}{p^2_\perp} \Big) - \frac{1}{8\pi} \Big(\frac{1}{2}\ln \frac{ 4e^{-2\gamma_E} }{p^2_\perp b^2_\perp } + 1 \Big) e^{-i k_\perp z_\perp }
\Big\} \mathcal{D}_{f}(z_\perp; \mu', \rho)
+ \mathcal{O}(x) + \mathcal{O}(\alpha^2_s)\,.
\nonumber
\end{eqnarray}
A similar matching equation can be for the Sivers function.

Finally, comparing Eq.~(\ref{eq:fig3bare-renormIRform2-rap-rew-ren}), derived with high-energy factorization imposed at the lower scales $\sigma'$, with the analogous expression obtained using MSTT factorization (\ref{eq:fig3bare-renorm-result-fin}), we find the following relation between the two factorization schemes for TMDPDFs:
\begin{eqnarray}
&&\int \frac{dz^-}{2\pi} e^{-2ixp^+z^-} \mathcal{M}^{[\gamma^+]}_{f}(z^-, -z^-, \frac{1}{2}b_\perp; \mu', \nu') + \frac{g^2 C_F}{\pi}  \int^1_0 du \int d^{2}z_\perp \mathcal{K}(u, z_\perp, b_\perp)
\label{eq:MSTT-he-match}\\
&&\times \int \frac{dz^-}{2\pi} e^{-2ixp^+z^-} \mathcal{M}^{[\gamma^+]}_{f}(uz^-, -uz^-, \frac{1}{2}z_\perp; \mu', \nu') = \mathcal{W}^{[\gamma^+]}_{f}(b_\perp; \mu', \rho) - \frac{g^2 C_F }{8\pi^2}  \int \dhd^{2 } k_{\perp} \int d^{2}z_\perp e^{i k_{\perp} z_\perp }
\nonumber\\
&&\times \Big[  - \frac{1}{2} (L^{\mu'}_b)^2  + L^{\mu'}_b \Big( \ln\frac{\rho}{xp^+} - \ln\frac{ k^2_\perp}{\mu'^2}- 1 \Big)
- \frac{\pi^2}{12} + 4\pi\int \dhd^{2} p_{\perp} e^{ip_{\perp} b_\perp } \frac{1}{p^2_\perp} \ln \frac{k^2_\perp}{p^2_\perp} \Big(  e^{-i p_\perp z_\perp } -  1 \Big) \Big]
\mathcal{W}^{[\gamma^+]}_{f}(z_\perp; \mu', \rho)
\nonumber\\
&&  + \frac{g^2 C_F}{\pi} \int \dhd^{2} p_{\perp} e^{ip_{\perp} b_\perp }
\int d^{2}z_\perp e^{-i ( p_\perp - k_{\perp}) z_\perp } \Big\{ \int^1_0 du   \int \dhd^{2} k_{\perp}  \frac{1}{ (1-u) p^2_{\perp} + u k^2_{\perp} }
+ \int  \frac{\dhd^2 k_{\perp}}{ k^2_{\perp} } \Big(1 - e^{-i k_{\perp} b_\perp } \Big)
\nonumber\\
&&\times \Big(1 + \frac{1 }{1 - k^2_\perp/p^2_\perp}\ln \frac{k^2_\perp}{p^2_\perp} \Big) - \frac{1}{8\pi} \Big(\frac{1}{2}\ln \frac{ 4e^{-2\gamma_E} }{p^2_\perp b^2_\perp } + 1 \Big) e^{-i k_\perp z_\perp }
\Big\} \mathcal{W}^{[\gamma^+]}_{f}(z_\perp; \mu', \rho)  + \mathcal{O}(x) + \mathcal{O}(\alpha^2_s)
\nonumber
\end{eqnarray}
Thus, the leading-eikonal term of MSTT  incorporates the leading-twist eikonal contribution of the high-energy factorization at small $x$.
However, MSTT also partially contains leading-twist sub-eikonal contributions that are suppressed by powers of $x$ in the small-$x$ region.

Eq.~(\ref{eq:MSTT-he-match}) and Eq.~(\ref{eq:MSTT-col-match-tw3}) show MSTT scheme is capable of describing the factorization structure over a broad kinematic range, incorporating not only the leading-twist contribution that dominates the collinear expansion, but also the leading eikonal contribution that governs in the small-$x$ region.

\subsection{Rapidity-only scheme\label{sec:rapidity-only}}
The quantum-background separation at scale $\sigma'$ also can be achieved through rapidity-only cutoff scheme. Similar scheme was employed for the gluon TMDPDF in Ref.~\cite{Balitsky:2015qba}. First, it is useful to address how MSTT and rapidity-only schemes will give rise to two different quantum-background separations at scale $\sigma'$. As discussed before, the quark TMD operator in Eqs.~ (\ref{eq:fig3bare-renorm-result}) and (\ref{eq:fig3-uv-new-regUV}) is infrared finite. However, separation between the quantum and the background modes at the $\sigma'$ scale can still be introduced explicitly splitting the finite perturbative coefficients in Eqs.~(\ref{eq:fig3bare-renorm-result}) and (\ref{eq:fig3-uv-new-regUV}) into contributions associated with the quantum and background fields. In MSTT scheme this is achieved by designating the entire finite perturbative coefficient to the contribution of the quantum modes. In contrast, the rapidity-only scheme splits the finite perturbative coefficient at the cutoff scale $\sigma'$; part of the finite perturbative coefficient above the scale $\sigma'$ (and below $\sigma$) is associated with contributions of the quantum mode and the other part finite perturbative coefficient below the scale $\sigma'$ (and below $\sigma$) is associated with contributions of the background field. Thus, unlike the rapidity-only scheme, MSTT scheme preserves the complete perturbative information obtained in the NLO beam-function calculation.

To implement rapidity-only cutoff scheme at the $\sigma'$ scale, let us introduce the longitudinal momentum variable (\ref{eq:long-var}) in the finite terms of the renormalized result in Eq. (\ref{eq:fig3bare-renormIRform2}). The Fourier transform of Eq. (\ref{eq:fig3bare-renormIRform2}) then gives
\begin{eqnarray}
&&\int \frac{dz^-}{2\pi} e^{-2ixp^+z^-}\mathcal{T}^{[\gamma^+]}_{f}(z^-, -z^-, \frac{1}{2}b_\perp; \mu, \zeta) = \int \frac{dz^-}{2\pi} e^{-2ixp^+z^-}\mathcal{U}^{[\gamma^+]}_{f;{\rm bg}}(z^-, -z^-, \frac{1}{2}b_\perp)
\nonumber\\
&&+ \frac{g^2 C_F}{8\pi^2} \Big\{ -  \frac{1}{2} (L^\mu_b)^2 + \Big( \frac{3}{2}  +  \ln\frac{2\mu^2}{\zeta} \Big) L^{\mu}_b + \frac{1}{2}  - \frac{\pi ^2 }{12} \Big\} \int \frac{dz^-}{2\pi} e^{-2ixp^+z^-}\mathcal{U}^{[\gamma^+]}_{f;{\rm bg}}(z^-, -z^-, \frac{1}{2}b_\perp)
\nonumber\\
&&+ \frac{g^2 C_F}{2\pi} \int_0^\infty \frac{dk^-}{ k^-} \int \dhd^2 p_{\perp} e^{ip_{\perp} b_\perp } \int d^2z_\perp e^{-i (p_{\perp} - k_\perp ) z_\perp }\Big\{ \int \dhd^2 k_{\perp} \frac{1}{ 2xp^+ k^- + p^2_{\perp} } \Big( \frac{1}{k^2_\perp}  \frac{(2xp^+ k^- + k^2_\perp)^2 + (2xp^+ k^-)^2}{2xp^+ k^- + k^2_\perp}
\nonumber\\
&&- \frac{k^2_\perp}{2xp^+ k^- + k^2_\perp} + \frac{p^2_\perp}{ 2xp^+ k^- + p^2_{\perp} } \Big)  \int \frac{dz^-}{2\pi} e^{-2i(x + \frac{k^2_\perp}{2p^+ k^-})p^+z^-} \mathcal{U}^{[\gamma^+]}_{f;{\rm bg}}(z^-, -z^-, \frac{1}{2}z_\perp)
\nonumber\\
&&+ \int \frac{\dhd^2 k_\perp}{k^2_\perp} \Big( \frac{2xp^+k^- }{2xp^+ k^- + k^2_\perp} \Big(1 - e^{i k_\perp b_\perp } \Big) e^{-ik_\perp z_\perp} - \frac{2xp^+k^-}{ 2xp^+ k^- + p^2_{\perp} } e^{-i k_{\perp} b_\perp } \Big)
\nonumber\\
&&\times \frac{(2xp^+ k^- + k^2_\perp)^2 + (2xp^+ k^-)^2}{(2xp^+ k^- + k^2_\perp)^2}  \int \frac{dz^-}{2\pi} e^{-2ixp^+z^-} \mathcal{U}^{[\gamma^+]}_{f;{\rm  bg}}(z^-, -z^-, \frac{1}{2}z_\perp)
\Big\} +\mathcal{O}(\alpha^2_s) \,,
\label{eq:fig3bare-renormIRform2-coff}
\end{eqnarray}
where the operators on the right-hand side are bare beam functions:
\begin{eqnarray}
&&\int \frac{dz^-}{2\pi} e^{-2ix p^+z^-} \mathcal{U}^{[\gamma^+]}_{f;{\rm bg}}(z^-, -z^-, \frac{1}{2}b_\perp)\,.
\label{eq:ronly-inf}
\end{eqnarray}
The rapidity-only scheme is introduced by imposing a cutoff on the integral over the variable $k^-$
\begin{eqnarray}
&&\int_0^\infty \frac{dk^-}{ k^-} = \int_{\sigma'}^\infty \frac{dk^-}{ k^-} + \int_0^{\sigma'} \frac{dk^-}{ k^-}\,.
\label{eq:rap-only-cut-off}
\end{eqnarray}
The resulting two terms should then be associated with the quantum- and background-field modes, respectively. The latter must consequently be absorbed into the renormalization of the bare background-field operators (\ref{eq:ronly-inf}):
\begin{eqnarray}
&&\int \frac{dz^-}{2\pi} e^{-2ixp^+z^-}\mathcal{T}^{[\gamma^+]}_{f}(z^-, -z^-, \frac{1}{2}b_\perp; \mu, \zeta) = \int \frac{dz^-}{2\pi} e^{-2ixp^+z^-}\mathcal{E}^{[\gamma^+]}_{f}(z^-, -z^-, \frac{1}{2}b_\perp;\sigma')
\label{eq:fig3bare-renormIRform2-coff-renorm}\\
&&+ \frac{g^2 C_F}{8\pi^2} \Big\{ -  \frac{1}{2} (L^\mu_b)^2 + \Big( \frac{3}{2}  +  \ln\frac{2\mu^2}{\zeta} \Big) L^{\mu}_b + \frac{1}{2}  - \frac{\pi ^2 }{12} \Big\} \int \frac{dz^-}{2\pi} e^{-2ixp^+z^-}\mathcal{E}^{[\gamma^+]}_{f}(z^-, -z^-, \frac{1}{2}b_\perp;\sigma')
\nonumber\\
&&+ \frac{g^2 C_F}{2\pi} \int_{\sigma'}^\infty \frac{dk^-}{ k^-} \int \dhd^2 p_{\perp} e^{ip_{\perp} b_\perp }\,\int d^2z_\perp e^{-i (p_{\perp} - k_\perp ) z_\perp } \,\Big\{ \int \dhd^2 k_{\perp} \frac{1}{ 2xp^+ k^- + p^2_{\perp} } \Big( \frac{1}{k^2_\perp}  \frac{(2xp^+ k^- + k^2_\perp)^2 + (2xp^+ k^-)^2}{2xp^+ k^- + k^2_\perp}
\nonumber\\
&&- \frac{k^2_\perp}{2xp^+ k^- + k^2_\perp} + \frac{p^2_\perp}{ 2xp^+ k^- + p^2_{\perp} } \Big) \int \frac{dz^-}{2\pi} e^{-2i(x + \frac{k^2_\perp}{2p^+ k^-})p^+z^-} \mathcal{E}^{[\gamma^+]}_{f}(z^-, -z^-, \frac{1}{2}z_\perp;\sigma')
\nonumber\\
&&+ \int \frac{\dhd^2 k_\perp}{k^2_\perp} \Big( \frac{2xp^+k^- }{2xp^+ k^- + k^2_\perp} \Big(1 - e^{i k_\perp b_\perp } \Big) e^{-ik_\perp z_\perp} - \frac{2xp^+k^-}{ 2xp^+ k^- + p^2_{\perp} } e^{-i k_{\perp} b_\perp } \Big)
\nonumber\\
&&\times \frac{(2xp^+ k^- + k^2_\perp)^2 + (2xp^+ k^-)^2}{(2xp^+ k^- + k^2_\perp)^2} \int \frac{dz^-}{2\pi} e^{-2ixp^+z^-} \mathcal{E}^{[\gamma^+]}_{f}(z^-, -z^-, \frac{1}{2}z_\perp;\sigma')
\Big\} +\mathcal{O}(\alpha^2_s) \,,
\nonumber
\end{eqnarray}
where operators on the right-hand side are now defined within the rapidity-only scheme, implemented by introducing the cutoff scale $\sigma'$ for the kinematic variable $k^-$.
Starting from Eq. (\ref{eq:fig3-uv-new-regUV}), one can obtains a similar equation for the rapidity-only UV-renormalized beam function at scale $\sigma$.

To derive the evolution equation with respect to the scale $\sigma'$ for the rapidity-only operators in Eq.~(\ref{eq:fig3bare-renormIRform2-coff-renorm}), we note that the full expression on the right-hand side is independent of the factorization scale $\sigma'$. Differentiating it with respect to $\sigma'$, we obtain, at next-to-leading order accuracy,
\begin{eqnarray}
\label{eq:fig3bare-renormIRform2-coff-renorm-evol}
&&\sigma'\frac{d}{d\sigma'}\int \frac{dz^-}{2\pi} e^{-2ix p^+z^-} \mathcal{E}^{[\gamma^+]}_{f}(z^-, -z^-, \frac{1}{2}b_\perp;\sigma') = \frac{g^2 C_F}{2\pi} \int \dhd^2 p_{\perp} e^{ip_{\perp} b_\perp } \Big\{ \int \dhd^2 k_{\perp} \frac{1}{ 2xp^+ \sigma' + p^2_{\perp} } \Big( \frac{p^2_\perp}{ 2xp^+ \sigma' + p^2_{\perp} }
\nonumber\\
&&+ \frac{1}{k^2_\perp} \frac{(2xp^+ \sigma' + k^2_\perp)^2 + (2xp^+ \sigma')^2}{2xp^+ \sigma' + k^2_\perp} - \frac{k^2_\perp}{2xp^+ \sigma' + k^2_\perp} \Big) \int d^2z_\perp e^{-i (p_\perp - k_{\perp}) z_\perp } \int \frac{dz^-}{2\pi} e^{-2i(x + \frac{k^2_\perp}{2p^+ \sigma'})p^+z^-}
\nonumber\\
&&\times \mathcal{E}^{[\gamma^+]}_{f}(z^-, -z^-, \frac{1}{2}z_\perp;\sigma') + \int \frac{\dhd^2 k_\perp}{k^2_\perp} \Big( \frac{2xp^+\sigma' }{2xp^+ \sigma' + k^2_\perp} \Big(1 - e^{i k_\perp b_\perp } \Big) e^{-ik_\perp z_\perp} - \frac{2xp^+\sigma'}{ 2xp^+ \sigma' + p^2_{\perp} } e^{-i k_{\perp} b_\perp } \Big)
\nonumber\\
&&\times \frac{(2xp^+ \sigma' + k^2_\perp)^2 + (2xp^+ \sigma')^2}{(2xp^+ \sigma' + k^2_\perp)^2} \int d^2z_\perp e^{-i (p_{\perp} - k_\perp ) z_\perp } \int \frac{dz^-}{2\pi} e^{-2ixp^+z^-} \mathcal{E}^{[\gamma^+]}_{f}(z^-, -z^-, \frac{1}{2}z_\perp;\sigma')
\Big\} +\mathcal{O}(\alpha^2_s) \,.
\end{eqnarray}
Unsurprisingly, the evolution in Eq.~\eqref{eq:fig3bare-renormIRform2-coff-renorm-evol} is clearly different from the renormalization evolution equation at the rapidity-only UV scale $\sigma$ in Eq.~(\ref{eq:RFTMDevol}), which is contains the double-logarithmic structure akin to the CSS evolution only.

Let us now show that the evolution of the rapidity-only operators described by Eq.~(\ref{eq:fig3bare-renormIRform2-coff-renorm-evol}) is consistent with both DGLAP and small-$x$ evolution in the corresponding kinematic limits, thereby providing a smooth transition between the two regimes. A similar analysis of rapidity-only gluon TMDPDFs was performed in Refs. \cite{Balitsky:2015qba,Balitsky:2026lop,Cai:2026lsz}.

\subsubsection{ Large transverse momentum/DGLAP limit}
We begin with the small-$b_\perp$ region, $b_\perp \ll \Lambda^{-1}_{\rm QCD}$. In this limit, the rapidity-only evolution equation simplifies to
\begin{eqnarray}
&&\sigma'\frac{d}{d\sigma'}\int \frac{dz^-}{2\pi} e^{-2ix p^+z^-} \mathcal{E}^{[\gamma^+]}_{f}(z^-, -z^-, \frac{1}{2}b_\perp;\sigma')\Big|_{b_\perp\to0} = \frac{g^2 C_F}{2\pi} \int \dhd^2 p_{\perp} \Big\{ \int \dhd^2 k_{\perp} \frac{1}{ 2xp^+ \sigma' + (p+k)^2_{\perp} }
\nonumber\\
&&\times\Big( \frac{(p+k)^2_\perp}{ 2xp^+ \sigma' + (p+k)^2_{\perp} } + \frac{1}{k^2_\perp} \frac{(2xp^+ \sigma' + k^2_\perp)^2 + (2xp^+ \sigma')^2}{2xp^+ \sigma' + k^2_\perp} - \frac{k^2_\perp}{2xp^+ \sigma' + k^2_\perp} \Big) \int d^2z_\perp e^{-i p_\perp z_\perp } \int \frac{dz^-}{2\pi} e^{-2i(x + \frac{k^2_\perp}{2p^+ \sigma'})p^+z^-}
\nonumber\\
&&\times \mathcal{E}^{[\gamma^+]}_{f}(z^-, -z^-, \frac{1}{2}z_\perp;\sigma') - \int \frac{\dhd^2 k_\perp}{k^2_\perp} \frac{2xp^+\sigma'}{ 2xp^+ \sigma' + (p+k)^2_{\perp} } \frac{(2xp^+ \sigma' + k^2_\perp)^2 + (2xp^+ \sigma')^2}{(2xp^+ \sigma' + k^2_\perp)^2} \int d^2z_\perp e^{-i p_{\perp} z_\perp }
\nonumber\\
&&\times \int \frac{dz^-}{2\pi} e^{-2ixp^+z^-}  \mathcal{E}^{[\gamma^+]}_{f}(z^-, -z^-, \frac{1}{2}z_\perp;\sigma')
\Big\} +\mathcal{O}(\alpha^2_s) \,,
\label{eq:fig3bare-renormIRform2-coff-renorm-evol-smx}
\end{eqnarray}
where we shifted the integration variable $p_\perp$.

Equation~\eqref{eq:fig3bare-renormIRform2-coff-renorm-evol-smx} is written for TMD-like operators and therefore contains contributions from all collinear twists. To isolate the leading-twist contribution, and hence the corresponding part of the evolution, we focus on the integration region $k_\perp \gg p_\perp$:
\begin{eqnarray}
&&\sigma'\frac{d}{d\sigma'}\int \frac{dz^-}{2\pi} e^{-2ix p^+z^-} \mathcal{E}^{[\gamma^+]}_{f}(z^-, -z^-, \frac{1}{2}b_\perp;\sigma')\Big|^{\rm lead.tw.}_{b_\perp\to0} = \frac{g^2 C_F}{2\pi} \int \dhd^2 p_{\perp} \int d^2z_\perp e^{-i p_\perp z_\perp }
\nonumber\\
&&\times \int \frac{\dhd^2 k_{\perp}}{k^2_\perp} \frac{(2xp^+ \sigma' + k^2_\perp)^2 + (2xp^+ \sigma')^2}{(2xp^+ \sigma' + k^2_\perp)^2} \Big\{ \int \frac{dz^-}{2\pi} e^{-2i(x + \frac{k^2_\perp}{2p^+ \sigma'})p^+z^-} \mathcal{E}^{[\gamma^+]}_{f}(z^-, -z^-, \frac{1}{2}z_\perp;\sigma')
\nonumber\\
&&- \frac{2xp^+\sigma'}{ 2xp^+ \sigma' + k^2_{\perp} } \int \frac{dz^-}{2\pi} e^{-2ixp^+z^-} \mathcal{E}^{[\gamma^+]}_{f}(z^-, -z^-, \frac{1}{2}z_\perp;\sigma')
\Big\} + \mathcal{O}(\alpha^2_s)\,.
\end{eqnarray}
Introducing the variable $\xi$ in place of the transverse momentum variable $k^2_\perp$,
\begin{eqnarray}
&&\xi \equiv \frac{x}{x + \frac{k^2_\perp}{2 p^+ \sigma'}}\,,
\end{eqnarray}
we obtain
\begin{eqnarray}
&&\sigma'\frac{d}{d\sigma'}\int \frac{dz^-}{2\pi} e^{-2ix p^+z^-} \mathcal{E}^{[\gamma^+]}_{f}(z^-, -z^-, \frac{1}{2}b_\perp;\sigma')\Big|^{\rm lead.tw.}_{b_\perp\to0} 
\nonumber\\
&&= \frac{g^2 C_F}{8\pi^2} \int_0^1 \frac{d\xi}{\xi} \Big[\frac{1 + \xi^2}{1-\xi}\Big]_+ \int \frac{dz^-}{2\pi} e^{-2i\frac{x}{\xi}p^+z^-} \mathcal{E}^{[\gamma^+]}_{f}(z^-, -z^-, z_\perp;\sigma') \Big|_{z_\perp\to0}
+\mathcal{O}(\alpha^2_s)\,,
\end{eqnarray}
which reproduces the DGLAP evolution equation for the leading-twist collinear quark PDF. Thus, the general evolution equation in Eq.~\eqref{eq:fig3bare-renormIRform2-coff-renorm-evol} for the rapidity-only TMD operator $\mathcal{E}^{[\gamma^+]}_{f}$ is consistent with the large-$x$ DGLAP evolution.

\subsubsection{Small-x/eikonal limit}
We next check whether the evolution equation in Eq.~(\ref{eq:fig3bare-renormIRform2-coff-renorm-evol}) reproduces the double-logarithmic evolution in the small-$x$ limit. In this limit, the evolution equation becomes
\begin{eqnarray}
&&\sigma'\frac{d}{d\sigma'}\int \frac{dz^-}{2\pi} e^{-2ix p^+z^-} \mathcal{E}^{[\gamma^+]}_{f}(z^-, -z^-, \frac{1}{2}b_\perp;\sigma')\Big|_{x\to0} = \frac{g^2 C_F}{2\pi} \int \dhd^2 p_{\perp} e^{ip_{\perp} b_\perp } \int \dhd^2 k_{\perp} \frac{1}{  p^2_{\perp} }
\nonumber\\
&&\times \int d^2z_\perp e^{-i (p_\perp - k_{\perp}) z_\perp } \int \frac{dz^-}{2\pi} e^{-2i \frac{k^2_\perp}{2p^+ \sigma'} p^+z^-} \mathcal{E}^{[\gamma^+]}_{f}(z^-, -z^-, \frac{1}{2}z_\perp;\sigma')
+\mathcal{O}(\alpha^2_s) \,.
\label{eq:eik-expansion-ronly}
\end{eqnarray}
Note that this expression corresponds to the diagram shown in Fig.~\ref{fig:fig3}c.

Equation~(\ref{eq:eik-expansion-ronly}) contains contributions from all terms of the eikonal expansion, Eq.~(\ref{eq:eikonal-expansion-Fourier}), to the rapidity-only evolution kernel. To isolate the leading-eikonal contribution, we neglect the exponential factor in the background-field operator and write
\begin{eqnarray}
&&\sigma'\frac{d}{d\sigma'}\int \frac{dz^-}{2\pi} e^{-2ix p^+z^-} \mathcal{E}^{[\gamma^+]}_{f}(z^-, -z^-, \frac{1}{2}b_\perp;\sigma')\Big|^{\rm lead.eik.}_{x\to0} = \frac{g^2 C_F}{2\pi} \int \frac{\dhd^2 p_{\perp}}{  p^2_{\perp} } e^{ip_{\perp} b_\perp }
\nonumber\\
&&\times \int \dhd^2 k_{\perp} \int d^2z_\perp e^{i  k_{\perp} z_\perp } \int \frac{dz^-}{2\pi} \mathcal{E}^{[\gamma^+]}_{f}(z^-, -z^-, \frac{1}{2}z_\perp;\sigma')
+\mathcal{O}(\alpha^2_s) \,,
\label{eq:ev-ronly-smallx}
\end{eqnarray}
where we shifted integration variable $k_\perp \to k_\perp + p_\perp$.

We thus find that the rapidity-only evolution equation (\ref{eq:fig3bare-renormIRform2-coff-renorm-evol}) is consistent with the small-$x$ result in Eq.~(\ref{eq:devolv}), which leads to double-logarithmic evolution of the distributions within high-energy factorization. Indeed, the kernel in Eq.~(\ref{eq:ev-ronly-smallx}) contains an IR-divergent integral over $p_\perp$. In the full rapidity-only evolution equation, Eq.~(\ref{eq:fig3bare-renormIRform2-coff-renorm-evol}), this divergence is regulated at nonzero $x$ by the rapidity cutoff $\sigma'$. Therefore, we can approximately write
\begin{eqnarray}
&&\sigma'\frac{d}{d\sigma'}\int \frac{dz^-}{2\pi} e^{-2ix p^+z^-} \mathcal{E}^{[\gamma^+]}_{f}(z^-, -z^-, \frac{1}{2}b_\perp;\sigma')\Big|^{\rm lead.eik.}_{x\to0}
\nonumber\\
&&\sim \frac{g^2 C_F}{8\pi^2} \ln b^2_\perp\sigma'^2  \int \frac{dz^-}{2\pi} \mathcal{E}^{[\gamma^+]}_{f}(z^-, -z^-, z_\perp;\sigma')\Big|_{z_\perp\to0}
+\mathcal{O}(\alpha^2_s) \,,
\label{eq:ev-ronly-smallx-2}
\end{eqnarray}
which is consistent with Eq.~(\ref{eq:devolv}), taking into account that in the rapidity-only scheme all divergences are regulated by the single factorization scale $\sigma'$, whereas in the high-energy factorization framework two independent scales were introduced. 

\subsubsection{Matching to MSTT scheme}
Finally, comparing Eqs. (\ref{eq:fig3bare-renormIRform2-coff-renorm}) and (\ref{eq:fig3bare-renorm-result-fin}) we can obtain a relation between MSTT and rapidity-only factorization schemes introduced at the lower scales $\sigma'$ for TMDPDFs:
\begin{eqnarray}
&&\int \frac{dz^-}{2\pi} e^{-2ixp^+z^-}\mathcal{M}^{[\gamma^+]}_{f}(z^-, -z^-, \frac{1}{2}b_\perp; \mu', \nu' ) + \frac{g^2 C_F}{\pi}  \int^1_0 du \int d^{2}z_\perp \int \frac{dz^-}{2\pi} e^{-2ixp^+z^-} \mathcal{K}(u, z_\perp, b_\perp)
\\
&&\times \mathcal{M}^{[\gamma^+]}_{f}(uz^-, -uz^-, \frac{1}{2}z_\perp; \mu', \nu' ) = \int \frac{dz^-}{2\pi} e^{-2ixp^+z^-}\mathcal{E}^{[\gamma^+]}_{f}(z^-, -z^-, \frac{1}{2}b_\perp;\sigma')
\nonumber\\
&&+ \frac{g^2 C_F}{2\pi} \int_{\sigma'}^\infty \frac{dk^-}{ k^-} \int \dhd^2 p_{\perp} e^{ip_{\perp} b_\perp } \Big\{ \int \dhd^2 k_{\perp} \frac{1}{ 2xp^+ k^- + p^2_{\perp} } \Big( \frac{1}{k^2_\perp}  \frac{(2xp^+ k^- + k^2_\perp)^2 + (2xp^+ k^-)^2}{2xp^+ k^- + k^2_\perp}
\nonumber\\
&&- \frac{k^2_\perp}{2xp^+ k^- + k^2_\perp} + \frac{p^2_\perp}{ 2xp^+ k^- + p^2_{\perp} } \Big) \int d^2z_\perp e^{-i (p_\perp - k_{\perp}) z_\perp } \int \frac{dz^-}{2\pi} e^{-2i(x + \frac{k^2_\perp}{2p^+ k^-})p^+z^-} \mathcal{E}^{[\gamma^+]}_{f}(z^-, -z^-, \frac{1}{2}z_\perp;\sigma')
\nonumber\\
&&+ \int \frac{\dhd^2 k_\perp}{k^2_\perp} \Big( \frac{2xp^+k^- }{2xp^+ k^- + k^2_\perp} \Big(1 - e^{i k_\perp b_\perp } \Big) e^{-ik_\perp z_\perp} - \frac{2xp^+k^-}{ 2xp^+ k^- + p^2_{\perp} } e^{-i k_{\perp} b_\perp } \Big)
\nonumber\\
&&\times \frac{(2xp^+ k^- + k^2_\perp)^2 + (2xp^+ k^-)^2}{(2xp^+ k^- + k^2_\perp)^2} \int d^2z_\perp e^{-i (p_{\perp} - k_\perp ) z_\perp } \int \frac{dz^-}{2\pi} e^{-2ixp^+z^-} \mathcal{E}^{[\gamma^+]}_{f}(z^-, -z^-, \frac{1}{2}z_\perp;\sigma')
\Big\} +\mathcal{O}(\alpha^2_s) \,.
\nonumber
\end{eqnarray}
\section{Conclusions}
We have analyzed the perturbative structure of the leading-power quark TMDPDF at next-to-leading order within the background-field formulation of QCD. Our aim was to construct a common description of its large- and small-$x$ dynamics while retaining the longitudinal and transverse momenta of the background fields without imposing collinear or eikonal ordering.

The background-field formulation separates two logically distinct mode boundaries. The upper scale set $\sigma$ defines and renormalizes the TMD operator and separates the modes assigned to the hard coefficient from the quantum modes. The lower scale set $\sigma'$ separates the perturbative quantum modes from the non-perturbative background modes and also specifies the renormalization of the lower-scale background-field operators. Integrating over the quantum modes between these scale sets leads to the operator relation in Eq.~(\ref{eq:fact-op}), while allowing the upper and lower schemes to be chosen independently.

At the upper scales, the standard TMD scheme gives the usual Collins-Soper-Sterman (CSS) evolution. We also formulated an upper rapidity-only scheme in which a single longitudinal cutoff regulates the ultraviolet divergences associated with the transverse-momentum and rapidity dependence of the TMD operator, reducing the upper-scale evolution to a single equation. An explicit next-to-leading-order calculation of the soft factor in this scheme shows that its contribution is already contained in the renormalized beam function. Consequently, introducing an additional soft factor would double count the soft contribution at this order.

At the lower scales, we implemented the background-field construction developed for gluon TMDPDFs in Ref.~\cite{Mukherjee:2023snp} for the quark case and referred to it as the MSTT scheme. The lower-scale operators are leading-power bilocal quark TMD operators built from background fields with unrestricted ordering between the longitudinal and transverse momenta. Starting from the upper-renormalized beam function, the finite perturbative remainder is assigned to the coefficient functions, while terms associated with the lower-scale separation are assigned to the renormalization of the background-field operators. For quark TMDPDFs at NLO, this lower-scale renormalization is trivial because the unexpanded TMDPDF operator contains no genuine infrared divergence. This contrasts with the gluon case, where an infrared rapidity divergence produces nontrivial lower-scale evolution related to BFKL dynamics.

To determine the perturbative content of this construction, we retained the leading term in Eq.~(\ref{eq:fact-op}), namely the term whose lower-scale operator is the leading-power bilocal quark TMD operator. Compared with the collinear and high-energy expansions, this term reproduces the DGLAP logarithm of collinear matching and the double-logarithmic contribution governing the quark TMDPDF at small $x$ in their respective limits. At finite $b_\perp$ and arbitrary $x$, it also retains portions of the higher-twist collinear and sub-eikonal structure generated by the leading-power TMD operator.

We further introduced a lower rapidity-only scheme by applying a single longitudinal cutoff to the finite perturbative terms of the upper-renormalized result. This cutoff defines a different separation between the quantum and background fields and should not be confused with the upper rapidity-only renormalization. The resulting lower-scale evolution reduces to DGLAP evolution in the collinear limit and to the characteristic double-logarithmic quark evolution at small $x$. Matching the lower rapidity-only and MSTT schemes shows that they describe the same underlying operator dynamics, while distributing the perturbative information differently between the coefficient functions and the background-field operators.

Thus, the MSTT scheme, combined with standard TMD renormalization, provides a unified leading-power description of quark TMDPDFs: CSS evolution governs the upper-scale renormalization, while the leading
operator term reproduces the DGLAP logarithms of collinear matching and the small-$x$ double-logarithmic dynamics in their respective limits.
\section{Acknowledgments}
This material is based upon work supported by the U.S. Department of Energy, Office of Science, Office of Nuclear Physics through Contract Nos.~DE-SC0012704 and DE-SC0020081, and within the framework of Saturated Glue (SURGE) Topical Collaboration in Nuclear Theory.
A.T. thanks the Aspen Center for Physics, supported by National Science Foundation grant PHY-2210452, where part of this work was performed.
\bibliography{main}
\end{document}